\documentclass[a4paper,11pt]{article}
\usepackage{iftex}
\ifpTeX
  \PassOptionsToPackage{dvipdfmx}{geometry}
  \PassOptionsToPackage{dvipdfmx}{graphicx}
  \PassOptionsToPackage{dvipdfmx}{xcolor}
  \PassOptionsToPackage{dvipdfmx}{hyperref}
\fi
\usepackage[margin=25mm]{geometry}
\usepackage{amsmath,amssymb,booktabs,array}
\usepackage{graphicx,placeins}
\usepackage{ifpdf}
\ifpdf
  \usepackage{epstopdf}
\fi
\DeclareGraphicsExtensions{.eps}
\usepackage[T1]{fontenc}
\usepackage{newtxtext,newtxmath}
\usepackage{xcolor}
\usepackage[hidelinks]{hyperref}
\usepackage{xurl}
\numberwithin{equation}{section}
\newif\ifmarkedrevision
\markedrevisionfalse
\definecolor{RetainedBlue}{RGB}{0,70,160}
\colorlet{RevisionColor}{black}
\ifmarkedrevision\colorlet{OriginalColor}{RetainedBlue}\else\colorlet{OriginalColor}{black}\fi
\newenvironment{revisionred}{\begingroup\color{RevisionColor}}{\par\endgroup}
\newcommand{\RevisionAddition}[1]{{\color{RevisionColor}#1}}
\newcommand{\RefEntryColor}[1]{\ifcsname revisedref@#1\endcsname\color{RevisionColor}\else\color{OriginalColor}\fi}
\newcommand{\RefDOI}[2]{\begingroup
\ifcsname revisedref@#1\endcsname\color{RevisionColor}\else
\ifcsname originaldoi@#1\endcsname\color{OriginalColor}\else\color{RevisionColor}\fi\fi
\unskip\newline\url{https://doi.org/#2}\endgroup}
\expandafter\def\csname revisedref@bachmayr2025variationally\endcsname{}
\expandafter\def\csname revisedref@cassia2025godunov\endcsname{}
\expandafter\def\csname revisedref@harten1983upstream\endcsname{}
\expandafter\def\csname revisedref@jagtap2020conservative\endcsname{}
\expandafter\def\csname revisedref@patel2022thermodynamically\endcsname{}
\expandafter\def\csname revisedref@pytorchautograd\endcsname{}
\expandafter\def\csname revisedref@pytorchgrad\endcsname{}
\expandafter\def\csname originaldoi@chaumet2022improving\endcsname{}
\expandafter\def\csname originaldoi@dafermos2016hyperbolic\endcsname{}
\expandafter\def\csname originaldoi@jagtap2022physics\endcsname{}
\expandafter\def\csname originaldoi@mizuno2026physics\endcsname{}
\expandafter\def\csname originaldoi@wassing2024physics\endcsname{}
\expandafter\def\csname originaldoi@liu2024discontinuity\endcsname{}
\expandafter\def\csname originaldoi@ferrer2024gradient\endcsname{}

\expandafter\def\csname revisedref@laxwendroff1960\endcsname{1}
\expandafter\def\csname revisedref@pytorchreproducibility\endcsname{1}

\hypersetup{pdftitle={Data-Free Finite-Volume Convolutional Neural Networks for Steady Compressible Euler Flow around a Circular Cylinder},pdfauthor={Ryosuke Yano}}
\title{\RevisionAddition{Data-Free Finite-Volume Convolutional Neural Networks for Steady Compressible Euler Flow around a Circular Cylinder}}
\author{Ryosuke Yano\\[5pt]
{\small Tokio Marine dR Co., Ltd.}\\
{\small 1-5-1, Otemachi, Chiyoda-ku, Tokyo Japan.}\\
{\small Tel.: +81-3-5288-6234.}\\
{\small\href{mailto:ryosuke.yano@tokio-dr.co.jp}{ryosuke.yano@tokio-dr.co.jp}}}
\date{}
\begin{document}
\color{OriginalColor}
\maketitle
\begin{abstract}
\begin{revisionred}
A data-free finite-volume convolutional neural network (FV-CNN) is formulated for steady Euler flow around a cylinder at eight Mach numbers, 2--15. Shared-face fluxes define conservative spatial balances, while automatic differentiation supplies parameter gradients. The derivative mismatch in the earlier automatic-differentiation-based convolutional neural network (AD-CNN) is derived. No computational fluid dynamics (CFD) or experimental field targets enter training; reference-independent selection of coefficients, seeds, and checkpoints is not established. Billig-based upstream fixing conditions shock-location agreement.

For eight adopted fields, density and pressure relative $L_2$ discrepancies from finite-resolution CFD are $5.43$--$18.69\%$ and $6.25$--$19.70\%$. Closed-band defects are smaller than AD-CNN in 32/32 self-centered and 22--31/32 CFD-centered comparisons over seven half-widths, 0.06--0.12 cylinder diameters. AD-CNN retains smaller enthalpy root-mean-square (RMS) discrepancies in 7/8 cases, shock-position RMS at Mach 11 and 15, and temperature and transverse-velocity discrepancies at Mach 3.

Ten new Mach-15 initializations and adopted seed 42 yield density discrepancies of $12.04$--$43.70\%$ (mean $24.34\%$, sample standard deviation $11.78$ percentage points); the new ten alone give $25.04\%$ and $12.17$ points. Dispersion exceeds the historical adopted-field gap of $3.40$ points; six of eleven outcomes are below AD-CNN. Transition-width ratios of AD-CNN to FV-CNN remain $4.07$--$4.66$, and all 44 self-centered seed--component balances at half-width 0.10 cylinder diameters are smaller for FV-CNN. Local sharpness and this balance metric persist, while initialization-robust accuracy superiority is not established. Component interventions remain one-seed conditional responses; Mach 2 uses a separate budget-limited protocol.
\end{revisionred}
\end{abstract}

\providecommand{\FVADReference}{\ref{sec:fv_ad_failure}}
\section{Introduction}
\label{sec:introduction}

\begin{revisionred}
Physics-informed neural networks (PINNs) approximate solutions of partial differential equations (PDEs) by optimizing a neural representation against governing equations and prescribed information~\cite{raissi2019physics}. For compressible Euler flows, a detached bow shock presents coupled difficulties: the solution has a discontinuity, the shock position is unknown, and the compressed flow must satisfy the wall and stagnation constraints. The conservation equations must hold in the weak sense across the shock~\cite{dafermos2016hyperbolic}. A small pointwise training loss, or a visually sharp transition, is therefore insufficient on its own to establish the physical accuracy of a learned flow field. Approximation properties, the spatial residual operator, and optimization must be considered separately.
\end{revisionred}

\begin{revisionred}
In this context, the observed low-frequency preference commonly termed \emph{spectral bias} \cite{rahaman2019spectral} is one possible difficulty for a standard Multi-Layer Perceptron (MLP) near a sharp shock. It does not by itself establish the cause of smearing or optimization failure in the present nonlinear PINN calculations.
\end{revisionred}

\begin{revisionred}
Previous studies have addressed compressible-flow learning through several complementary formulations. Mao et al.~\cite{mao2020physics} considered forward and inverse problems for one- and two-dimensional Euler flows. Wassing et al.~\cite{wassing2024physics} introduced adaptive artificial-viscosity reduction for stationary parametric Euler problems, while Mizuno et al.~\cite{mizuno2026physics} examined the thermodynamic consistency of output variables and the equation of state in one-dimensional shock and two-dimensional oblique-shock calculations. Jagtap et al.~\cite{jagtap2022physics} studied inverse supersonic problems using additional flow information. These works involve different combinations of boundary information, reference observations, and regularization; their results should be interpreted within those respective settings. Liu et al.~\cite{liu2024discontinuity} combine equation weighting with jump and global conservation constraints. Ferrer-S\'{a}nchez et al.~\cite{ferrer2024gradient} use gradient-dependent residual weighting for relativistic Riemann problems.
\end{revisionred}

\begin{revisionred}
Conservation-aware neural formulations also provide an established alternative to relying solely on strong-form collocation. Conservative PINNs impose flux continuity across decomposed subdomains~\cite{jagtap2020conservative}; control-volume PINNs use integrated balances and permit thermodynamic constraints~\cite{patel2022thermodynamically}; and weak PINNs address discontinuous solutions through weak residual norms~\cite{chaumet2022improving}. More recently, Cassia and Kerswell~\cite{cassia2025godunov} incorporated approximate-Riemann-solver fluxes into a finite-volume loss. These developments motivate treating the discretization of conservation separately from the choice of neural architecture. The present work examines their combination for steady cylinder bow shocks over a broad Mach-number range.
\end{revisionred}

\begin{revisionred}
A methodological distinction becomes essential when a convolutional neural network (CNN) predicts an entire grid field. In a pointwise coordinate network, each output depends on its own coordinate input. Convolutions instead couple neighboring grid inputs and outputs. Consequently, differentiating the sum of these coupled outputs with respect to the input coordinates generally gives a column sum of the input Jacobian, rather than the required local spatial derivative. The earlier automatic-differentiation-based convolutional neural network (AD-CNN) implementation used this sum-based derivative construction. Section~\FVADReference{} identifies the mismatch mathematically. This is a property of that residual implementation: automatic differentiation (AD) remains valid for the computational graph it differentiates, and is also used here to obtain gradients with respect to network parameters.
\end{revisionred}

\begin{revisionred}
A finite-volume (FV) convolutional neural network (FV-CNN) is therefore formulated, with the predicted field and its spatial residual treated as distinct components. A body-fitted grid supplies dual control volumes, and a single numerical flux on each shared face enters the adjacent balances with opposite signs. The flux uses a Harten--Lax--van Leer (HLL)-type construction~\cite{harten1983upstream}, with the executed dissipation and energy-jump modifications specified explicitly in the methodology. Shock regularization is supplied by this numerical flux; the former pressure-gradient artificial-viscosity residual is not used in finite-volume mode. Shared-face cancellation is an algebraic conservation property of the discrete operator. The remaining balance error depends on optimization, so this property alone does not establish a converged steady Euler solution or entropy admissibility. Table~\ref{tab:pinn_summary_extended} compares the related formulations and their reported problem classes with the present FV-CNN.
\begin{table}[!p]
\centering\color{RevisionColor}\small
\caption{Qualitative comparison of neural formulations for shock-containing compressible flows and conservation laws. The rows describe the cited methods and problem classes; numerical accuracy and timing are not ranked across different benchmarks. The present row describes the FV-CNN formulation used here.}
\label{tab:pinn_summary_extended}
\medskip
\setlength{\tabcolsep}{3pt}
\renewcommand{\arraystretch}{1.15}
\begin{tabular}{@{}>{\raggedright\arraybackslash}p{2.25cm}>{\raggedright\arraybackslash}p{3.65cm}>{\raggedright\arraybackslash}p{4.6cm}>{\raggedright\arraybackslash}p{4.8cm}@{}}
\toprule
Study & Spatial formulation & Prescribed information or additional constraints & Reported problem class\\
\midrule
Mao et al.~\cite{mao2020physics} & Euler residual collocation & Initial/boundary states; additional observations for inverse problems & One- and two-dimensional Euler flows, including shock-tube and oblique-shock problems\\
\addlinespace[3pt]
Jagtap et al.~\cite{jagtap2022physics} & Euler residuals, including domain-decomposed networks & Density-gradient and boundary observations; entropy and positivity constraints & Inverse supersonic flows: expansion, oblique shocks, and detached bow shocks\\
\addlinespace[3pt]
Liu et al.~\cite{liu2024discontinuity} & Equation-weighted differential residuals & Rankine--Hugoniot jump and global conservation constraints & Burgers equation and one- and two-dimensional Euler discontinuities\\
\addlinespace[3pt]
Wassing et al.~\cite{wassing2024physics} & Euler residuals with adaptive artificial viscosity & Boundary data; viscosity-reduction and training schedules & Parametric stationary two-dimensional subsonic and supersonic Euler flows\\
\addlinespace[3pt]
Ferrer-S\'{a}nchez et al.~\cite{ferrer2024gradient} & Gradient-dependent residual weighting & Initial/boundary data; reduced residual weighting near discontinuities & Riemann problems in relativistic hydrodynamics\\
\addlinespace[3pt]
Chaumet and Giesselmann~\cite{chaumet2022improving} & Weak residual norms with learned test functions & Initial/boundary and entropy conditions & Scalar conservation laws and the Euler Sod shock tube\\
\addlinespace[3pt]
Mizuno et al.~\cite{mizuno2026physics} & Euler residuals with explicit equation-of-state treatment & Initial/boundary states; thermodynamic output-variable choices & One-dimensional shock tube and two-dimensional oblique shock\\
\midrule
Jagtap et al.~\cite{jagtap2020conservative} & Domain-decomposed residuals with interface flux continuity & Subdomain interface constraints; forward/inverse problem data & Nonlinear conservation laws, including compressible Euler systems\\
\addlinespace[3pt]
Patel et al.~\cite{patel2022thermodynamically} & Space--time control-volume balances & Entropy and variation control; thermodynamic data in inverse examples & Hyperbolic systems and shock hydrodynamics\\
\addlinespace[3pt]
Cassia and Kerswell~\cite{cassia2025godunov} & Finite-volume loss with Godunov-type intercell fluxes & Time-stepping and super-resolution reconstruction protocols & Two-dimensional Euler Riemann problems\\
\midrule
Present \mbox{FV-CNN} & Grid-based convolutional representation; shared-face finite-volume residual & No CFD field targets in the objective; boundary data and explicit fixing, stagnation, enthalpy, and entropy priors & Steady cylinder bow shocks at Mach 2--15; Mach 2 has a distinct training protocol\\
\bottomrule
\end{tabular}
\end{table}

\end{revisionred}

\begin{revisionred}
The network retains radial and azimuthal convolutional couplings on the body-fitted grid. Geometry-adapted convolutional representations have precedent in physics-informed learning~\cite{gao2021phygeonet}; here the directional couplings provide a representation whose role can be tested without changing the finite-volume residual. Fourier encoding is included as a separate representational component, motivated by its ability to enrich high-frequency coordinate features~\cite{tancik2020fourier}. Its contribution and those of the two convolutional directions are assessed through explicit ablations. The formulation consequently permits separate examination of the conservation operator and the spatial couplings within the network.
\end{revisionred}

\begin{revisionred}
Training uses freestream and boundary conditions together with stated physical priors: an upstream fixing region, stagnation-point relations, total-enthalpy and nose constraints, and density total-variation (TV) regularization. For the recorded continuation stages, an upstream entropy penalty suppresses departures from the freestream entropy proxy ahead of the shock. This local upstream prior is distinct from enforcing a global entropy inequality. In this paper, \emph{data-free} means that no computational fluid dynamics (CFD) or experimental reference values enter the loss or gradient updates. The upstream fixing boundary is prescribed from a Mach-dependent shock-detachment estimate based on Billig's correlation~\cite{billig1967shock}; it is not fitted to the present CFD shock position. The loss coefficients are prescribed empirically and their sensitivity is evaluated explicitly. The analytical and geometric information entering the objective is specified explicitly. Mach-number-guided weights act on the squared momentum and energy residuals. They address component imbalance empirically, in the broader context of gradient imbalance in PINNs~\cite{wang2021understanding}, without assuming that they determine the physical solution or guarantee favorable conditioning.
\end{revisionred}

\begin{revisionred}
The main computations comprise independently trained cases at $Ma_\infty=2,3,5,7,9,11,13,15$. Their recorded conditions are collected in a single table. The resulting FV-CNN fields are compared with CFD using contour overlays, stagnation-line profiles, and quantitative discrepancies; earlier AD-CNN fields are identified separately in the comparisons. Temperature profiles are normalized by the freestream temperature. Operator-consistent Euler residuals and closed-band flux-balance diagnostics examine both non-shock regions and finite-width shock bands, with the evaluated regions shown explicitly. Closed-boundary physical-flux balances provide a finite-width weak-conservation diagnostic across the curved shock. This evaluation distinguishes agreement with a finite-resolution CFD reference from satisfaction of conservation, and avoids equating either with a low weighted training loss.
\end{revisionred}

\begin{revisionred}
The main flow-field evaluation is separated from the component studies. A $\mathrm{Ma}=15$ sensitivity study starts from a common saved initialization and compares each factor with its baseline on the stagnation streamline. It tests numerical dissipation, TV regularization, residual and prior weights, the fixing region, Fourier encoding, and directional convolution. The pointwise FV-trained variant removes both spatial couplings while retaining the remaining network construction. It supplies the within-family control for neighbor coupling, with the accompanying change in active weights stated explicitly. Scaling-on/off comparisons at $\mathrm{Ma}=2,3,15$ are reported separately and interpreted according to their recorded settings. These initial-stage tests are distinguished from the continued-training fields used for the main validation. Their purpose is to establish the sensitivity of the executed configuration, rather than to infer universal necessity from a single-seed comparison.
\end{revisionred}

\begin{revisionred}
The contribution is thus a specified and testable combination of shared-face FV residuals, a directional convolutional representation, and explicit physical priors for this steady bow-shock benchmark. The numerical study separates reference-field accuracy, conservation diagnostics, and component sensitivity. Its conclusions concern the tested grids, training procedures, and Mach numbers; they do not assume a general advantage over conventional CFD or establish an architecture ranking independently of the residual formulation.
\end{revisionred}

\paragraph{\textcolor{OriginalColor}{Organization of the paper.}}
\begin{revisionred}
The methodology first defines the Euler benchmark, explains the earlier spatial-derivative mismatch, and specifies the FV residual, network, objective, and training stages. The results then present the main comparisons across Mach number and the complementary conservation diagnostics. Separate component studies examine sensitivity and Mach scaling. Optimization histories and computational cost are interpreted alongside physical accuracy, followed by a discussion of scope and the conclusions.
\end{revisionred}

\begin{revisionred}
\providecommand{\llbracket}{\mathopen{[\![}}
\providecommand{\rrbracket}{\mathclose{]\!]}}
\begingroup\color{OriginalColor}
\section{Methodology}
\label{sec:fv_methodology}
\endgroup

The network representation and the spatial residual operator are specified separately. A convolutional network predicts a flow field on a fixed body-fitted grid, and a finite-volume (FV) operator evaluates the balance of numerical fluxes on shared faces. Automatic differentiation (AD) differentiates this discrete training objective with respect to the network parameters. It is not used as a substitute for spatial differentiation of the coupled grid outputs. FV-CNN denotes this combination of a convolutional representation and an FV residual. The previous AD-CNN denotes specifically the residual implementation analyzed below. Initial training uses adaptive moment estimation with decoupled weight decay (AdamW); limited-memory Broyden--Fletcher--Goldfarb--Shanno (L-BFGS) refinement is used only where recorded.

\subsection{Euler benchmark and weak conservation}
\label{sec:fv_governing}

Let $\Omega\subset\mathbb R^2$ denote the fluid domain outside a circular cylinder of radius $R_{\rm cyl}=0.5$ and diameter $D=2R_{\rm cyl}=1$. All lengths below are nondimensionalized by the cylinder diameter. The Cartesian coordinates are $\mathbf{x}=(x,y)$, the velocity is $\mathbf{u}=(u,v)$, and the ratio of specific heats is $\gamma=1.4$. The conserved variables and calorically perfect-gas closure are
\begin{equation}
 U=(\rho,\rho u,\rho v,E)^T,\qquad
 E=\frac{p}{\gamma-1}+\frac{\rho}{2}(u^2+v^2),\qquad T=\frac{p}{\rho}.
 \label{eq:fv_state}
\end{equation}
Here $\rho$, $p$, $T$, and $E$ denote density, pressure, temperature, and total energy per unit volume; superscript $T$ denotes transpose. Subscript $\infty$ denotes the freestream. The nondimensional freestream state is
\begin{equation}
 (\rho_\infty,u_\infty,v_\infty,p_\infty,T_\infty)
 =(1,Ma_\infty,0,\gamma^{-1},\gamma^{-1}).
 \label{eq:fv_freestream}
\end{equation}
Thus the freestream sound speed $c_\infty$ is unity and $Ma_\infty=u_\infty/c_\infty$ is the freestream Mach number. The local sound speed and Mach number are $c=\sqrt{\gamma p/\rho}$ and $M=\sqrt{u^2+v^2}/c$. The steady Euler system is
\begin{align}
 &\partial_xF_x(U)+\partial_yF_y(U)=0, \label{eq:fv_euler}\\
 &F_x(U)=(\rho u,\rho u^2+p,\rho uv,(E+p)u)^T,\nonumber\\
 &F_y(U)=(\rho v,\rho uv,\rho v^2+p,(E+p)v)^T. \nonumber
\end{align}
The ideal-gas model defines a numerical benchmark, including at $Ma_\infty=15$. Mach number alone does not determine the onset of real-gas effects; the model does not describe vibrational excitation, dissociation, or thermochemical nonequilibrium.

For a scalar test function $\varphi\in C_c^1(\Omega)$ (continuously differentiable with compact support in $\Omega$), the weak steady conservation law is the componentwise identity
\begin{equation}
 \int_\Omega [F_x(U)\partial_x\varphi+
              F_y(U)\partial_y\varphi]\,d\mathbf{x}=0.
 \label{eq:fv_weak}
\end{equation}
For a piecewise smooth state separated by a stationary shock surface $\Sigma$, orient its unit normal $\mathbf n$ from the minus to the plus side. The distributional divergence is
\begin{equation}
 \nabla\cdot F(U)
 =(\nabla\cdot F(U))_{\rm reg}
 +\llbracket F_n(U)\rrbracket\delta_\Sigma,\qquad
 F_n=n_xF_x+n_yF_y.
 \label{eq:fv_distribution}
\end{equation}
Here $F=(F_x,F_y)$ collects the two Cartesian fluxes, subscript $\mathrm{reg}$ denotes the regular divergence away from the shock, $\llbracket a\rrbracket=a_+-a_-$, and $\delta_\Sigma$ is the surface Dirac distribution. A steady weak solution therefore requires the Rankine--Hugoniot condition
\begin{equation}
 \llbracket F_n(U)\rrbracket=0.
 \label{eq:fv_RH}
\end{equation}
Small strong-form residuals on the two smooth sides alone do not establish Eq.~\eqref{eq:fv_RH}. Physical admissibility also requires an entropy inequality, expressed using physical specific entropy $s$ as $\nabla\cdot(\rho s\mathbf u)\geq0$ in the distributional sense~\cite{dafermos2016hyperbolic}. Discrete flux conservation, convergence to a weak solution, and entropy admissibility are distinct properties. The classical Lax--Wendroff result concerns conservative, consistent approximations under convergence assumptions~\cite{laxwendroff1960}; it does not supply a convergence theorem for the optimized neural fields considered here.

\subsection{Spatial coupling and the previous AD residual}
\label{sec:fv_ad_failure}

For a pointwise coordinate network, an output component at node $j$ has the form $q_j=q_\vartheta(\mathbf{x}_j)$, where $\vartheta$ denotes the trainable parameters. It depends only on its own node coordinates. In the convolutional network,
\begin{equation}
 q_j=\mathcal N_{\vartheta,j}(X),\qquad
 X=(\mathbf{x}_1,\ldots,\mathbf{x}_N),\qquad
 J^x_{ji}=\frac{\partial q_j}{\partial x_i}.
 \label{eq:fv_input_jacobian}
\end{equation}
Here $N$ is the number of grid nodes, $X$ collects their coordinates, and $i,j$ are flattened node indices in this subsection. The matrix $J^x$ generally has off-diagonal entries because convolution couples spatial nodes.

The previous implementation used
\texttt{autograd.grad(q,X,grad\_outputs=ones\_like(q))}.
For a scalar output component, the returned $x$ component at input node $i$ is
\begin{equation}
 g_i^x=\frac{\partial}{\partial x_i}\sum_jq_j(X)
      =\sum_jJ^x_{ji}
      =[(J^x)^T\mathbf1]_i.
 \label{eq:fv_vjp}
\end{equation}
Here $\mathbf1$ is the all-ones vector. This is the vector--Jacobian product specified by the AD operation~\cite{pytorchgrad}; the computation graph is differentiated correctly. For a pointwise network,
$J^x_{ji}=\delta_{ji}\partial_xq_\vartheta(\mathbf{x}_i)$,
where $\delta_{ji}$ is the Kronecker delta, and Eq.~\eqref{eq:fv_vjp} gives the desired derivative. Spatial coupling removes this identity. Applied to the two Cartesian flux arrays, the previous procedure yields, for each conservation equation,
\begin{equation}
 \widetilde R_i^{\rm AD}
 =\sum_j\frac{\partial F_{x,j}(X)}{\partial x_i}
  +\sum_j\frac{\partial F_{y,j}(X)}{\partial y_i}.
 \label{eq:fv_old_residual}
\end{equation}
This is a sensitivity of sums of flux predictions to an input coordinate. It is not, in general, the spatial divergence of the predicted field on the fixed grid.

For example, on a locally uniform one-dimensional grid of spacing $h$, let
\begin{equation}
 q_i=a\phi(x_{i-1})+b\phi(x_i)+c\phi(x_{i+1}),\qquad \phi(x)=x^2.
 \label{eq:fv_counterexample}
\end{equation}
Away from array boundaries, Eq.~\eqref{eq:fv_vjp} gives $g_i=2(a+b+c)x_i$. The constants $a,b,c$ are stencil coefficients. The natural translated-stencil reconstruction
$q_h(x)=a\phi(x-h)+b\phi(x)+c\phi(x+h)$ instead gives
\begin{equation}
 q_h'(x_i)=2(a+b+c)x_i+2h(c-a).
 \label{eq:fv_counterexample_difference}
\end{equation}
The example demonstrates the absence of a general identity, not a discrepancy for every field or every convolution. Extracting only the diagonal of $J^x$ also does not by itself define a physical-space derivative for an arbitrary coupled representation. A continuous reconstruction or a spatial discrete operator must be specified.

The issue concerns the previous residual implementation, not an inability of AD to solve Euler problems. AD can differentiate a correctly defined pointwise representation, a continuous local reconstruction, or a conservative discrete residual. In FV-CNN, it is retained for parameter optimization:
\begin{equation}
 \nabla_\vartheta\mathcal L
 =\left(\frac{\partial U_\vartheta}{\partial\vartheta}\right)^T
   \nabla_U\mathcal L.
 \label{eq:fv_parameter_AD}
\end{equation}
Here $U_\vartheta$ is the decoded network state and $\mathcal L$ is its training objective. The parameter derivative in Eq.~\eqref{eq:fv_parameter_AD} is distinct from the input-coordinate derivative in Eq.~\eqref{eq:fv_input_jacobian}.

For a spatially variable artificial-viscosity coefficient $\nu(\mathbf x)$, with $\Delta$ denoting the Laplacian,
\begin{equation}
 \nabla\cdot(\nu\nabla U)=\nu\Delta U+\nabla\nu\cdot\nabla U.
 \label{eq:fv_variable_viscosity}
\end{equation}
Dropping the second term does not retain conservative diffusion. The previous index-space regularizer supplies no weak-conservation guarantee. In FV mode its coefficient is set to zero; dissipation enters through shared face fluxes as defined below.

\subsection{Body-fitted grid and convolutional representation}
\label{sec:fv_network}

Figure~\ref{fig:fig1}(a) summarizes the field representation, Fig.~\ref{fig:fig1}(b) expands one of the six hybrid blocks, and Fig.~\ref{fig:fig1}(c) shows how the predicted field enters the FV training objective. The convolutional layers mix grid-node features; the control-volume flux operator evaluates spatial conservation. Backpropagation differentiates the resulting discrete objective with respect to the trainable parameters. These roles are kept separate in both the diagram and the implementation.
\begin{figure}[!htbp]
\centering
\color{RevisionColor}
\includegraphics[width=\linewidth]{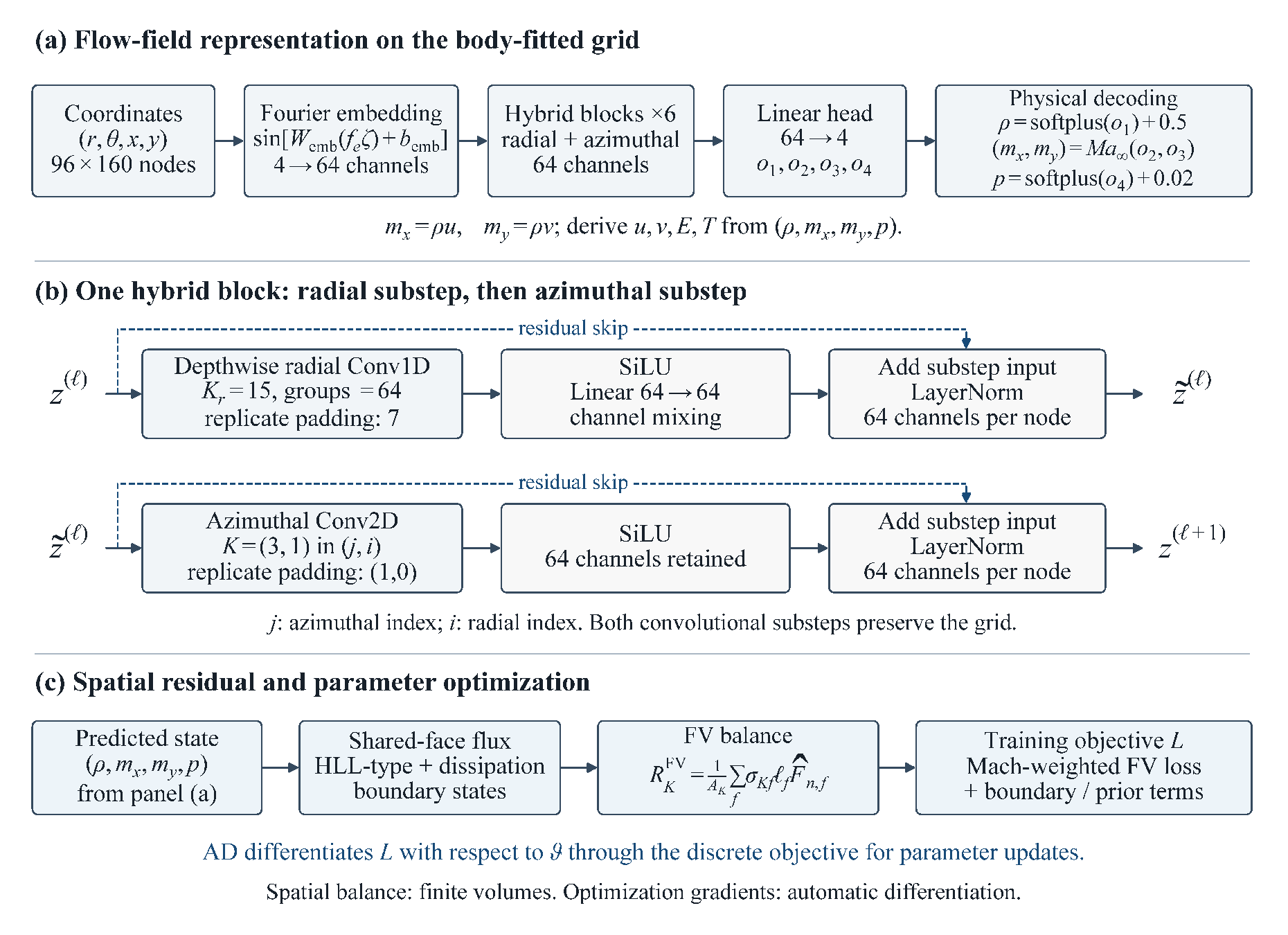}
\caption{FV-CNN architecture and residual evaluation. (a) Nodewise coordinates are Fourier-encoded into 64 channels, passed through six hybrid blocks, and mapped by one linear head to four raw outputs. Decoding gives density, two Cartesian momenta, and pressure; velocity, energy, and temperature are derived. (b) A depthwise radial convolution, the sigmoid linear unit (SiLU), and channel mixing precede a residual addition and layer normalization (LN, labeled LayerNorm); an azimuthal convolution and SiLU precede a second residual addition and LayerNorm. The kernel sizes are $K_r=15$ and $K_\theta=(3,1)$; both convolutions use replicate padding. The indices $(j,i)$ denote azimuthal and radial directions. (c) Shared-face fluxes define the FV residual and training objective: $A_K$ is cell area, $\ell_f$ is face length, and $\sigma_{Kf}$ accounts for orientation. AD gives parameter gradients. Flux modifications, residual weights and physical priors are specified in the text.}
\label{fig:fig1}
\end{figure}

Let $j=0,\ldots,N_\theta-1$ and $i=0,\ldots,N_r-1$ denote azimuthal and radial indices, with $(N_\theta,N_r)=(96,160)$. Set $\xi_i=i/(N_r-1)$ and $\eta_j=j/(N_\theta-1)$. The coordinate map is
\begin{align}
 \theta_j&=\frac{\pi}{2}+\frac{\pi}{2}\eta_j,\nonumber\\
 R_{\rm out}(\theta)&=R_{\rm side}[1-\sin(\theta-\pi/2)]
                  +R_{\rm nose}\sin(\theta-\pi/2),\nonumber\\
 r_{ji}&=(1-\xi_i)R_{\rm cyl}+\xi_iR_{\rm out}(\theta_j),\nonumber\\
 \mathbf x_{ji}&=(r_{ji}\cos\theta_j,r_{ji}\sin\theta_j).
 \label{eq:fv_grid_map}
\end{align}
Here $R_{\rm side}=R_{\rm out}(\pi/2)$ is the side-boundary radius and $R_{\rm nose}=R_{\rm out}(\pi)=R_{\rm cyl}+d$, where $d$ is the prescribed upstream domain extent from the cylinder nose, not an imposed shock position. The $\mathrm{Ma}=2$ comparison uses $(d,R_{\rm side})=(1.5,4)$, whereas the standard higher-Mach grid uses $(0.6,2)$. Equal array sizes therefore do not imply equal physical grids. The wall, outer boundary, symmetry line, and outflow edge correspond to $i=0$, $i=N_r-1$, $j=N_\theta-1$, and $j=0$, respectively.

In Fig.~\ref{fig:fig1}(a), each node has input features $\zeta_{ji}=(r_{ji},\theta_j,x_{ji},y_{ji})^T$. The Fourier-enabled encoding is
\begin{equation}
 z^{(0)}_{ji}
 =\sin\!\left(W_{\rm emb}[f_e\zeta_{ji}]+b_{\rm emb}\right)
 \in\mathbb R^{64}.
 \label{eq:fv_embedding}
\end{equation}
The sine acts componentwise, $f_e$ is a prescribed frequency scale at AdamW update $e$, and $W_{\rm emb},b_{\rm emb}$ are trainable. The frequency scale multiplies the features before the affine map, not its bias. The affine embedding uses the standard \texttt{Linear(4,64)} initialization: each weight and bias is sampled uniformly on $[-1/2,1/2]$. It is trainable rather than a frozen random-frequency matrix.

The six repeated blocks in Fig.~\ref{fig:fig1}(b) apply a depthwise radial convolution of width 15, channel mixing, and a three-point azimuthal convolution, with layer normalization (LN) after each substep:
\begin{align}
 \widetilde z^{(\ell)}_{ji}
 &=\operatorname{LN}\!\left[z^{(\ell)}_{ji}
   +W_r^{(\ell)}\phi((C_r^{(\ell)}z^{(\ell)})_{ji})
   +b_r^{(\ell)}\right],\nonumber\\
 z^{(\ell+1)}_{ji}
 &=\operatorname{LN}\!\left[\widetilde z^{(\ell)}_{ji}
   +\phi((C_\theta^{(\ell)}\widetilde z^{(\ell)})_{ji})\right].
 \label{eq:fv_blocks}
\end{align}
Here $z^{(\ell)}$ is the 64-channel block input, $\widetilde z^{(\ell)}$ is its radial-substep output, and $\ell=0,\ldots,5$ indexes the blocks. $W_r^{(\ell)}$ and $b_r^{(\ell)}$ are the trainable channel-mixing matrix and bias. The activation $\phi(a)=a/(1+\exp(-a))$ is the sigmoid linear unit (SiLU), and $\operatorname{LN}$ normalizes the channels at each node. Convolution biases are included in $C_r,C_\theta$. The radial operator acts separately on each channel before $W_r$ mixes channels. The azimuthal operator mixes channels and has tensor kernel size $(3,1)$ in the $(j,i)$ directions. Both use replicate padding. They encode neighbor relations and are not the Euler differential operators.

After the sixth block, the pointwise linear head in Fig.~\ref{fig:fig1}(a) maps 64 channels to four raw outputs $o_{ji}$, decoded as
\begin{align}
 \rho_{ji}&=\operatorname{softplus}(o_{ji,1})+0.5,\nonumber\\
 (\rho u)_{ji}&=Ma_\infty o_{ji,2},\qquad
 (\rho v)_{ji}=Ma_\infty o_{ji,3},\nonumber\\
 p_{ji}&=\operatorname{softplus}(o_{ji,4})+0.02.
 \label{eq:fv_output}
\end{align}
Here $\operatorname{softplus}(a)=\log(1+\exp a)$. The offsets impose the explicit priors $\rho>0.5$ and $p>0.02$, rather than positivity alone. Near either offset, the derivative of softplus with respect to its raw channel becomes small; saturation can therefore accompany an optimization failure. Temperature and energy follow from Eq.~\eqref{eq:fv_state}, rather than being independent outputs. The two middle raw channels predict the Cartesian momenta $m_x=\rho u$ and $m_y=\rho v$, rather than velocities. Velocities are recovered by division by density. Thus the output stage consists of one affine map and physical decoding, with no additional multilayer decoder. The decoded state supplies the shared-face fluxes in Fig.~\ref{fig:fig1}(c).

In the architectural tests, removing radial or azimuthal coupling retains only the respective central spatial tap. Removing both leaves Fourier encoding, channel transformations, residual connections, and nodewise LayerNorm. This pointwise residual network is the FV-trained, no-neighbor control used in this study. Its effective operations at each node are affine channel maps, activations, residual connections, and channel normalization. The parameter tensor shapes are preserved, although the number of active spatial weights changes; the comparison tests the removal of connections within this network family. The Fourier-free configuration replaces Eq.~\eqref{eq:fv_embedding} by $W_{\rm emb}\zeta_{ji}+b_{\rm emb}$, removing both the sine nonlinearity and its prescribed frequency schedule.

\FloatBarrier

\subsection{Shared-face finite-volume residual}
\label{sec:fv_discretization}

\subsubsection{Dual control volumes}

A polygonal dual control volume $K_{ji}$ is associated with each node. An interior dual corner is
\begin{equation}
 \mathbf c_{j+1/2,i+1/2}
 =\tfrac14(\mathbf x_{ji}+\mathbf x_{j+1,i}
                 +\mathbf x_{j,i+1}+\mathbf x_{j+1,i+1}).
 \label{eq:fv_corners}
\end{equation}
Boundary corners use adjacent boundary-node midpoints; the domain corners remain at their boundary nodes. The volumes stop on a polygonal approximation to the physical boundaries and have half widths in boundary-normal grid directions. Their areas $A_K=|K|$ are computed by the polygon shoelace formula.

Every shared face has one straight edge, one length $\ell_f>0$, and one chosen unit normal $\mathbf n_f$, used by both neighboring volumes. Geometry is constructed in float64 before casting to field precision. The predictions are nodal states associated with these volumes, rather than analytically evaluated cell averages.

\subsubsection{Numerical flux and dissipation}

For a face of oriented unit normal $\mathbf n=(n_x,n_y)$, let $U_L,U_R$ be its left and right states. In the first-order configuration described here, interior face states are the adjacent predictions. Define
\begin{align}
 v_{n,L/R}&=u_{L/R}n_x+v_{L/R}n_y,\qquad
 c_{L/R}=\sqrt{\gamma p_{L/R}/\rho_{L/R}},\nonumber\\
 S_L&=\min(v_{n,L}-c_L,v_{n,R}-c_R),\nonumber\\
 S_R&=\max(v_{n,L}+c_L,v_{n,R}+c_R).
 \label{eq:fv_signal_speeds}
\end{align}
The quantities $v_{n,L/R}$ are normal velocities, $c_{L/R}$ are sound speeds, and $S_L,S_R$ bound the left- and right-going wave speeds. The base HLL flux uses a two-wave construction~\cite{harten1983upstream}. An optional enthalpy-based modification changes its dissipative energy jump. For a switch $\kappa\in\{0,1\}$, define
\begin{equation}
 \Delta Q_\kappa=(\Delta\rho,\Delta(\rho u),\Delta(\rho v),
                       \Delta E+\kappa\Delta p)^T,\qquad
 \Delta a=a_R-a_L.
 \label{eq:fv_modified_jump}
\end{equation}
The flux evaluated before extra dissipation is
\begin{equation}
 \widehat F^{(\kappa)}_{\rm HLL}=
 \begin{cases}
 F_n(U_L),& S_L\geq0,\\[2pt]
 \displaystyle\frac{S_RF_n(U_L)-S_LF_n(U_R)+S_LS_R\Delta Q_\kappa}
                   {S_R-S_L},&S_L<0<S_R,\\[7pt]
 F_n(U_R),&S_R\leq0.
 \end{cases}
 \label{eq:fv_hll}
\end{equation}
The code safeguards the denominator using $\max(S_R-S_L,10^{-6})$. For $\kappa=0$ this is standard HLL. For $\kappa=1$ the dissipative energy jump becomes $\Delta(E+p)=\Delta(\rho H)$, with $H=(E+p)/\rho$. This modification is not identified with an unchanged standard HLL solver.

With dissipation multiplier $\alpha(e)\geq1$, the final face flux is
\begin{align}
 \widehat F_n&=\widehat F^{(\kappa)}_{\rm HLL}
        -\frac{\alpha(e)-1}{2}a_{\max}\Delta Q_\kappa,\nonumber\\
 a_{\max}&=\max(|v_{n,L}|+c_L,\ |v_{n,R}|+c_R).
 \label{eq:fv_extra_dissipation}
\end{align}
The correction applies to the selected flux branch, without the old AD pressure-gradient sensor. Its prescribed schedule is
\begin{equation}
 \alpha(e)=\alpha_{\rm f}
     \left(\frac{\alpha_0}{\alpha_{\rm f}}\right)^{1-\chi_e},
 \qquad \chi_e=\min(e/N_{\rm ann},1).
 \label{eq:fv_alpha_schedule}
\end{equation}
Here $\alpha_0$ and $\alpha_{\rm f}$ are the initial and terminal multipliers, $e$ is the zero-based AdamW update index, $N_{\rm ann}$ is the annealing duration in updates, and $\chi_e$ is the clipped progress fraction. Setting $\alpha_0=\alpha_{\rm f}$ keeps dissipation fixed, as in the $\mathrm{Ma}=15$ controlled sensitivity cases.

For $\kappa=1$ and equal exact total enthalpy $H_L=H_R=H_*$, the physical energy fluxes and $\Delta(E+p)$ are $H_*$ times their mass counterparts. Thus
\begin{equation}
 H_L=H_R=H_*,\ \kappa=1
 \ \Longrightarrow\ [\widehat F_n]_E=H_*[\widehat F_n]_\rho.
 \label{eq:fv_enthalpy_flux_property}
\end{equation}
This property applies to the HLL branches and extra dissipation. It is a local consistency property, not a proof of global enthalpy accuracy or entropy stability.

\subsubsection{Discrete balance and boundary fluxes}

Let $\sigma_{Kf}=+1$ if the chosen normal $\mathbf n_f$ points out of $K$, and $-1$ otherwise. The residual is
\begin{equation}
 R_K^{\rm FV}=\frac1{A_K}
       \sum_{f\subset\partial K}\sigma_{Kf}\ell_f\widehat F_{n,f}.
 \label{eq:fv_residual}
\end{equation}
The radial and azimuthal face arrays are each formed once, and their adjacent differences give its four components. For any collection $\mathcal S$ of volumes,
\begin{equation}
 \sum_{K\in\mathcal S}A_KR_K^{\rm FV}
 =\sum_{f\subset\partial(\cup_{K\in\mathcal S}K)}
                 \sigma_{\mathcal S f}\ell_f\widehat F_{n,f}.
 \label{eq:fv_telescoping}
\end{equation}
Here $\sigma_{\mathcal S f}$ orients the face normal outward from the union of volumes, as $\sigma_{Kf}$ does for a single volume. Every internal face cancels algebraically, before or after training, including for $\kappa=1$ and $\alpha>1$. Equal left and right states give the physical flux. Equation~\eqref{eq:fv_telescoping} does not assert that optimized residuals vanish.

At the wall, the primitive interior state is the componentwise average
\begin{equation}
 P_I^w=\tfrac12(P_{j,0}+P_{j,1}),\qquad P=(\rho,u,v,p).
 \label{eq:fv_wall_average}
\end{equation}
At the symmetry face it is $P_I^s=\tfrac12(P_{N_\theta-1,i}+P_{N_\theta-2,i})$. The ghost state is
\begin{equation}
 \rho_G=\rho_I,\quad p_G=p_I,\qquad
 \mathbf u_G=\mathbf u_I
       -2(\mathbf u_I\cdot\mathbf n)\mathbf n.
 \label{eq:fv_reflection}
\end{equation}
Subscripts $I,G$ denote interior and ghost states; superscripts $w,s$ identify wall and symmetry faces. The left/right ordering follows the chosen face orientation. These midpoint states are part of the implemented half-width boundary-volume treatment; they are not replaced by reflection of the boundary node alone.

At the outer and outflow boundaries the ghost state is copied from the adjacent boundary prediction. The outer freestream target is enforced by the inflow loss below, rather than inserted as an exact external state into the copied flux. Outflow uses a zero-gradient closure. At $\theta=\pi/2$, its outward normal is $(1,0)$ and the normal Mach number is $M_n=u/\sqrt{\gamma p/\rho}$. In the adopted $\mathrm{Ma}=2$ FV-CNN field, 76 of the 158 non-endpoint outflow nodes have $0<M_n<1$ and the minimum is 0.525; the $\mathrm{Ma}=3$ minimum is 1.234. The $\mathrm{Ma}=2$ copied-state boundary therefore cannot be justified as an everywhere-supersonic outflow. Its possible boundary influence is retained as a limitation of that distinct protocol. Additional wall and symmetry penalties act on predicted velocities.

\subsection{Training loss and physical priors}
\label{sec:fv_loss}

\subsubsection{Node sets and residual weights}

For the boundary-inclusive ghost configuration, define
\begin{align}
 \mathcal C_P&=\{(j,i):0\leq j<N_\theta,\ 0\leq i<N_r-1\},\nonumber\\
 \mathcal C_F&=\{(j,i):0\leq j<N_\theta-1,\ 0\leq i<N_r-1\},\nonumber\\
 \mathcal C_N&=\{(j,i)\in\mathcal C_F:\theta_j>0.9\pi\},\nonumber\\
 \mathcal C_{\rm in}&=\{(j,N_r-1)\},\quad
 \mathcal C_w=\{(j,0)\},\quad
 \mathcal C_s=\{(N_\theta-1,i)\},\nonumber\\
 \mathcal C_m&=\{(j,i):x_{ji}<x_{\rm fix}\}.
 \label{eq:fv_node_sets}
\end{align}
Free indices in the boundary sets range over their complete radial or azimuthal array. The notation $|\mathcal C|$ denotes the number of nodes in a set, and sums over $\mathcal C$ run over its associated nodal volumes. Thus $|\mathcal C_P|=15264$ and $|\mathcal C_F|=15105$. The nose set selects an angular sector; the fixing set selects a prescribed upstream region. Its boundary $x_{\rm fix}$ is based on Billig's empirical cylinder correlation~\cite{billig1967shock}. With cylinder radius $R=D/2$ and nose at $x=-D/2$, the estimate and the upstream allowance implied by each executed boundary are
\begin{equation}
 \frac{\Delta_B}{R}=0.386\exp\!\left(\frac{4.67}{Ma_\infty^2}\right),\qquad
 \frac{x_{\rm fix}}{D}=-\frac12-\frac{\Delta_B}{D}-b.
 \label{eq:fv_billig_fixing}
\end{equation}
Here $\Delta_B$ is the estimated axis shock stand-off distance from the cylinder nose and $b$ is a dimensionless allowance beyond the estimated shock, not part of Billig's correlation. Table~\ref{tab:fv_main_settings} gives both the executed boundaries and their implied allowances. The latter are approximately $0.08$ at $\mathrm{Ma}\geq5$ and $0.18$ at $\mathrm{Ma}=2,3$; these approximate descriptions are not an exact uniform-buffer generation rule. The larger low-Mach allowances move the constrained region farther upstream of the estimated shock. The historical reason for choosing those larger allowances was not retained. This prior locates the freestream-penalty region; $x_{\rm fix}$ is neither a CFD-fitted shock coordinate nor the shock position measured from the learned field.

The PDE weight $\omega^P_K$ is unity at ordinary interior nodes and equals the recorded coefficient $\beta_b$ at wall/outflow rows. On the symmetry line excluding wall and outer corners, it equals a separate coefficient $\beta_s$. The symmetry--wall corner retains $\beta_b$, and outer nodes are excluded from $\mathcal C_P$. Auxiliary weights $\omega^F_K,\omega^N_K$ equal $\beta_b$ at wall/outflow rows and unity elsewhere in their sets. These are penalty weights, not areas.

Writing $R_K^{\rm FV}=(R_{\rho,K},R_{x,K},R_{y,K},R_{E,K})^T$ for mass, $x$-momentum, $y$-momentum, and energy residuals, respectively, set
\begin{align}
 \Phi_K&=R_{\rho,K}^2+w_m(R_{x,K}^2+R_{y,K}^2)+w_ER_{E,K}^2,\nonumber\\
 (w_m,w_E)&=
 \begin{cases}
 (Ma_\infty^{2},Ma_\infty^{4}),&Ma_\infty=2,\\
 (Ma_\infty^{-2},Ma_\infty^{-4}),&Ma_\infty\geq3
                                     \text{ in the tested cases}.
 \end{cases}
 \label{eq:fv_mach_weights}
\end{align}
Unscaled controls use $(w_m,w_E)=(1,1)$. These weights multiply squared residuals. The corresponding unsquared residual is multiplied by
$D_{Ma}=\operatorname{diag}(1,\sqrt{w_m},\sqrt{w_m},\sqrt{w_E})$.
The tested rules are empirical component balancing. A dimensional comparison follows from the freestream physical-flux scales in Eq.~\eqref{eq:res_flux_scales}. Normalizing each residual by its component flux scale and multiplying the resulting squared norm by the mass scale squared would give
\begin{equation}
 \widetilde w_m=\left(\frac{Ma_\infty}{Ma_\infty^2+\gamma^{-1}}\right)^2,
 \qquad
 \widetilde w_E=\left(\frac{1}{(\gamma-1)^{-1}+Ma_\infty^2/2}\right)^2.
 \label{eq:fv_flux_balancing_rationale}
\end{equation}
These are a comparison rule, not the executed weights. Their high-Mach limits are $Ma_\infty^{-2}$ and $4Ma_\infty^{-4}$; the latter retains a factor four relative to the implemented energy rule. At $\mathrm{Ma}=3$ they are $(0.0954,0.0204)$, and at $\mathrm{Ma}=15$ $(0.00442,7.56\times10^{-5})$. Thus freestream scales motivate the orders of the higher-Mach balancing, but do not derive its exact coefficients. The $\mathrm{Ma}=2$ choice $(4,16)$ follows a separate historical protocol and is opposite to the corresponding flux-normalized pair $(0.180,0.0494)$. No common continuous Mach-weight law is inferred from these case-specific calculations. Shock-jump magnitudes alone do not establish a Hessian condition-number bound.

The PDE and nose losses are
\begin{equation}
 \mathcal L_P=\frac1{|\mathcal C_P|}\sum_{\mathcal C_P}\omega^P_K\Phi_K,
 \qquad
 \mathcal L_N=\frac1{|\mathcal C_N|}\sum_{\mathcal C_N}\omega^N_K\Phi_K.
 \label{eq:fv_pde_nose_loss}
\end{equation}
The logged symmetry-line FV diagnostic is already represented in $\mathcal L_P$ and is not added as another independent loss.

The joint-objective configurations use
\begin{align}
 \mathcal L={}&\lambda_P\mathcal L_P+\lambda_H\mathcal L_H
  +\lambda_N\mathcal L_N+\lambda_{\rm in}\mathcal L_{\rm in}
  +\lambda_w\mathcal L_w\nonumber\\
 &+\lambda_0\mathcal L_0+\lambda_s\mathcal L_s+\lambda_m\mathcal L_m
  +\lambda_{\rm TV}\mathcal L_{\rm TV}+\lambda_{\rm ent}\mathcal L_{\rm ent}.
 \label{eq:fv_total_loss}
\end{align}
Coefficients are those executed for the particular case and training stage; zero removes a contribution. Empty fixing or upstream-entropy sets contribute zero.

Although Eq.~\eqref{eq:fv_residual} divides by physical volume area, Eq.~\eqref{eq:fv_pde_nose_loss} averages squared residuals over nodes without area quadrature. It is not a physical-area $L^2(\Omega)$ residual norm, a certified variational residual norm, or an a priori solution-error estimator. In particular, no residual-to-solution-error equivalence of the type sought in variationally correct residual formulations is established here~\cite{bachmayr2025variationally}. Conservative flux balances address the spatial operator; they do not by themselves settle residual-norm or optimization questions.

For a precise comparison with area averaging, let $\Phi_h=\Phi_K$ on each polygonal control volume $K\in\mathcal C_P$, and let $\Omega_P^h$ be their union. The corresponding piecewise-constant physical-area average is
\begin{equation}
 \frac{1}{|\Omega_P^h|}\int_{\Omega_P^h}\Phi_h\,dA
 =\frac{\sum_{K\in\mathcal C_P}A_K\Phi_K}
        {\sum_{K\in\mathcal C_P}A_K}.
 \label{eq:fv_area_residual_average}
\end{equation}
This equality defines the area average of the reconstructed squared residual; it is not the implemented node mean with penalty weights in Eq.~\eqref{eq:fv_pde_nose_loss}. The areas follow from the grid geometry and polygon construction in Eqs.~\eqref{eq:fv_grid_map} and \eqref{eq:fv_corners}. Neither choice alone yields a stability estimate relating the training loss to the error of a discontinuous Euler solution.

The effect of Mach weighting on parameter derivatives can also be stated without a shock-amplitude assumption. Let $\mathbf r(\vartheta)$ stack the four unweighted FV residual components over $\mathcal C_P$, and let $W$ be block diagonal with blocks $(\omega_K^P/|\mathcal C_P|)D_{Ma}^TD_{Ma}$. Thus $\mathcal L_P=\mathbf r^TW\mathbf r$. With fixed case weights, on a differentiable branch of the discrete residual,
\begin{align}
 \nabla_\vartheta\mathcal L_P
 &=2J_\vartheta^TW\mathbf r,
 \qquad J_\vartheta=\frac{\partial\mathbf r}{\partial\vartheta},
 \label{eq:fv_weighted_gradient}\\
 \nabla_\vartheta^2\mathcal L_P
 &=2J_\vartheta^TWJ_\vartheta
   +2\sum_a(W\mathbf r)_a\nabla_\vartheta^2r_a.
 \label{eq:fv_weighted_hessian}
\end{align}
Here $a$ indexes entries of the stacked residual vector $\mathbf r$. The first Hessian term is the Gauss--Newton matrix. A common scalar multiplication of all residuals changes its norm but not its condition number when it is positive definite; component weighting can alter its spectrum. The magnitude of $\mathbf r$ or a shock jump does not determine $J_\vartheta$. No gradient-spectrum or Hessian-spectrum measurements are reported, so improved conditioning is not established. The scaling evidence below concerns the resulting physical profiles under the recorded interventions.

\subsubsection{Boundary, enthalpy, and density-variation penalties}

Define the primitive freestream discrepancy
\begin{equation}
 \Psi_K=(\rho_K-1)^2+(u_K-Ma_\infty)^2+v_K^2
                                      +(p_K-\gamma^{-1})^2.
 \label{eq:fv_freestream_penalty}
\end{equation}
The boundary and fixing losses are
\begin{align}
 \mathcal L_{\rm in}&=\frac1{|\mathcal C_{\rm in}|}
                           \sum_{\mathcal C_{\rm in}}\Psi_K,\qquad
 \mathcal L_m=\frac1{|\mathcal C_m|}\sum_{\mathcal C_m}\Psi_K,\nonumber\\
 \mathcal L_w&=\frac1{|\mathcal C_w|}
                    \sum_{\mathcal C_w}(\mathbf u_K\cdot\mathbf n_w)^2,
 \qquad
 \mathcal L_s=\frac1{|\mathcal C_s|}\sum_{\mathcal C_s}v_K^2.
 \label{eq:fv_boundary_losses}
\end{align}
Here $\mathbf n_w$ is a cylinder-normal unit vector. The symmetry penalty imposes $v=0$; even-symmetry derivatives of $\rho,u,p$ are not separate losses.

For $\epsilon_\rho=10^{-6}$, the implemented enthalpy quantity and target are
\begin{equation}
 H_{\epsilon,K}=\frac{E_K+p_K}{\rho_K+\epsilon_\rho},\qquad
 H_\infty=\frac1{\gamma-1}+\frac{Ma_\infty^2}{2}.
 \label{eq:fv_enthalpy_quantities}
\end{equation}
The enthalpy and azimuthal density-variation losses are
\begin{align}
 \mathcal L_H&=\frac1{|\mathcal C_F|}
          \sum_{\mathcal C_F}\omega^F_K(H_{\epsilon,K}-H_\infty)^2,\nonumber\\
 \mathcal L_{\rm TV}
 &=\frac1{(N_\theta-1)N_r}
       \sum_{j=0}^{N_\theta-2}\sum_{i=0}^{N_r-1}
                       |\rho_{j+1,i}-\rho_{j,i}|.
 \label{eq:fv_enthalpy_TV_losses}
\end{align}
The exact $H$ in Eq.~\eqref{eq:fv_enthalpy_flux_property} is distinct from the safeguarded $H_\epsilon$. The TV term is an index-space regularizer, without division by angular spacing or a physical Jacobian. Its coefficient can differ between stages.

\subsubsection{Stagnation-point prior}

A normal shock followed by isentropic deceleration gives the stagnation target. Let subscript 2 denote the post-normal-shock state:
\begin{align}
 \rho_2&=\frac{(\gamma+1)Ma_\infty^2}{(\gamma-1)Ma_\infty^2+2},\nonumber\\
 p_2&=p_\infty\frac{2\gamma Ma_\infty^2-(\gamma-1)}{\gamma+1},\nonumber\\
 M_2^2&=\frac{(\gamma-1)Ma_\infty^2+2}{2\gamma Ma_\infty^2-(\gamma-1)},
 \qquad B_2=1+\frac{\gamma-1}{2}M_2^2,\nonumber\\
 p_0&=p_2B_2^{\gamma/(\gamma-1)},\qquad
 \rho_0=\rho_2B_2^{1/(\gamma-1)}.
 \label{eq:fv_stagnation_targets}
\end{align}
At the wall--symmetry node $K_0=(N_\theta-1,0)$, the relative formulation is
\begin{equation}
 \mathcal L_0=
 \left[\max\!\left(\frac{|\rho_{K_0}-\rho_0|}{\rho_0}-\tau_0,0\right)\right]^2
 +\left[\max\!\left(\frac{|p_{K_0}-p_0|}{p_0}-\tau_0,0\right)\right]^2.
 \label{eq:fv_stagnation_loss}
\end{equation}
Here $M_2$ is the post-normal-shock Mach number, $B_2$ is the isentropic deceleration factor, and subscript 0 denotes the stagnation target. The recorded $\tau_0$ is a relative dead band. With absolute normalization, each term is instead $\max(|q-q_0|-\tau_0q_0,0)^2$, where $(q,q_0)$ is either $(\rho_{K_0},\rho_0)$ or $(p_{K_0},p_0)$.

\subsubsection{Upstream entropy continuation}

Where entropy cleanup is used, it penalizes a departure from the freestream entropy proxy
\begin{equation}
 Z_K=\frac{p_K}{p_\infty}
     \left(\frac{\rho_\infty}{\rho_K}\right)^\gamma
     =\gamma p_K\rho_K^{-\gamma}.
 \label{eq:fv_entropy_proxy}
\end{equation}
Physical specific entropy satisfies $s-s_\infty=c_v\log Z$, where $c_v>0$ is the constant-volume specific heat. The penalty uses $(Z-1)^2$, not $s$ itself.

For each ray $j$, the code finds $i_j^*=\max\{i:p_{ji}>p_{{\rm th},j}\}$, if this set is nonempty. The threshold is either $p_{{\rm th},j}=k_{\rm ent}p_\infty$, or
$p_{{\rm th},j}=p_\infty+r_{\rm ent}(\max_i p_{ji}-p_\infty)$
when the relative option is active. Define
\begin{equation}
 m_{ji}=\mathbf1_{\{i_j^*\ {\rm exists}\}}
              \mathbf1_{\{i\geq i_j^*+d_{\rm ent}\}},\qquad
 \mathcal L_{\rm ent}
 =\frac{\sum_{ji}m_{ji}(Z_{ji}-1)^2}{\max(\sum_{ji}m_{ji},1)}.
 \label{eq:fv_entropy_loss}
\end{equation}
Here $\mathbf1_A$ is one when condition $A$ holds and zero otherwise; $k_{\rm ent}>1$, $r_{\rm ent}>0$ when used, and $d_{\rm ent}$ is a radial-index offset. Dynamic mask construction is detached from parameter differentiation. The recorded $\mathrm{Ma}=5$ continuation uses a mask frozen at its starting state. This upstream physical prior suppresses a spurious entropy precursor; it is not an enforcement of the global shock entropy inequality. The pressure threshold is an operational mask locator, not an exact shock surface.

The term \emph{data-free} is restricted to the training objective and parameter updates: no CFD or experimental target values enter them. The fixing boundary is specified from the shock-detachment estimate above, not fitted to the present CFD field. Case-specific loss coefficients are prescribed empirically. Stagnation relations, freestream targets, geometry, upstream fixing, and the upstream entropy prior supply additional problem-specific information. The CFD comparisons assess the resulting fields under these stated priors and coefficients.

\subsection{Training stages and saved-field evaluation}
\label{sec:fv_protocol}

\subsubsection{Main computations at all Mach numbers}
Table~\ref{tab:fv_main_settings} collects the conditions of the main computations at all eight Mach numbers in one table. It identifies both the initial stage and the later refinement that produced each adopted field. The fixing boundaries follow the shock-detachment estimate described above, and the coefficients are prescribed empirical settings. The case-specific coefficients, seeds, and adopted endpoints are recorded, but the complete sequence of candidate trials, rejected checkpoints, and selection decisions is unavailable. The number of trials and the extent to which CFD comparisons informed coefficient or checkpoint selection cannot be reconstructed from these records; the historical $\mathrm{Ma}=11$ seed rationale is also unavailable. Thus reference-independent hyperparameter or checkpoint selection is not established. The nonmonotonic coefficients in the table are executed empirical choices, not a derived Mach-number law. The data-free designation concerns only the objective and parameter updates, not independence of all model-selection decisions. The sensitivity baseline is a documented comparison state; the factor tests assess departures from it without claiming that it is optimal. Sensitivity and scaling comparisons have their own tables below.
\begin{table}[!htb]
\centering
\color{RevisionColor}
\small
\setlength{\tabcolsep}{4pt}
\caption{Conditions of the main FV-CNN computations at all eight Mach numbers. The columns identify the runs used in the flow-field and stagnation-line comparisons. Initial training and later refinement are distinguished within this single table; common settings are specified below.}
\label{tab:fv_main_settings}
\begin{tabular}{@{}l*{8}{c}@{}}
\toprule
$Ma_\infty$ & $2$ & $3$ & $5$ & $7$ & $9$ & $11$ & $13$ & $15$\\
\midrule
$d$ & $1.5$ & $0.6$ & $0.6$ & $0.6$ & $0.6$ & $0.6$ & $0.6$ & $0.6$\\
$R_{\rm side}$ & $4$ & $2$ & $2$ & $2$ & $2$ & $2$ & $2$ & $2$\\
$x_{\rm fix}$ & $-1.3$ & $-1.0$ & $-0.815$ & $-0.792$ & $-0.784$ & $-0.781$ & $-0.778$ & $-0.777$\\
$b$ (implied) & $0.1797$ & $0.1757$ & $0.0824$ & $0.0797$ & $0.0795$ & $0.0804$ & $0.0796$ & $0.0800$\\
$\lambda_P$ & $1$ & $60$ & $120$ & $200$ & $50$ & $20$ & $20$ & $50$\\
$\lambda_0$ & $2\times10^6$ & $10^4$ & $10^4$ & $10^4$ & $2000$ & $10^4$ & $7500$ & $10^4$\\
$\tau_0$ & $0$ & $0$ & $0.005$ & $0.01$ & $0.01$ & $0.01$ & $0.01$ & $0.01$\\
$\lambda_s$ & $20$ & $20$ & $20$ & $200$ & $200$ & $200$ & $200$ & $200$\\
$s_N$ & $1$ & $1$ & $1$ & $0.5$ & $1$ & $1$ & $1$ & $1$\\
$\kappa$ & $0$ & $1$ & $1$ & $1$ & $1$ & $1$ & $1$ & $1$\\
$\alpha_0$ & $16$ & $16$ & $16$ & $16$ & $16$ & $16$ & $1.5$ & $1.5$\\
$\alpha_{\rm f}$ & $1.5$ & $1.5$ & $1.5$ & $1.5$ & $1.5$ & $1.5$ & $1.5$ & $1.5$\\
Seed & $42$ & $42$ & $42$ & $42$ & $42$ & $1$ & $42$ & $42$\\
\midrule
Initial AdamW updates & $15000$ & $15000$ & $15000$ & $15000$ & $15000$ & $15000$ & $15000$ & $15000$\\
Base extension updates & $0$ & $0$ & $5000$ & $0$ & $0$ & $0$ & $0$ & $0$\\
Entropy continuation updates & $0$ & $0$ & $2500^*$ & $10000$ & $10000$ & $10000$ & $10000$ & $10000$\\
$\lambda_{\rm ent}$, cleanup & $0$ & $0$ & $10^*$ & $20000$ & $20000$ & $20000$ & $20000$ & $20000$\\
Cleanup learning rate & --- & --- & adaptive$^*$ & $10^{-5}$ & $10^{-5}$ & $10^{-5}$ & $10^{-5}$ & $10^{-5}$\\
L-BFGS closure calls & $10002$ & $0$ & $0$ & $0$ & $0$ & $0$ & $0$ & $0$\\
\bottomrule
\end{tabular}
\par\medskip
\begin{minipage}{\linewidth}
\footnotesize
Billig's cylinder estimate is $\Delta_B/R=0.386\exp(4.67/Ma_\infty^2)$ with $R=D/2$; $x_{\rm fix}/D=-1/2-\Delta_B/D-b$. The tabulated implied $b$ is about $0.08$ for $\mathrm{Ma}\geq5$ and $0.18$ for $\mathrm{Ma}=2,3$, placing the low-Mach fixing region farther upstream; the historical reason for the larger low-Mach allowance is unavailable. Executed boundaries, rather than an exact uniform allowance, define the runs. Loss coefficients are empirical, and the shock position is measured from each field. $\mathrm{Ma}=2$ is a separate protocol, with its larger domain, reversed Mach-weight rule, $\kappa=0$, stronger stagnation prior, and L-BFGS stage. Cross-Mach aggregates describe these adopted runs.\par\smallskip
Common settings: $(N_\theta,N_r)=(96,160)$, six 64-channel blocks, radial kernel 15, azimuthal kernel $(3,1)$; first-order shared-face FV residual; $\lambda_H=0.5$, $\lambda_{\rm in}=20$, $\lambda_w=0$, $\lambda_m=1000$, $\lambda_{\rm TV}=0.1$, and $\beta_b=\beta_s=0.01$. All use the Mach weights in Eq.~\eqref{eq:fv_mach_weights}, the relative stagnation penalty, and $N_{\rm ann}=10000$. Initial AdamW uses $10^{-3}\to10^{-5}$ over updates 5000--10000, and $f_e:1\to3$. Later stages retain the terminal frequency, dissipation, and nose coefficient. $\mathrm{Ma}=2$ also uses $\lambda_{\rm TV}=0.1$ in L-BFGS and is the historical nondeterministic run.
\par\smallskip
$^*$The $\mathrm{Ma}=5$ training history comprises 15000 initial, 5000 further base-objective, and 2500 entropy-continuation updates. The latter uses a frozen mask, a separate entropy optimizer with coefficient ramping to 10, and an adaptive step size. Its recorded effective learning rate at the compared state is $6.04\times10^{-14}$. This executed schedule differs from the fixed-$20000$ joint-objective continuation used at $\mathrm{Ma}=7$--$15$.
\end{minipage}
\end{table}

The initial stage uses AdamW with first- and second-moment decay factors $\beta_1=0.9$, $\beta_2=0.999$, denominator safeguard $\epsilon=10^{-8}$, and weight decay $10^{-6}$. Its full Euclidean parameter-gradient norm is clipped at 1.0 before each update. Subsequent continuation and optional L-BFGS refinement follow the case-specific procedures recorded for each adopted result. The main $\mathrm{Ma}=11$ run uses seed 1 while the other main runs use 42. The reason for that historical seed change was not retained, and no matched seed-42 $\mathrm{Ma}=11$ counterpart is included in the verified records. Each adopted field is one run, so the Mach sequence is not an ensemble reproducibility study. Runtime metadata are not uniform across the historical main runs. The retained initialization records for $\mathrm{Ma}=3,5,7,15$ state deterministic convolutions using the Compute Unified Device Architecture (CUDA) Deep Neural Network library (cuDNN), with benchmarking and TensorFloat-32 (TF32) disabled, but strict deterministic algorithms disabled. The $\mathrm{Ma}=2$ source uses the speed-priority cuDNN settings. Equivalent initial-runtime snapshots were not retained for every other main case; seed values alone are therefore not presented as a deterministic-execution guarantee. For a fresh run the standard frequency schedule is $f_e=1+2\chi_e$. The prescribed cosine learning rate is
\begin{equation}
 \eta(e)=\eta_{\rm f}
 +\frac{\eta_0-\eta_{\rm f}}2[1+\cos(\pi t_e)],\qquad
 t_e=\min\!\left(\max\!\left(\frac{e-e_a}{e_b-e_a},0\right),1\right),
 \label{eq:fv_learning_rate}
\end{equation}
Here $\eta_0,\eta_{\rm f}$ are the initial and terminal learning rates, $e_a<e_b$ specify the decay interval, and $t_e$ is its clipped progress fraction. During initial AdamW training,
\begin{equation}
 \lambda_N(e)=s_N[2+8\min(e/30000,1)]
 \label{eq:fv_nose_schedule}
\end{equation}
with recorded multiplier $s_N$. A completed 15000-update initial stage ends at $\lambda_N=5.999733\ldots\,s_N$, using zero-based update $e=14999$.

Entropy cleanup starts from the completed initial-stage checkpoint. Its inheritance procedure retains the executed geometry, boundary treatment, residual and prior weights, terminal dissipation, Fourier scale, and terminal nose coefficient. For $\mathrm{Ma}=7,9,11,13,15$, the fixed-coefficient protocol uses AdamW with learning rate $10^{-5}$ and $\lambda_{\rm ent}=20000$ from the first continuation update, without an entropy ramp or further nose growth. These settings are not substituted into the separate initial-stage sensitivity tests.

For $\mathrm{Ma}=5$, the recorded training history comprises 15000 initial updates, 5000 further base-objective updates at $10^{-5}$, and 2500 entropy-continuation updates for the field used in the comparisons. The latter stage uses a frozen mask and an entropy coefficient ramping to 10 over 2000 updates. Its separate entropy AdamW optimizer uses a nominal cosine schedule $10^{-4}\to10^{-5}$ over 10000 updates, multiplied by the initial ramp and an adaptive backoff factor; the direct base-optimizer learning rate is zero. The implementation projects proposed entropy displacements to remove positive first-order changes of the base objective, and restores rejected update blocks before reducing the step size. These are recorded optimizer settings for this run. The full settings and history are retained in the reproducibility records. $\mathrm{Ma}=5$ enters the same flow-field and stagnation-line comparisons as the other Mach numbers.

\subsubsection{Sensitivity experiments}
\begin{table}[!htb]
\centering
\color{RevisionColor}
\small
\setlength{\tabcolsep}{7pt}
\caption{$\mathrm{Ma}=15$ sensitivity experiments. Each row varies the named factor against the same baseline while retaining the other settings. All use 15000 AdamW updates from the shared seed-42 untrained state; no L-BFGS or entropy cleanup is used. Mach-scaling controls are listed separately in Table~\ref{tab:fv_scaling_settings}.}
\label{tab:fv_sensitivity_settings}
\begin{tabular}{@{}lll@{}}
\toprule
Factor & Baseline & Tested alternatives\\
\midrule
Dissipation $\alpha_0=\alpha_{\rm f}$ & $1.5$ & $1.0,\ 2.0$\\
TV coefficient $\lambda_{\rm TV}$ & $0.1$ & $0,\ 0.01$\\
PDE coefficient $\lambda_P$ & $50$ & $25,\ 100$\\
Stagnation coefficient $\lambda_0$ & $10000$ & $0,\ 5000,\ 20000$\\
Fixing boundary $x_{\rm fix}$ & $-0.777$ & $-0.9,\ -1.0$\\
Fixing coefficient $\lambda_m$ & $1000$ & $0$\\
Nose multiplier $s_N$ & $1$ & $0$\\
Enthalpy coefficient $\lambda_H$ & $0.5$ & $0$\\
Fourier encoding & on & off\\
Spatial convolution & radial + azimuthal &
 \shortstack[l]{radial only; azimuthal only;\\pointwise (both spatial couplings off)}\\
\bottomrule
\end{tabular}
\par\medskip
\begin{minipage}{\linewidth}
\footnotesize
Baseline: $(d,R_{\rm side})=(0.6,2)$, $\kappa=1$, $(\alpha_0,\alpha_{\rm f})=(1.5,1.5)$, $x_{\rm fix}=-0.777$, $\lambda_P=50$, $\lambda_0=10000$, $\tau_0=0.01$, $\lambda_s=200$, $\lambda_H=0.5$, $\lambda_{\rm in}=20$, $\lambda_w=0$, $\lambda_m=1000$, $\lambda_{\rm TV}=0.1$, $s_N=1$, and $\beta_b=\beta_s=0.01$. The grid, architecture, frequency and learning-rate schedules match the corresponding definitions in Methods. Dissipation is constant within each dissipation test. The table contains 18 non-scaling alternatives plus their baseline; the twentieth run of the full suite is the scaling-off case in Table~\ref{tab:fv_scaling_settings}.
\end{minipage}
\end{table}

The 20 $\mathrm{Ma}=15$ sensitivity cases use a common saved untrained initialization, seed 42, and 15000 AdamW updates, without entropy cleanup or L-BFGS. Each factor is compared with its shared baseline while retaining the other executed settings. The dissipation multiplier, TV, PDE coefficient, stagnation anchor, mask, and network couplings are reported as separate tests. This single-seed study does not estimate across-seed variability.

\subsubsection{Scaling comparisons}
\begin{table}[!htb]
\centering
\color{RevisionColor}
\small
\setlength{\tabcolsep}{6pt}
\caption{Separate conditions for Mach-scaling comparisons. The coefficients $w_m,w_E$ multiply squared residuals. All rows have 15000 initial AdamW updates and no entropy cleanup. The $\mathrm{Ma}=15$ rows use the sensitivity baseline, not the final production field in Table~\ref{tab:fv_main_settings}.}
\label{tab:fv_scaling_settings}
\begin{tabular}{@{}cccccc@{}}
\toprule
$Ma_\infty$ & Scaling & $w_m$ & $w_E$ & $\lambda_P$ & L-BFGS closure calls\\
\midrule
$2$ & on & $4$ & $16$ & $1$ & $10002$\\
$2$ & off & $1$ & $1$ & $1$ & $10002$\\
$3$ & on & $1/9$ & $1/81$ & $60$ & $0$\\
$3$ & off & $1$ & $1$ & $1$ & $0$\\
$15$ & on & $1/225$ & $1/50625$ & $50$ & $0$\\
$15$ & off & $1$ & $1$ & $50$ & $0$\\
\bottomrule
\end{tabular}
\par\medskip
\begin{minipage}{\linewidth}
\footnotesize
$\mathrm{Ma}=2$: other recorded conditions are inherited from the main run. Reconstructed seed-42 initial tensors match the saved scaling-off initialization exactly; the AdamW cuDNN flags remain nondeterministic. See Sec.~\ref{sec:mach_scaling_ablation}.
$\mathrm{Ma}=3$: the overall PDE coefficient also differs ($60$ versus $1$), so this is an auxiliary final-result comparison and cannot isolate the effect of Mach weighting.
$\mathrm{Ma}=15$: the common saved initialization and remaining sensitivity-baseline settings are retained; only Mach weighting changes.
\end{minipage}
\end{table}

The $\mathrm{Ma}=2$ unscaled calculation inherits the scaled run settings, changing momentum and energy squared-residual weights from $(4,16)$ to $(1,1)$. The original run did not enforce deterministic cuDNN execution and its initial checkpoint was not archived. It is a comparison under inherited nominal conditions, rather than a verified deterministic paired trajectory.

For $\mathrm{Ma}=3$, the available final-result comparison also changes $\lambda_P$ and is therefore auxiliary. For $\mathrm{Ma}=15$, the scaling comparison uses the shared sensitivity initialization and has no entropy-cleanup stage.

If L-BFGS is used, closure evaluations, accepted iterations, arithmetic precision, and stopping conditions are reported separately. Changes in objective weights or entropy masks across stages are specified. Losses from differently weighted objectives are not direct measures of comparative physical accuracy.

\subsubsection{Saved-field evaluation}
No reference targets enter the loss or updates. For a field $q\in\{\rho,p,T,u,v\}$, let $q_K$ and $q_K^{\rm CFD}$ be its coordinate-matched neural and CFD values. Define
\begin{equation}
 e_{2,q}=
 \left[\frac{\sum_KJ_K(q_K-q_K^{\rm CFD})^2}
              {\sum_KJ_K(q_K^{\rm CFD})^2}\right]^{1/2},
 \qquad
 e_{\infty,q}=\frac{\max_K|q_K-q_K^{\rm CFD}|}
                         {\max_K|q_K^{\rm CFD}|}.
 \label{eq:fv_evaluation_norms}
\end{equation}
Here $J_K=A_K$ is the physical area of the node-centered dual control volume used for evaluation, not a training penalty weight. Percent values multiply the ratios by 100. Where a five-percent trimmed norm is given, nodes are omitted separately for each variable by the largest contributions $J_K(q_K-q_K^{\rm CFD})^2$. This statistical exclusion is not a geometrical shock mask.

Temperature profiles use $T/T_\infty$, with one common freestream normalization. The theoretical stagnation temperature is $T_0=T_\infty[1+(\gamma-1)Ma_\infty^2/2]$; individual curve maxima are not used for normalization. Comparisons use matched physical domains, including the distinct $\mathrm{Ma}=2$ geometry. Radial resampling is not used to conceal a grid mismatch.

Finite-resolution CFD comparisons are supplemented by operator-consistent centered-flux residuals and closed-band physical-flux quadrature. The centered residual shares the FV training mesh and nondissipative flux form, so its smooth-region ordering can reflect operator alignment. Explicit regions and a stride-two stencil check are supplied. Finite-width shock bands are tested through their complete boundaries, including lateral faces. Such balances are distinguished from zero-thickness Rankine--Hugoniot tests. They assess the computed field rather than simply repeating its HLL training loss. Algebraic conservation, finite optimization residuals, discretization error, and physical admissibility remain separate aspects of the method.

\end{revisionred}
\clearpage
\providecommand{\FVResultColor}{\color{RevisionColor}}
\providecommand{\FVMainTableReference}{Table~\ref{tab:fv_main_settings}}
\providecommand{\FVNormReference}{Eq.~\eqref{eq:fv_evaluation_norms}}
\providecommand{\FVADSectionReference}{Section~\ref{sec:fv_ad_failure}}
\section{Baseline Validation Across Mach Numbers}
\label{sec:results}
\label{sec:baseline_results}

This section first establishes what the complete, fixed configuration achieves.
\begin{revisionred}
The main FV-CNN computations cover $Ma_\infty=2,3,5,7,9,11,13,15$, using the adopted fields and case-specific training histories in \FVMainTableReference. ``Fixed configuration'' here denotes each completed main computation; it does not mean identical loss coefficients or training stages across Mach numbers. The flow fields and stagnation-line profiles are first compared with CFD; field discrepancies are then distinguished from operator-consistent residuals and closed-band conservation diagnostics. Sensitivity and scaling tests are treated separately from these main results.
\end{revisionred}

\subsection{\textcolor{RevisionColor}{Reference fields and comparison protocol}}
\begin{revisionred}
No closed-form solution exists for the complete flow field around the cylinder.
The full-field quantities below are discrepancies with the specified numerical CFD reference, not errors against a closed-form Euler solution.
\end{revisionred}
\begin{revisionred}
The reference was generated by the archived custom compressible-Euler solver, not an analytical solution. Its source uses first-order left/right states with HLL flux differences on a body-fitted O-grid, without higher-order reconstruction, and the two-stage second-order strong-stability-preserving (SSP) Runge--Kutta update. The full mesh has 160 radial and 381 angular nodes, including a periodic endpoint; only the matched upstream quadrant is evaluated below. The outer radius is $R_{\rm side}+(R_{\rm side}-R_{\rm nose})\cos\theta$, with $(R_{\rm nose},R_{\rm side})=(1.1,2)$ for $\mathrm{Ma}\geq3$ and $(2,4)$ for the adopted $\mathrm{Ma}=2$ reference. The boundary treatment is periodic in angle, reflected normal velocity at the cylinder, and prescribed freestream in the outer ghost layers. Density and pressure are floored at $10^{-4}$ during primitive conversion. Initially the outer region is at freestream and the inner 80\% of radial indices has zero velocity.

The Courant--Friedrichs--Lewy (CFL) coefficient is 0.5, using the smaller of the radial spacing and the minimum angular neighbor distance, divided by the maximum local speed plus sound speed. For $\mathrm{Ma}=3$--15 the stopping diagnostic is the root-mean-square (RMS) value of the integrated residual before division by the metric Jacobian; it is checked every 100 steps and must fall below $10^{-4}$ times its value at iteration 500, with nondimensional physical time $t=40$ as a cap. The runs terminate after 30,900--33,800 steps. The separate $\mathrm{Ma}=2$ run reaches $t=40$ in 30,442 steps; the RMS of its final metric-divided steady right-hand side (RHS) is $1.68\times10^{-5}$, or $7.55\times10^{-7}$ of its initial value. The two logged residual definitions are kept distinct in the supplied summary. Solver sources, launch commands, complete fields, and convergence histories accompany the revision. These recorded residuals document the iterative stopping condition and attained residual level of the discrete CFD calculation. The reported discrepancies use this specified numerical reference; they are not continuum-solution error estimates.

The stored CFD fields provide a finite-resolution Euler comparator. For each main FV-CNN case, the evaluation uses $96\times160$ physically matched nodes in the upstream quadrant, $\pi/2\leq\theta\leq\pi$. The largest coordinate difference is $1.20\times10^{-7}D$; neither field is interpolated for the main FV-CNN--CFD norms. $\mathrm{Ma}=2$ uses its own larger domain, with upstream extent $x=-2D$ and side extent $y=4D$, rather than the geometry used at the other Mach numbers. A periodic endpoint in the full CFD grid is counted only once when extracting the quadrant. The wake is outside this comparison.

The earlier AD-CNN fields are retained as a historical comparator for the sum-based spatial derivative implementation discussed in \FVADSectionReference. The available legacy source uses $Ma_\infty$ to decode momenta during training but a hardcoded factor 7 during field export. This explains the dominant velocity-scale mismatch in archives whose outer streamwise velocity is near 7; the $\mathrm{Ma}=7$ archive instead has an outer value near 5 and its matching exporter is not traceable in the supplied source. To retain the established displayed-state convention, both velocity components are multiplied by
\begin{equation}
 c_u=\frac{Ma_\infty}{\operatorname{median}_{j}\,u^{\rm archive}_{j,N_r-1}}.
 \label{eq:res_ad_velocity}
\end{equation}
Density, pressure, and temperature are unchanged. The contour Mach number, total enthalpy, and physical fluxes are then evaluated from this displayed state. This convention reproduces the adopted stagnation-line data. It primarily corrects the export-scale mismatch and also normalizes small outer-boundary deviations; it is not an exact, source-verified inversion for every archive. The outer-boundary-derived conversion, especially its incomplete $\mathrm{Ma}=7$ provenance, remains a limitation of the historical comparator. It does not repair the earlier residual construction or supply a newly trained Euler solution.

The historical $\mathrm{Ma}=2$ AD-CNN domain is smaller than the main FV-CNN and CFD domains. Accordingly, comparisons between the two neural results use their common physical support. For every Mach number, AD-CNN values are interpolated bilinearly in polar angle and normalized radial coordinate onto the FV-CNN nodes; unsupported nodes are omitted from both models' comparative norms. The main FV-CNN norms still cover its entire matched CFD domain. These historical fields were not produced by a newly controlled, identical-initialization training experiment, so the comparison does not isolate architecture or residual formulation as the sole cause of every difference.
\end{revisionred}

\subsection{\textcolor{RevisionColor}{Bow-shock fields and stagnation-line profiles}}
\begin{revisionred}
Figures~\ref{fig:res_ma2}--\ref{fig:res_ma15} give one combined comparison for each Mach number. In the contour panels, the neural result is solid and CFD is dashed, with identical rainbow colors and contour levels for each variable in both rows. Thus displacement of a given physical level can be assessed directly. The profiles show density, streamwise velocity, normalized temperature $T/T_\infty$, and pressure along $y=0$, using $T_\infty=1/\gamma$ throughout.

The FV-CNN fields reproduce a detached bow shock, compression through the shock layer, and deceleration towards the cylinder. Their axis stand-off distance decreases from $0.686D$ at $\mathrm{Ma}=2$ to $0.192D$ at $\mathrm{Ma}=15$; the corresponding CFD values are $0.676D$ and $0.201D$. These measured locations are obtained with the shock-detachment-based fixing boundaries and other stated priors. For example, $\mathrm{Ma}=15$ has $x_{\rm fix}=-0.777D$ but its measured axis shock is at $x=-0.692D$; the two positions are distinct, although changing the prior can substantially displace the learned shock. The larger $\mathrm{Ma}=2$ domain is necessary to display its extended shock. The contours also show why agreement on the axis does not establish agreement over the whole shock: for example, the FV-CNN axis shock location at $\mathrm{Ma}=7$ differs from CFD by less than $0.001D$, while its angular RMS shock-position discrepancy is $0.030D$.

The FV-CNN shock transition is narrower than the historical AD-CNN transition in all eight cases under the local width definition below. Its position and post-shock profiles nevertheless retain visible discrepancies. $\mathrm{Ma}=5,9,11$ have relatively small aggregate discrepancies, whereas $\mathrm{Ma}=7,13,15$ retain larger differences near the curved front. The data do not show monotonic deterioration with Mach number, and the main case-specific settings prevent interpreting this sequence as a controlled Mach-number-only experiment.
\end{revisionred}
\begin{revisionred}
The stagnation-line profiles also display the analytical density and pressure values as circular markers.
\end{revisionred}
\begin{revisionred}
These values are imposed through the localized prior, so their agreement checks that constraint rather than providing an independent field prediction. In the adopted FV-CNN states, the largest absolute relative deviations from the density and pressure targets are $1.00\%$ and $1.09\%$, respectively. These maxima lie at or just beyond the $1\%$ dead-band edge used in the corresponding runs; they are not exact recovery of the imposed targets. The soft penalty permits a finite excess outside that edge. Pressure, temperature, and velocity through the finite shock layer must therefore be assessed from the full profiles and conservation tests, not from the two wall markers alone.
\end{revisionred}

\begin{revisionred}
\clearpage
\begin{figure}[p]
 \FVResultColor
 \centering
 \includegraphics[width=.96\linewidth,viewport=0.000000000000 476.872778299649 1000.000000000000 1141.259940036874,clip]{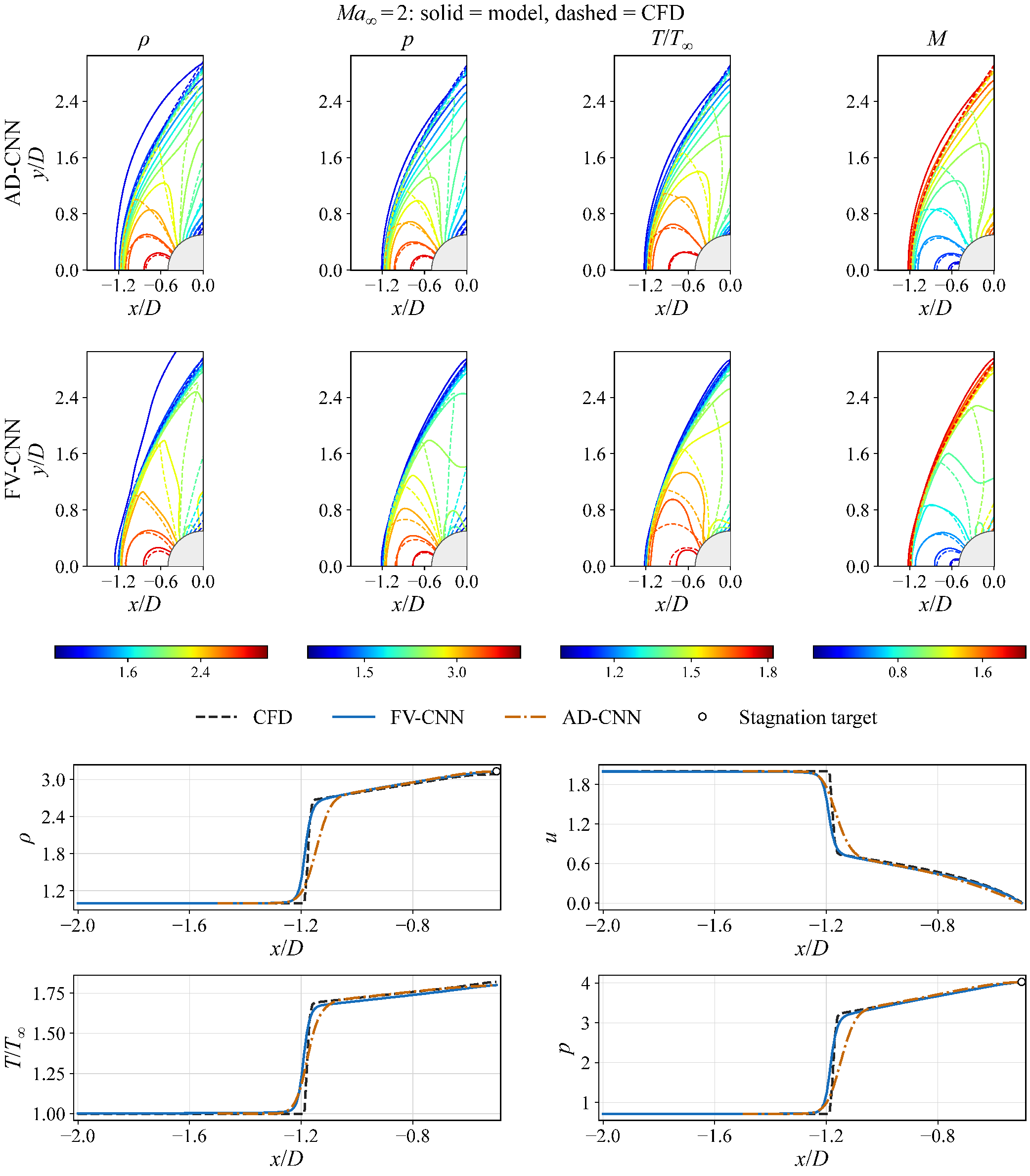}
 \par\smallskip
 \includegraphics[width=.96\linewidth,viewport=0.000000000000 0.000000000000 1000.000000000000 464.872778299649,clip]{Figure_2.eps}
 \caption{Flow-field comparison at $Ma_\infty=2$. Top: contours of $\rho$, $p$, $T/T_\infty$, and local Mach number $M$, with AD-CNN in the first row and FV-CNN in the second. Solid neural and dashed CFD contours share colors and levels. Bottom: stagnation-line profiles for CFD (black dashed), FV-CNN (blue solid), and AD-CNN (orange dash-dot); circles show the imposed analytical density and pressure targets. Temperature uses the common freestream normalization, and every profile axis retains vertical margins.}
 \label{fig:res_ma2}
\end{figure}
\clearpage
\begin{figure}[p]
 \FVResultColor
 \centering
 \includegraphics[width=.96\linewidth,viewport=0.000000000000 481.237002318835 1000.000000000000 1131.753975965406,clip]{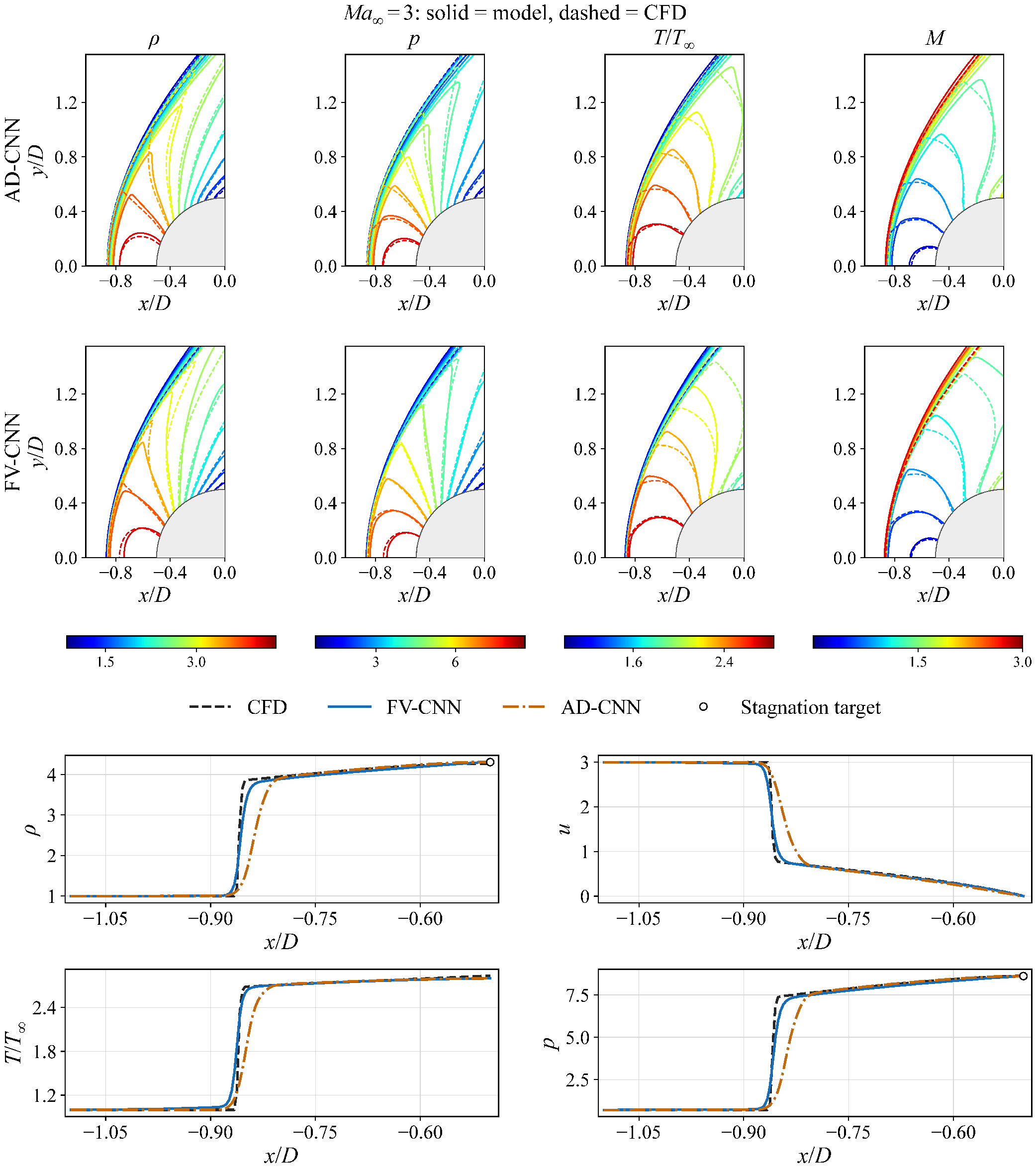}
 \par\smallskip
 \includegraphics[width=.96\linewidth,viewport=0.000000000000 0.000000000000 1000.000000000000 469.237002318835,clip]{Figure_3.eps}
 \caption{Flow-field comparison at $Ma_\infty=3$. Top: contours of $\rho$, $p$, $T/T_\infty$, and local Mach number $M$, with AD-CNN in the first row and FV-CNN in the second. Solid neural and dashed CFD contours share colors and levels. Bottom: stagnation-line profiles for CFD (black dashed), FV-CNN (blue solid), and AD-CNN (orange dash-dot); circles show the imposed analytical density and pressure targets. Temperature uses the common freestream normalization, and every profile axis retains vertical margins.}
 \label{fig:res_ma3}
\end{figure}
\clearpage
\begin{figure}[p]
 \FVResultColor
 \centering
 \includegraphics[width=.96\linewidth,viewport=0.000000000000 481.237002318835 1000.000000000000 1137.620591401312,clip]{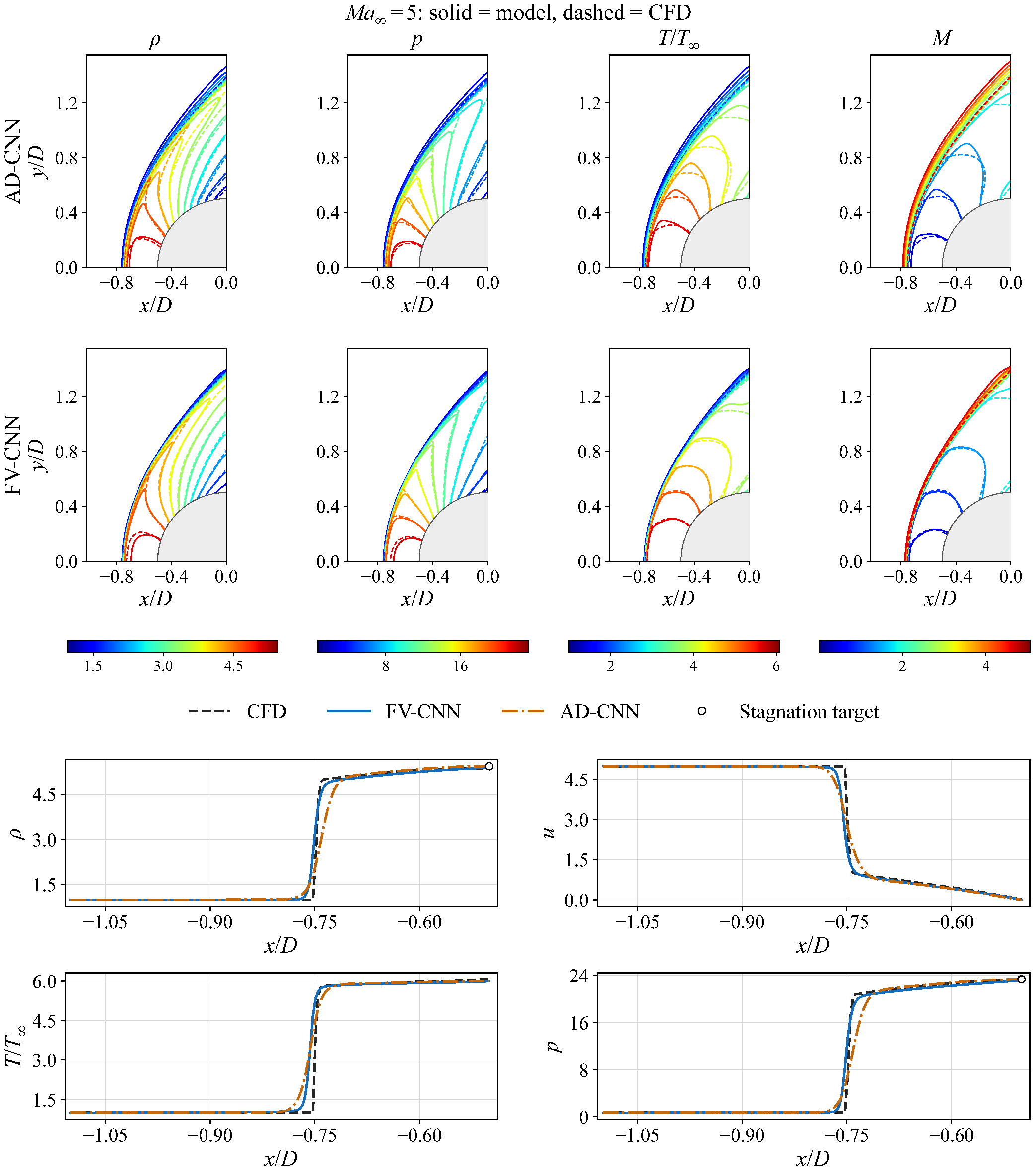}
 \par\smallskip
 \includegraphics[width=.96\linewidth,viewport=0.000000000000 0.000000000000 1000.000000000000 469.237002318835,clip]{Figure_4.eps}
 \caption{Flow-field comparison at $Ma_\infty=5$. Top: contours of $\rho$, $p$, $T/T_\infty$, and local Mach number $M$, with AD-CNN in the first row and FV-CNN in the second. Solid neural and dashed CFD contours share colors and levels. Bottom: stagnation-line profiles for CFD (black dashed), FV-CNN (blue solid), and AD-CNN (orange dash-dot); circles show the imposed analytical density and pressure targets. Temperature uses the common freestream normalization, and every profile axis retains vertical margins.}
 \label{fig:res_ma5}
\end{figure}
\clearpage
\begin{figure}[p]
 \FVResultColor
 \centering
 \includegraphics[width=.96\linewidth,viewport=0.000000000000 482.202621818857 1000.000000000000 1138.586210901334,clip]{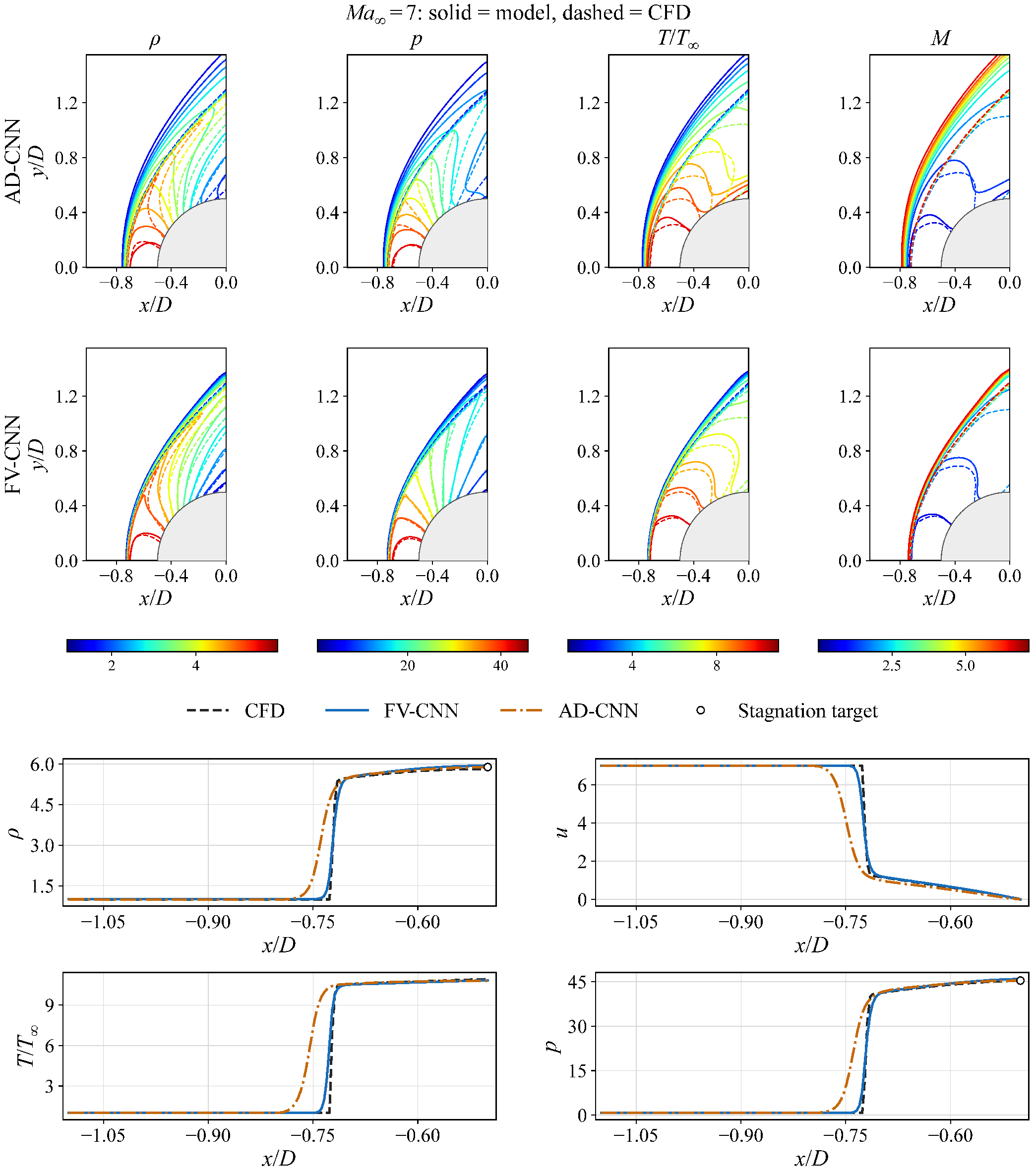}
 \par\smallskip
 \includegraphics[width=.96\linewidth,viewport=0.000000000000 0.000000000000 1000.000000000000 470.202621818857,clip]{Figure_5.eps}
 \caption{Flow-field comparison at $Ma_\infty=7$. Top: contours of $\rho$, $p$, $T/T_\infty$, and local Mach number $M$, with AD-CNN in the first row and FV-CNN in the second. Solid neural and dashed CFD contours share colors and levels. Bottom: stagnation-line profiles for CFD (black dashed), FV-CNN (blue solid), and AD-CNN (orange dash-dot); circles show the imposed analytical density and pressure targets. Temperature uses the common freestream normalization, and every profile axis retains vertical margins.}
 \label{fig:res_ma7}
\end{figure}
\clearpage
\begin{figure}[p]
 \FVResultColor
 \centering
 \includegraphics[width=.96\linewidth,viewport=0.000000000000 482.202621818857 1000.000000000000 1138.586210901334,clip]{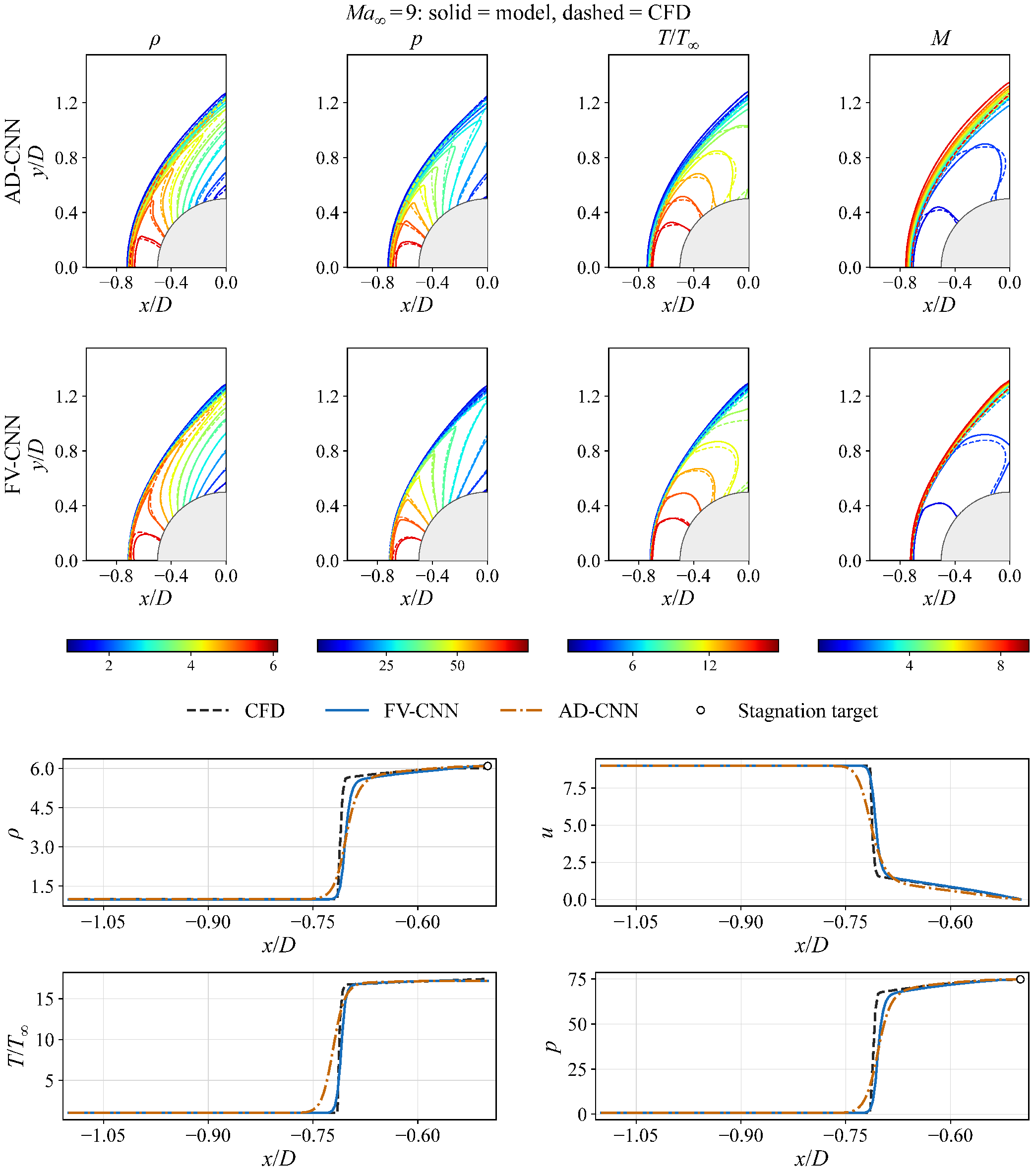}
 \par\smallskip
 \includegraphics[width=.96\linewidth,viewport=0.000000000000 0.000000000000 1000.000000000000 470.202621818857,clip]{Figure_6.eps}
 \caption{Flow-field comparison at $Ma_\infty=9$. Top: contours of $\rho$, $p$, $T/T_\infty$, and local Mach number $M$, with AD-CNN in the first row and FV-CNN in the second. Solid neural and dashed CFD contours share colors and levels. Bottom: stagnation-line profiles for CFD (black dashed), FV-CNN (blue solid), and AD-CNN (orange dash-dot); circles show the imposed analytical density and pressure targets. Temperature uses the common freestream normalization, and every profile axis retains vertical margins.}
 \label{fig:res_ma9}
\end{figure}
\clearpage
\begin{figure}[p]
 \FVResultColor
 \centering
 \includegraphics[width=.96\linewidth,viewport=0.000000000000 482.202621818857 1000.000000000000 1138.586210901334,clip]{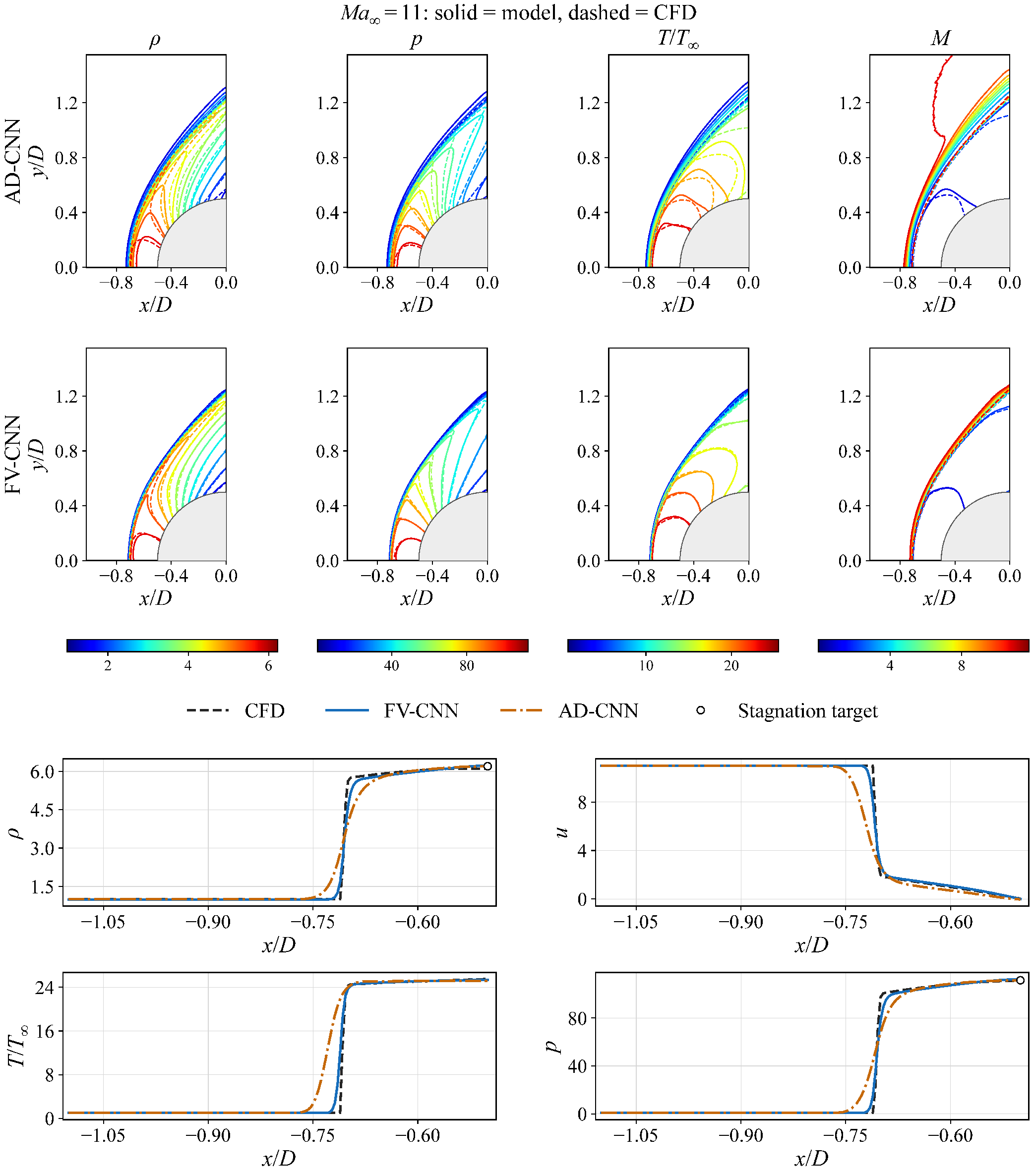}
 \par\smallskip
 \includegraphics[width=.96\linewidth,viewport=0.000000000000 0.000000000000 1000.000000000000 470.202621818857,clip]{Figure_7.eps}
 \caption{Flow-field comparison at $Ma_\infty=11$. Top: contours of $\rho$, $p$, $T/T_\infty$, and local Mach number $M$, with AD-CNN in the first row and FV-CNN in the second. Solid neural and dashed CFD contours share colors and levels. Bottom: stagnation-line profiles for CFD (black dashed), FV-CNN (blue solid), and AD-CNN (orange dash-dot); circles show the imposed analytical density and pressure targets. Temperature uses the common freestream normalization, and every profile axis retains vertical margins.}
 \label{fig:res_ma11}
\end{figure}
\clearpage
\begin{figure}[p]
 \FVResultColor
 \centering
 \includegraphics[width=.96\linewidth,viewport=0.000000000000 482.202621818857 1000.000000000000 1138.586210901334,clip]{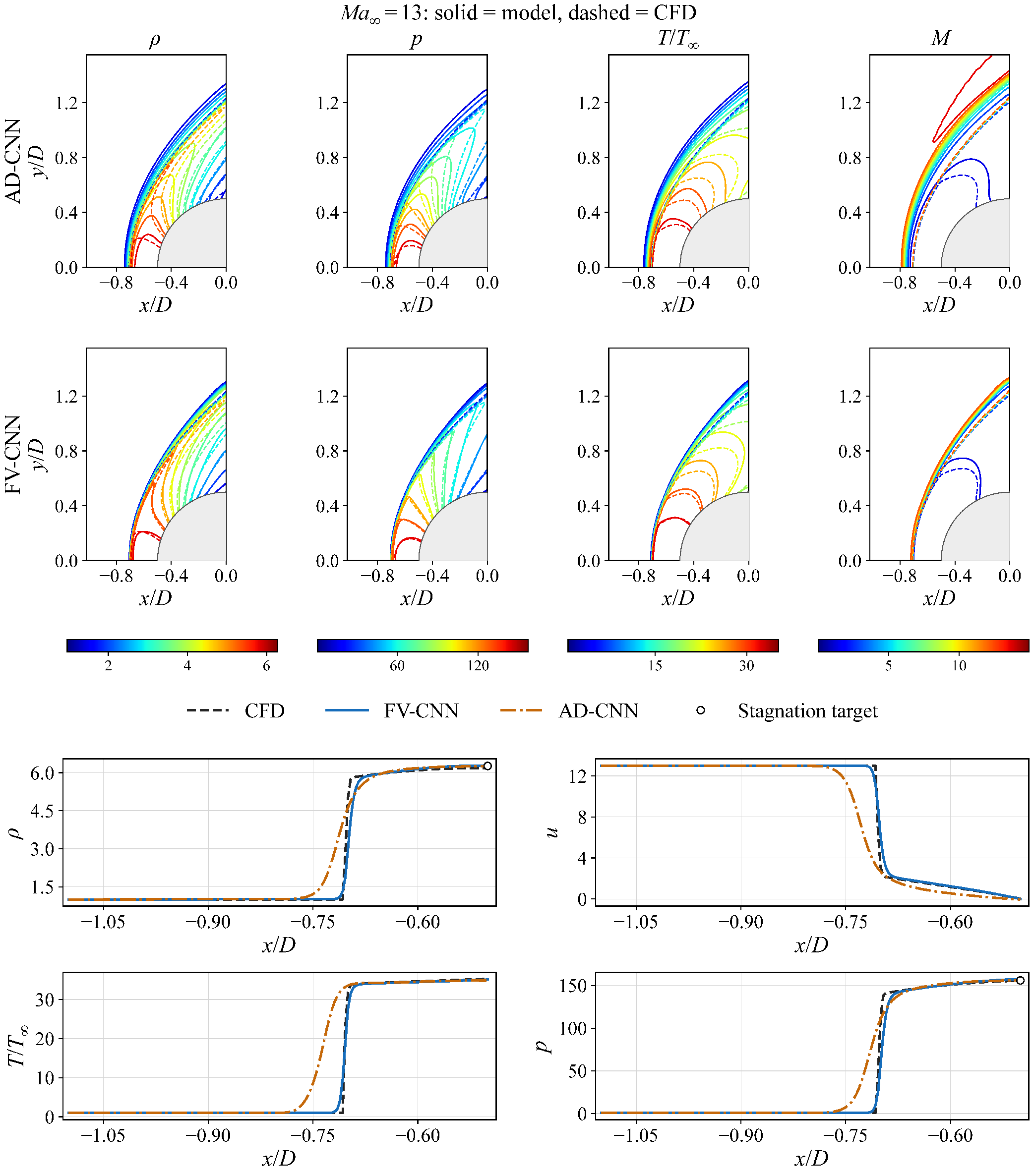}
 \par\smallskip
 \includegraphics[width=.96\linewidth,viewport=0.000000000000 0.000000000000 1000.000000000000 470.202621818857,clip]{Figure_8.eps}
 \caption{Flow-field comparison at $Ma_\infty=13$. Top: contours of $\rho$, $p$, $T/T_\infty$, and local Mach number $M$, with AD-CNN in the first row and FV-CNN in the second. Solid neural and dashed CFD contours share colors and levels. Bottom: stagnation-line profiles for CFD (black dashed), FV-CNN (blue solid), and AD-CNN (orange dash-dot); circles show the imposed analytical density and pressure targets. Temperature uses the common freestream normalization, and every profile axis retains vertical margins.}
 \label{fig:res_ma13}
\end{figure}
\clearpage
\begin{figure}[p]
 \FVResultColor
 \centering
 \includegraphics[width=.96\linewidth,viewport=0.000000000000 482.202621818857 1000.000000000000 1138.586210901334,clip]{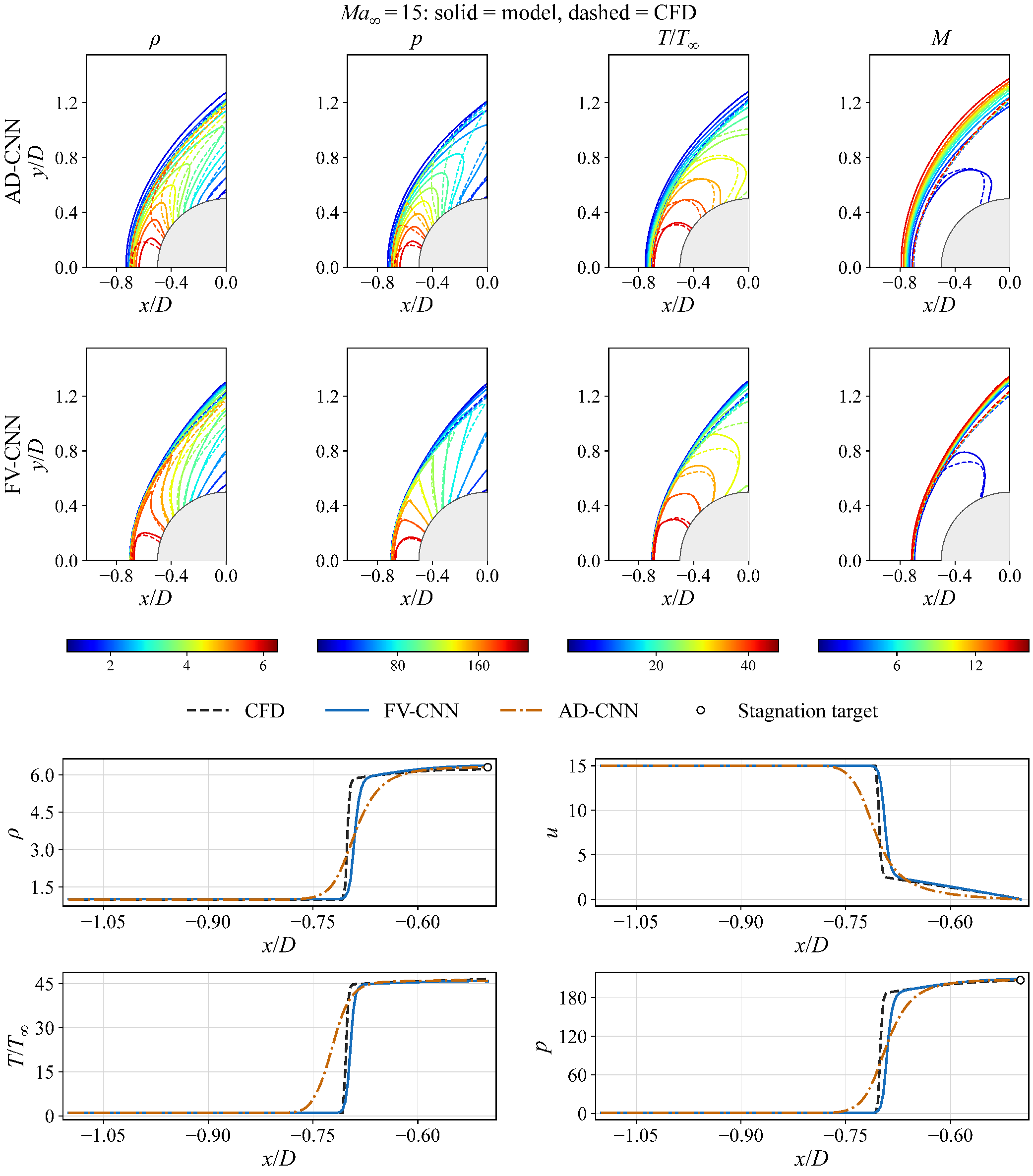}
 \par\smallskip
 \includegraphics[width=.96\linewidth,viewport=0.000000000000 0.000000000000 1000.000000000000 470.202621818857,clip]{Figure_9.eps}
 \caption{Flow-field comparison at $Ma_\infty=15$. Top: contours of $\rho$, $p$, $T/T_\infty$, and local Mach number $M$, with AD-CNN in the first row and FV-CNN in the second. Solid neural and dashed CFD contours share colors and levels. Bottom: stagnation-line profiles for CFD (black dashed), FV-CNN (blue solid), and AD-CNN (orange dash-dot); circles show the imposed analytical density and pressure targets. Temperature uses the common freestream normalization, and every profile axis retains vertical margins.}
 \label{fig:res_ma15}
\end{figure}
\clearpage

\end{revisionred}
\FloatBarrier

\subsection{\textcolor{RevisionColor}{Quantitative Error Assessment Against the CFD Reference}}
\label{sec:quant_err}
The field and profile comparisons identify where the agreement degrades.
\begin{revisionred}
Table~\ref{tab:res_field_errors} reports the area-weighted norms of \FVNormReference{} for the adopted FV-CNN fields. The physical weights are the node-centered dual-control-volume areas $A_K$. Across the eight cases, full-domain $e_2$ is $5.43$--$18.69\%$ for density, $6.25$--$19.70\%$ for pressure, and $4.02$--$21.71\%$ for temperature. The streamwise and transverse velocity discrepancies are $2.53$--$8.26\%$ and $8.25$--$34.95\%$, respectively. These are discrepancies over the matched upstream quadrant. They quantify differences from the specified numerical reference, including its finite-resolution shock representation.

The five-percent trimmed norms are much smaller in most cases. For density, the omitted five percent of nodes carry $73.0$--$98.1\%$ of the area-weighted squared discrepancy, and the retained norm is $1.01$--$5.30\%$. For pressure and temperature the retained ranges are $1.27$--$7.52\%$ and $0.52$--$3.78\%$. The excluded nodes are selected separately for each variable by their error contribution. They are not a prescribed shock region, and a trimmed norm does not bound the discrepancy everywhere else. The normalized $L_\infty$ values remain substantial, consistent with a finite displacement, broadening, or amplitude mismatch of a strong jump. Concentration on a small set of nodes does not identify displacement as the sole cause. Both FV-CNN and AD-CNN have relatively large density discrepancies at $\mathrm{Ma}=7,13$; the sequence is nonmonotonic, and random initialization, fixing priors, and reference discretization have not been isolated as causes. At $\mathrm{Ma}=15$ alone, the additional seed experiment spans $12.04$--$43.70\%$ in density discrepancy, wider than the $5.43$--$18.69\%$ range of the eight adopted fields (Sec.~\ref{sec:seed_ensemble}); a Mach-dependent accuracy trend cannot be inferred from this adopted-field sequence.
\begin{table}[!htb]
\centering\small\FVResultColor
\setlength{\tabcolsep}{4pt}
\caption{Main FV-CNN discrepancies on all $96\times160$ coordinate-matched CFD nodes. Entries for $\rho,p,T$ give $100e_2$ / five-percent trimmed $100e_2$ / $100e_\infty$; the two velocity columns give full-domain $100e_2$. All entries are percentages. $\mathrm{Ma}=2$ uses its distinct larger physical domain.}
\label{tab:res_field_errors}
\begin{tabular}{@{}c ccc cc@{}}
\toprule
$Ma_\infty$ & $\rho$ & $p$ & $T/T_\infty$ & $u$ & $v$\\
\midrule
2 & 9.2 / 4.9 / 55.4 & 12.8 / 7.5 / 46.4 & 4.0 / 2.6 / 21.1 & 5.5 & 13.3 \\
3 & 10.1 / 2.1 / 42.1 & 11.9 / 1.6 / 41.7 & 9.4 / 2.4 / 41.9 & 8.0 & 21.7 \\
5 & 5.4 / 1.0 / 28.6 & 6.2 / 1.4 / 33.5 & 6.1 / 1.3 / 61.0 & 2.5 & 8.2 \\
7 & 18.4 / 4.0 / 57.6 & 19.7 / 1.7 / 54.5 & 20.8 / 3.1 / 71.9 & 8.3 & 34.9 \\
9 & 8.5 / 1.3 / 48.4 & 9.5 / 1.3 / 54.1 & 9.8 / 1.5 / 42.0 & 3.3 & 15.1 \\
11 & 6.3 / 1.4 / 38.1 & 7.7 / 1.3 / 42.0 & 11.2 / 0.5 / 73.5 & 3.2 & 10.7 \\
13 & 18.7 / 3.5 / 58.6 & 19.1 / 2.2 / 50.1 & 21.7 / 3.8 / 68.0 & 7.2 & 34.4 \\
15 & 17.3 / 5.3 / 60.6 & 18.2 / 1.8 / 70.1 & 19.6 / 2.8 / 61.9 & 6.7 & 31.6 \\
\bottomrule
\end{tabular}
\end{table}

Figure~\ref{fig:res_common_errors} compares both neural fields on identical common-support nodes. Only at $\mathrm{Ma}=2$ does the smaller historical AD-CNN domain reduce the evaluated FV-CNN support: 10,579 rather than 15,360 nodes. Its common-support density, pressure, temperature, and streamwise-velocity norms are $10.16\%$, $13.40\%$, $4.83\%$, and $8.68\%$, whereas Table~\ref{tab:res_field_errors} gives $9.17\%$, $12.79\%$, $4.02\%$, and $5.52\%$ on the complete native FV-CNN quadrant. At $\mathrm{Ma}=3$--$15$ the supports coincide. FV-CNN has smaller density, pressure, and streamwise-velocity $e_2$ in all eight cases. At $\mathrm{Ma}=3$, however, its temperature discrepancy is $9.36\%$ versus $6.91\%$ for AD-CNN, and its transverse-velocity discrepancy is $21.72\%$ versus $12.65\%$. At $\mathrm{Ma}=15$, the streamwise-velocity norms are nearly equal ($6.66\%$ and $6.68\%$). The evidence therefore supports an overall improvement for these adopted fields, with explicit exceptions rather than uniform superiority in every variable; the initialization sensitivity in Sec.~\ref{sec:seed_ensemble} limits generalization of this ordering.

\begin{figure}[!htb]
 \FVResultColor
 \centering
 \includegraphics[width=.94\linewidth]{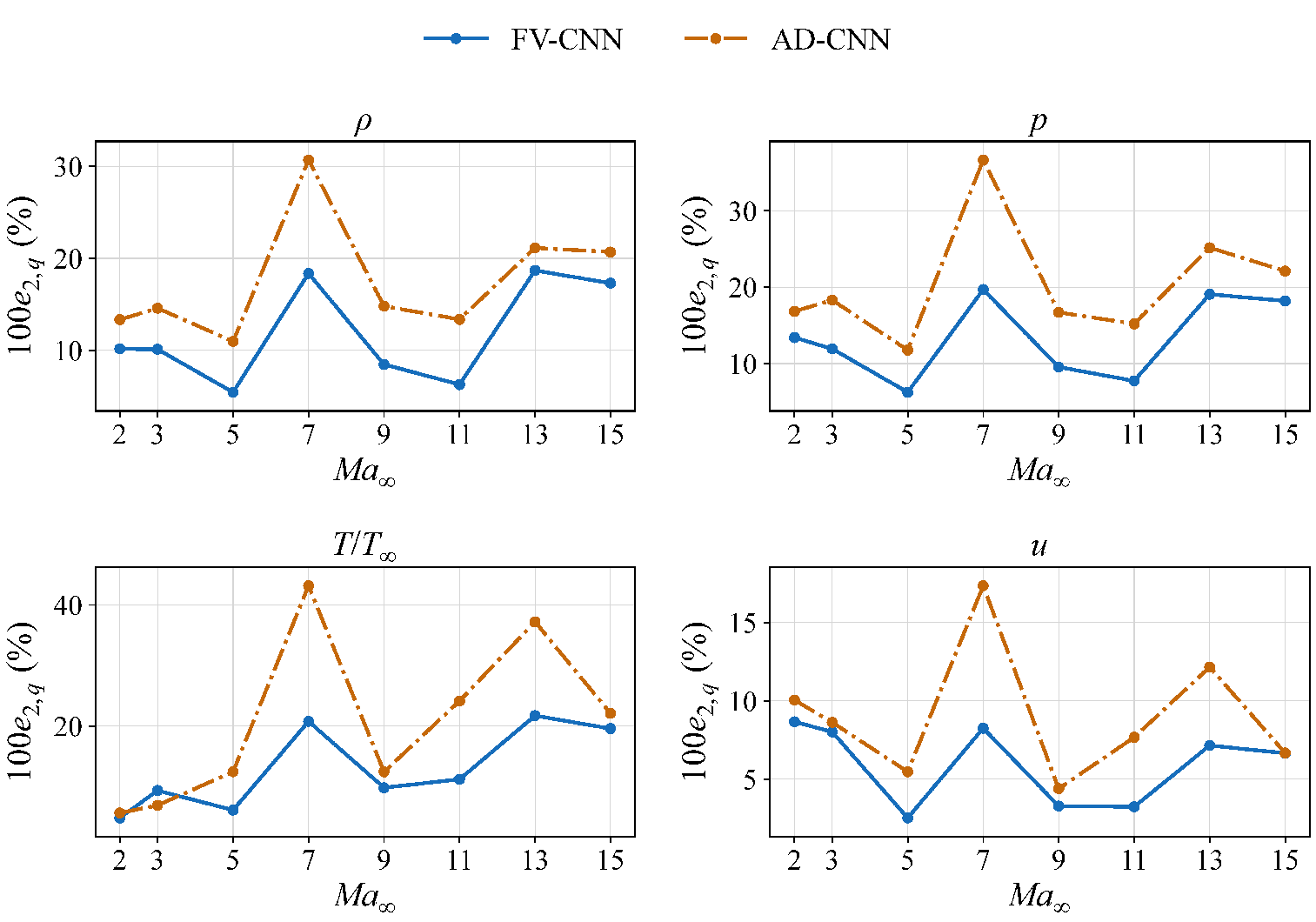}
 \caption{Area-weighted field discrepancies on common physical support. Each panel compares the same nodes and CFD reference for the two neural fields. At $\mathrm{Ma}=2$ only, the common support contains 10,579 nodes inside the smaller historical AD-CNN domain; Table~\ref{tab:res_field_errors} instead uses all 15,360 native FV-CNN nodes. At the other Mach numbers the supports coincide. Temperature uses $T/T_\infty$; its relative norm equals that for $T$ because the normalization is common. These historical comparisons include the velocity conversion in Eq.~\eqref{eq:res_ad_velocity}.}
 \label{fig:res_common_errors}
\end{figure}
\end{revisionred}
\FloatBarrier

\subsection{\textcolor{RevisionColor}{Operator-consistent physical-flux residuals in specified regions}}
\begin{revisionred}
Field agreement alone does not establish that a learned state satisfies the Euler equations. Physical Euler fluxes are therefore evaluated from each saved field without the HLL dissipative term. On the matched node-centered dual mesh, the diagnostic is
\begin{equation}
 D_{K,c}=\frac{1}{A_K}\sum_{f\subset\partial K}
 \frac{\mathbf F_c(\mathbf U_K)+\mathbf F_c(\mathbf U_{K_f})}{2}
 \mathbin{\cdot}\mathbf n_{Kf}|f|,
 \qquad \mathbf F_c=(F_{x,c},F_{y,c}),
 \label{eq:res_physical_operator}
\end{equation}
for interior nodes, where $c\in\{\rho,x,y,E\}$ labels mass, the two momentum components, and energy. Here $\mathbf U=U$, $K_f$ is the neighboring node across face $f$, $\mathbf n_{Kf}$ is the outward unit normal from $K$, and $|f|=\ell_f$ is the face length. No HLL dissipation is included. This centered flux balance uses the same dual mesh and central flux form as the nondissipative part of FV-CNN training. It is an operator-consistent diagnostic, rather than an independent spatial discretization. Historical AD-CNN is evaluated after interpolation on an operator it was not trained to minimize, so smooth-region differences can partly reflect this operator alignment. The four components are normalized with freestream flux scales
\begin{equation}
 \mathbf S=\left(\rho_\infty u_\infty,\;
 \rho_\infty u_\infty^2+p_\infty,\;
 \rho_\infty u_\infty^2+p_\infty,\;
 \rho_\infty u_\infty H_\infty\right),\qquad
 H_\infty=\frac{\gamma}{\gamma-1}T_\infty+\frac{u_\infty^2}{2}.
 \label{eq:res_flux_scales}
\end{equation}
The transverse-momentum scale is deliberately nonzero even though $v_\infty=0$. With dimensional diameter $D$, the normalized residual is $D D_{K,c}/S_c$; $D=1$ in the reported nondimensional data.

The reference shock radius $r_s^{\rm CFD}(\theta)$ is located by the largest compression-oriented radial pressure gradient, $-\partial p^{\rm CFD}/\partial r$, with local quadratic peak interpolation. Three node layers are removed at every domain boundary, and only common-support nodes are retained for all three methods. Fixed regions are then defined by
\begin{align}
 \Omega_{\rm up}&=\{K:r_K>r_s^{\rm CFD}(\theta_K)+h\},\nonumber\\
 \Omega_{\rm band}&=\{K:|r_K-r_s^{\rm CFD}(\theta_K)|\leq h\},\nonumber\\
 \Omega_{\rm down}&=\{K:r_K<r_s^{\rm CFD}(\theta_K)-h\},\qquad h=0.10D,
 \label{eq:res_regions}
\end{align}
intersected with this common interior support; $(r_K,\theta_K)$ are the polar coordinates of node $K$. Figure~\ref{fig:res_regions} displays the actual masks. The designation ``smooth'' means $\Omega_{\rm up}\cup\Omega_{\rm down}$ under this CFD-based partition. A displaced or broadened neural shock can extend beyond that band, so the designation does not guarantee smoothness of every model within the mask.

For any displayed region $\Omega$, the reported area-weighted RMS is
\begin{equation}
 \mathcal R_{c,\Omega}=
 \left[\frac{\sum_{K\in\Omega}A_K(D D_{K,c}/S_c)^2}
 {\sum_{K\in\Omega}A_K}\right]^{1/2}.
 \label{eq:res_regional_rms}
\end{equation}
Figure~\ref{fig:res_smooth} shows all four components outside the reference shock band. FV-CNN reduces this diagnostic relative to AD-CNN in 31 of the 32 case--component comparisons; the exception is streamwise momentum at $\mathrm{Ma}=3$ ($0.0993$ versus $0.0882$). The composite RMS, defined as the RMS of the four component values, falls from $0.1506$--$0.6094$ for AD-CNN to $0.0204$--$0.1551$ for FV-CNN across the cases. At $\mathrm{Ma}=15$, for example, mass and energy fall from $0.7354$ and $0.7361$ to $0.1731$ and $0.1766$.

These FV-CNN residuals remain above the CFD diagnostic, and the larger values at $\mathrm{Ma}=3,7,13,15$ are concentrated mainly in the reference-defined upstream region. At $\mathrm{Ma}=15$, the mass RMS is $0.1918$ upstream and $0.0432$ downstream; at $\mathrm{Ma}=9$ it is $0.0050$ and $0.0540$, respectively. The partition therefore reveals differences that an all-domain value would conceal. Complete upstream, band, downstream, and interior values are supplied with the numerical data.

To remove the particular effect of a displaced front entering a reference-defined non-band region, a second common mask excludes the union of the three centered bands: $|r-r_s^a(\theta)|>0.10D$ for every $a\in\{\mathrm{CFD},\mathrm{FV},\mathrm{AD}\}$. It uses the same retained nodes for every field. The solid curves in Fig.~\ref{fig:res_smooth} give this additional comparison; FV-CNN is smaller in all 32 case--component values. This common-exclusion result also holds at each of the seven tested half-widths from $0.06D$ to $0.12D$. Both masks are operational geometric exclusions: a broad transition can still extend outside a chosen width. The revised data include every mask definition and component value. A stride-two re-evaluation on rebuilt dual meshes is additionally reported in Appendix~\ref{sec:coarsened_diagnostic}; changing the stencil reduces exact training-grid alignment but does not make the centered-flux family fully independent. The CFD field norms and the separately quadrature-evaluated closed-band balances therefore carry the main comparative evidence.

\begin{figure}[!htb]
 \FVResultColor
 \centering
 \includegraphics[width=.94\linewidth,viewport=0.000000000000 655.765878849036 1000.000000000000 1291.668946355166,clip]{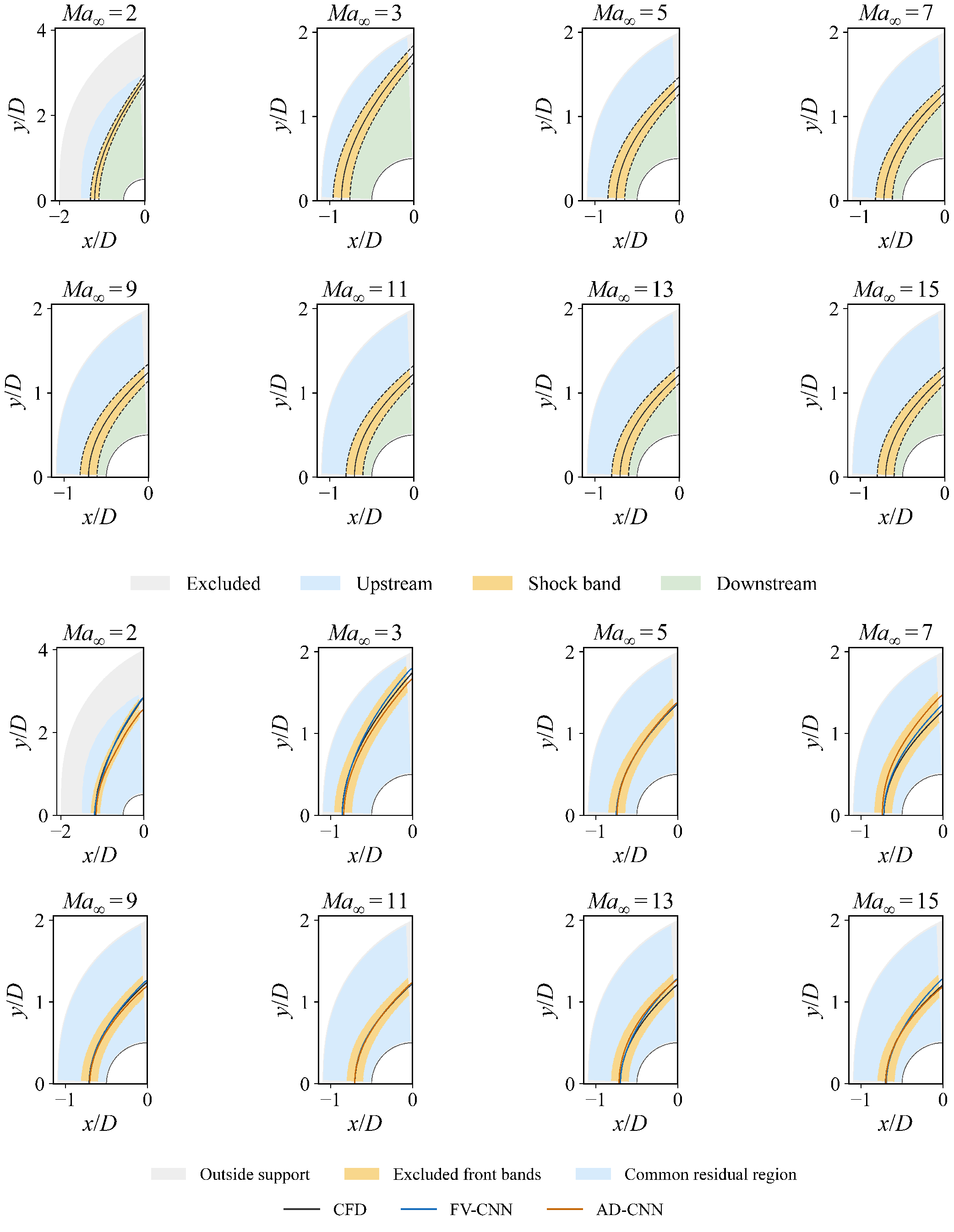}
 \par\medskip
 \includegraphics[width=.94\linewidth,viewport=0.000000000000 0.000000000000 1000.000000000000 643.765878849036,clip]{Figure_11.eps}
 \caption{Common physical support and CFD-based regions for the operator-consistent Euler diagnostic. The solid curve is $r_s^{\rm CFD}$ and the dashed curves are radial offsets $\pm0.10D$. Gray denotes excluded nodes, including the unsupported outer part of the $\mathrm{Ma}=2$ historical AD-CNN domain and three boundary layers. Upper panels: reference-defined upstream, band, and downstream classes. Lower panels: the additional common mask excludes all three located-front bands; blue nodes are retained for the solid curves in Fig.~\ref{fig:res_smooth}, yellow nodes are excluded by at least one front band, and gray nodes are outside the common interior support. The three colored curves mark the located centers. Within each diagnostic, exactly the same nodes are used for all methods.}
 \label{fig:res_regions}
\end{figure}
\begin{figure}[!htb]
 \FVResultColor
 \centering
 \includegraphics[width=.94\linewidth]{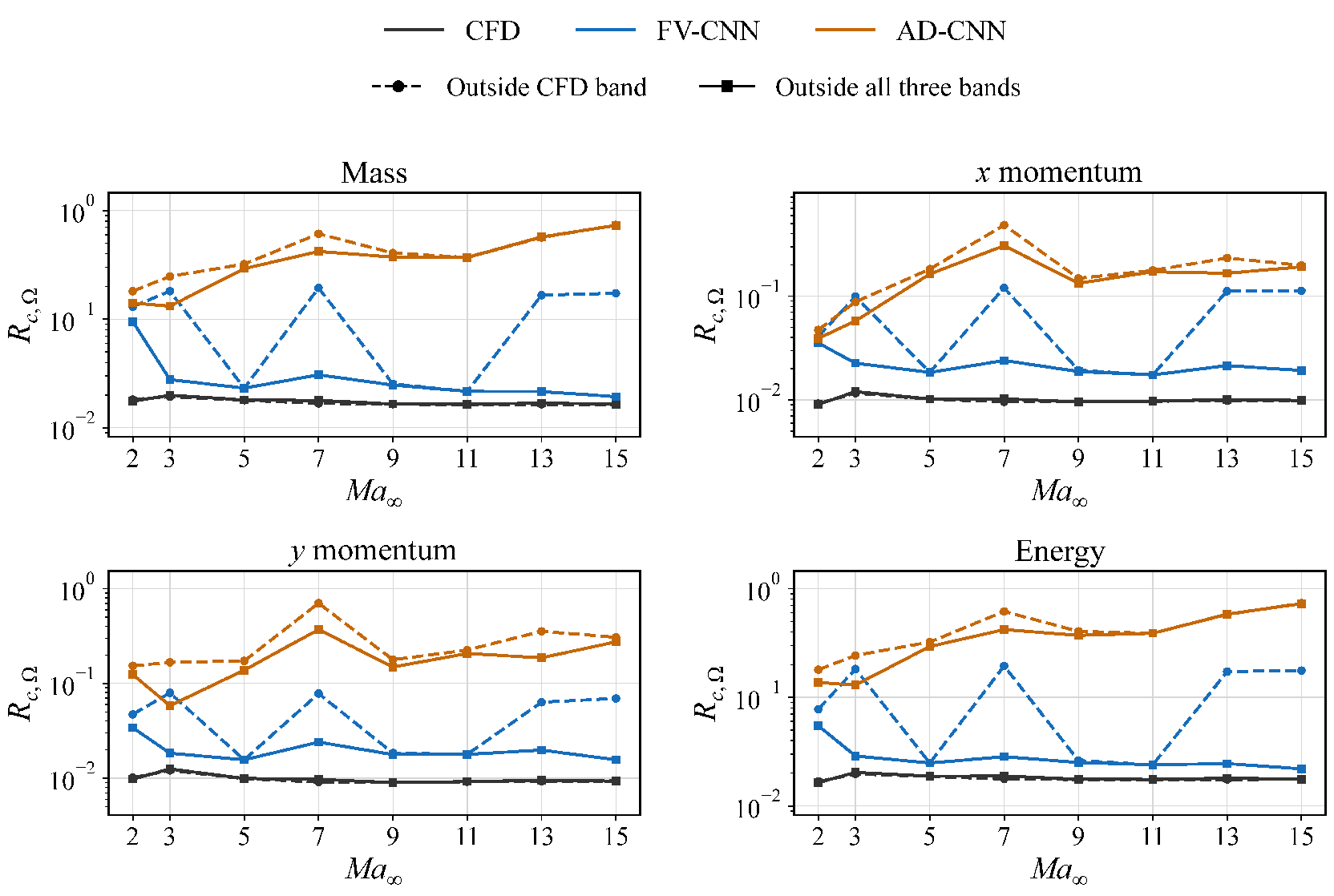}
 \caption{Operator-consistent normalized physical-flux residuals on common nodes. Dashed/circles: outside the CFD-centered band. Solid/squares: outside the union of bands centered on all three fields. Every band has radial half-width $0.10D$; the same mask is applied to all methods within each comparison. The vertical axes are logarithmic. The diagnostic differs from the CFD evolution operator, so its finite value is not a zero-error benchmark.}
 \label{fig:res_smooth}
\end{figure}

The physical-flux divergence inside a numerically resolved shock can be large even when an integrated balance is small. Conversely, a smeared field can reduce a local derivative norm without improving its shock transition. Accordingly, no pointwise residual within the band is interpreted as an independent zero-thickness shock condition. Shared-face cancellation in the training discretization, finite optimization error, and the present physical-flux diagnostic are distinct statements.
\end{revisionred}
\FloatBarrier

\subsection{\textcolor{RevisionColor}{Finite-width balances across the curved shock}}
\label{sec:res_conservation}
\begin{revisionred}
The physical flux is integrated over complete closed curved bands, retaining both a common-volume and a shock-centered assessment. Fifteen equal-angle wedges span the reference angular interval after excluding three rays at each end. For a chosen center $r_s^a(\theta)$, a wedge $V_m^a(h)$ lies between $r_s^a-h$ and $r_s^a+h$. All four faces, including the two radial lateral faces, enter
\begin{equation}
 b^a_{m,c}(h)=\frac{\displaystyle\oint_{\partial V_m^a(h)}
      [F_{x,c}\,dy-F_{y,c}\,dx]}{S_c\ell_m^a(h)},\qquad
 \mathcal B_c^a(h)=\left[\frac1{15}\sum_{m=1}^{15}(b^a_{m,c}(h))^2\right]^{1/2},
 \label{eq:res_band_balance}
\end{equation}
where $m=1,\ldots,15$ indexes the wedges and $a$ labels the band center. The flux in each integral is evaluated from the field being tested. The boundary orientation is counterclockwise and $\ell_m^a(h)$ is the absolute $y$-projection of the outer curved face. The center is a shape-preserving cubic interpolant of the same compression-gradient locator used for the shock-geometry diagnostic. In the common-volume test $a=\mathrm{CFD}$ for every field; in the self-centered test each field supplies its own $r_s^a$. The latter centers the volume on the located compression peak but compares different physical volumes; a finite transition need not be wholly enclosed at every width. Both definitions are therefore reported.

Each face uses 384 Gauss--Legendre nodes with bilinear interpolation of the method's native saved state. Integration uses the physical Euler flux with no HLL dissipation. For every Mach number and all three fields, $h/D=0.06,0.07,\ldots,0.12$ is evaluated. The former $h=0.10D$ common-volume values are reproduced to numerical precision. Constant-state closure is below $2.2\times10^{-15}$ in the additional checks. Doubling face quadrature to 768 nodes in the self-centered $\mathrm{Ma}=2,15$ checks at $h/D=0.06,0.10,0.12$ changes any component RMS by at most $9.85\times10^{-6}$.

At $h=0.10D$ on the CFD-centered volumes, FV-CNN has smaller imbalance in 30 of 32 case--component comparisons. The two exceptions are $\mathrm{Ma}=2$ mass and energy, with $0.0210$ and $0.00970$ versus $0.00776$ and $0.00909$ for AD-CNN. This count varies with half-width: 22, 24, 27, 29, 30, 31, and 31 of 32 over the seven tested widths. The common-volume diagnostic therefore mixes local balance defects with whether the displaced transition is enclosed.

After centering each band on its own field, FV-CNN has smaller imbalance than AD-CNN in all 32 comparisons at every tested width. This ordering also holds when all wedge integrals are normalized by the corresponding CFD-band projected length, ruling out the change of that denominator as its explanation. At $\mathrm{Ma}=15$, the self-centered mass RMS is 0.00580 at $h=0.06D$ and 0.00219 at $h=0.12D$, versus 0.2194 and 0.2618 for AD-CNN. Table~\ref{tab:res_band_width} retains the mass and energy values for both center choices. At $h=0.10D$ the self-centered FV-CNN vector is $(0.00238,0.00475,0.00217,0.00755)$, compared with $(0.2604,0.1439,0.0769,0.2604)$ for AD-CNN.

\begin{figure}[!htb]
 \FVResultColor
 \centering
 \includegraphics[width=.94\linewidth]{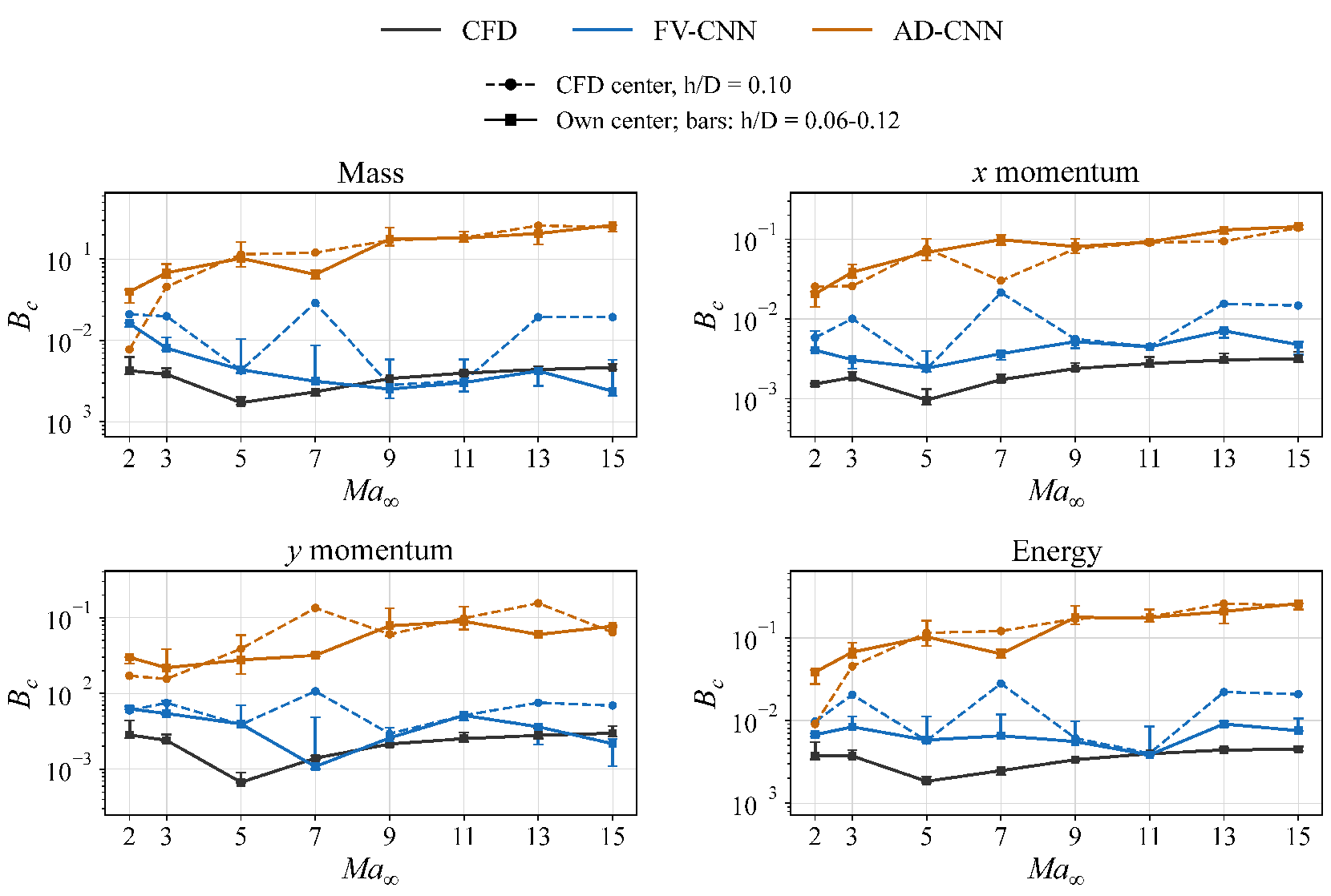}
 \caption{Complete-boundary physical-flux imbalance over 15 curved wedges. The plotted $\mathcal B_c$ is $\mathcal B_c^a$ in Eq.~\eqref{eq:res_band_balance} for the indicated band-center choice. Dashed/circles use the common CFD-centered volumes at $h=0.10D$; solid/squares use each field's own center at that width. Bars show the minimum and maximum self-centered values across the seven tested widths $0.06\leq h/D\leq0.12$, not statistical uncertainty. The self-centered volumes differ between fields. All four faces are included and the axes are logarithmic.}
 \label{fig:res_band}
\end{figure}
\begin{table}[!htb]
\centering\small\FVResultColor
\setlength{\tabcolsep}{6pt}
\caption{$\mathrm{Ma}=15$ finite-band dependence on center and radial half-width. Entries are mass / energy RMS from 15 complete wedges; these dimensionless ratios are not percentages. Both center choices use the same locator and quadrature.}
\label{tab:res_band_width}
\begin{tabular}{@{}c ccc@{}}
\toprule
$h/D$ & CFD & FV-CNN & AD-CNN\\\midrule
\multicolumn{4}{c}{Common CFD-centered bands}\\\midrule
0.06 & 0.0052 / 0.0048 & 0.1296 / 0.1305 & 0.2007 / 0.2007 \\
0.07 & 0.0051 / 0.0048 & 0.1078 / 0.1091 & 0.2175 / 0.2174 \\
0.08 & 0.0050 / 0.0047 & 0.0773 / 0.0788 & 0.2305 / 0.2305 \\
0.09 & 0.0048 / 0.0046 & 0.0435 / 0.0447 & 0.2400 / 0.2399 \\
0.10 & 0.0047 / 0.0045 & 0.0193 / 0.0208 & 0.2464 / 0.2461 \\
0.11 & 0.0045 / 0.0044 & 0.0075 / 0.0111 & 0.2499 / 0.2495 \\
0.12 & 0.0044 / 0.0043 & 0.0032 / 0.0093 & 0.2507 / 0.2502 \\
\midrule
\multicolumn{4}{c}{Each field centered on its own shock}\\\midrule
0.06 & 0.0052 / 0.0048 & 0.0058 / 0.0106 & 0.2194 / 0.2193 \\
0.07 & 0.0051 / 0.0048 & 0.0028 / 0.0082 & 0.2354 / 0.2353 \\
0.08 & 0.0050 / 0.0047 & 0.0021 / 0.0075 & 0.2473 / 0.2474 \\
0.09 & 0.0048 / 0.0046 & 0.0023 / 0.0075 & 0.2555 / 0.2556 \\
0.10 & 0.0047 / 0.0045 & 0.0024 / 0.0075 & 0.2604 / 0.2604 \\
0.11 & 0.0045 / 0.0044 & 0.0023 / 0.0076 & 0.2624 / 0.2621 \\
0.12 & 0.0044 / 0.0043 & 0.0022 / 0.0077 & 0.2618 / 0.2614 \\
\bottomrule
\end{tabular}
\end{table}

The two center choices resolve complementary questions. Common bands compare fluxes on identical physical boundaries; self-centered bands test balances around each field's own located transition. Their difference establishes substantial geometric dependence, rather than a width-independent conservation error. Signed net balances, all four components, and the full center/width sweep are supplied with the data. A finite curved band includes geometry and lateral transport and is not an exact two-state Rankine--Hugoniot jump. Its finite defects, global entropy admissibility, and refinement limits remain separate questions.
\end{revisionred}
\FloatBarrier

\subsection{\textcolor{RevisionColor}{Shock geometry and thermodynamic checks}}
\begin{revisionred}
Figure~\ref{fig:res_geometry} complements the balance tests with quantities that conservation alone does not determine. Shock-position RMS uses 90 interior angular rays and the same compression-gradient locator for every field. The density-transition width is measured locally on each ray about its own shock center: mean states in the outermost $0.01D$ of the interval $r_s\pm0.10D$ define the local jump, and the nearest 10\% and 90\% crossings define its width. These are operational finite-resolution widths based on local density crossings. Reference transition thickness is not an uncertainty estimate for the shock locator. Small location differences are reported descriptively against this grid and are not evidence of superior continuum shock-location accuracy.

The mean FV-CNN width is $0.0238$--$0.0374D$ for $\mathrm{Ma}=3$--$15$, compared with $0.0108$--$0.0144D$ for CFD. AD-CNN gives $0.0596$--$0.1062D$ over the same Mach range. At $\mathrm{Ma}=2$ the respective widths are $0.0818D$, $0.0376D$, and $0.1332D$. Narrowing does not imply a uniformly better shock position: at $\mathrm{Ma}=11$ and $15$, AD-CNN has the smaller position RMS ($0.00435D$ versus $0.00510D$, and $0.00552D$ versus $0.0253D$). This explains why geometric, field, and conservation assessments must be reported separately.

For the axis entropy check, choose the additive entropy constant so that $s_\infty=0$, define $s/c_v=\log(p/p_\infty)-\gamma\log(\rho/\rho_\infty)$ and take the post-minus-pre difference between the fixed CFD-centered radii $r_s^{\rm CFD}(\pi)\mp0.08D$. All FV-CNN values are positive, increasing from $0.113$ at $\mathrm{Ma}=2$ to $3.085$ at $\mathrm{Ma}=15$, compared with $0.132$ and $3.097$ for CFD. The $\mathrm{Ma}=2$ increase is about $14.3\%$ below CFD. Together with its $2.02\%$ downstream enthalpy RMS and the unmet L-BFGS gradient criterion in Sec.~\ref{sec:optimization_dynamics}, this makes $\mathrm{Ma}=2$ a particularly limited convergence and thermodynamic case, despite smaller field norms than some higher-Mach cases. This is a sampled finite-offset check on the stagnation line, not an entropy inequality verified throughout the domain. The upstream entropy penalty is also a training prior and is not independent evidence of admissibility.

The final panel evaluates $H=\gamma p/[(\gamma-1)\rho]+(u^2+v^2)/2$ in the common downstream region. FV-CNN has area-weighted RMS $H/H_\infty-1$ of $2.02\%$ at $\mathrm{Ma}=2$ and $0.207$--$0.552\%$ at $\mathrm{Ma}=3$--$15$. AD-CNN has the smaller total-enthalpy RMS in seven of the eight cases; FV-CNN is smaller only at $\mathrm{Ma}=11$. This metric uses the same CFD-defined downstream mask for both fields and therefore retains its geometric dependence. A small enthalpy discrepancy, like a visually plausible shock, is therefore insufficient to establish simultaneous satisfaction of all four conservation equations.

\begin{figure}[!htb]
 \FVResultColor
 \centering
 \includegraphics[width=.94\linewidth]{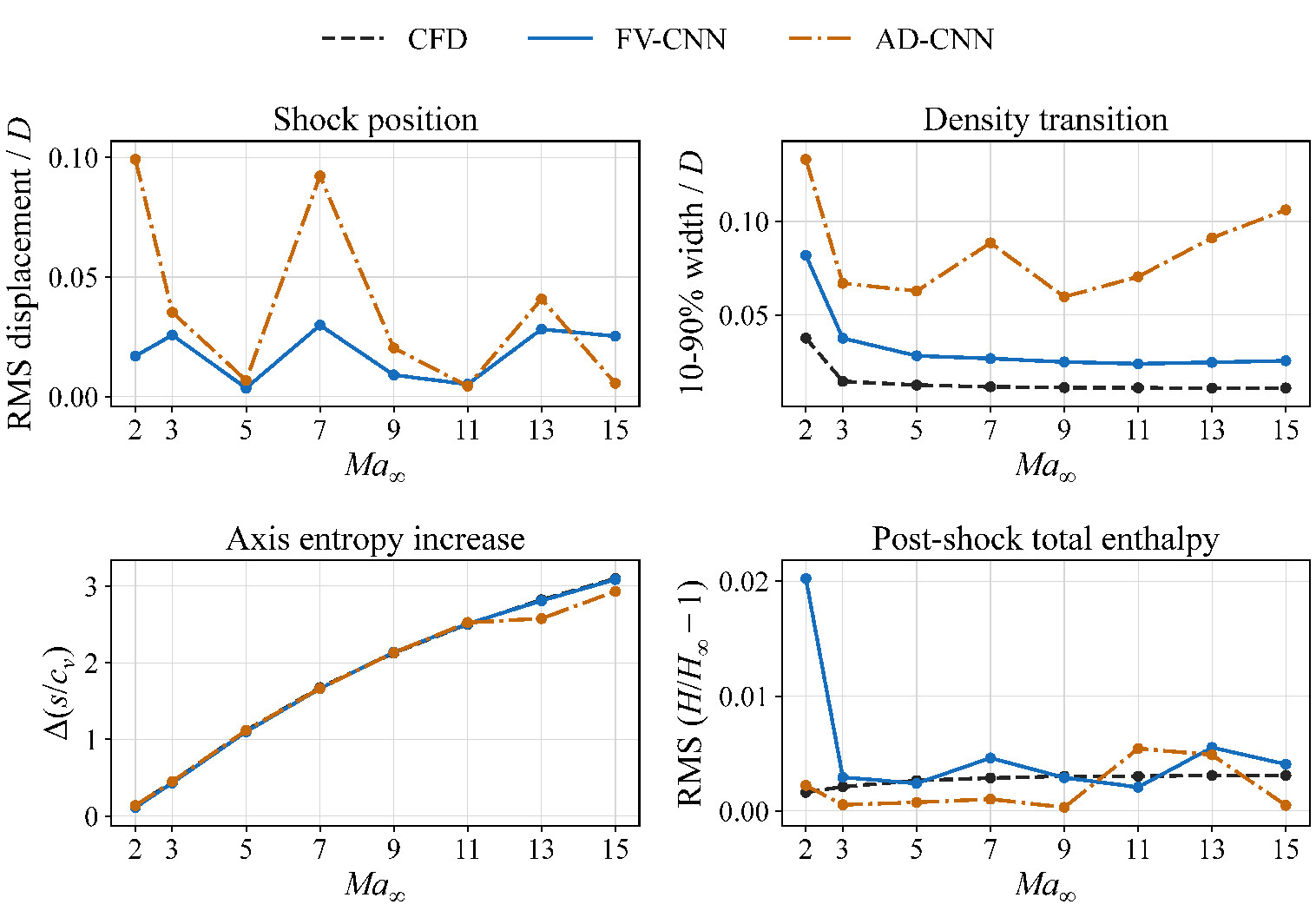}
 \caption{Complementary diagnostics of the adopted fields: angular RMS shock displacement from CFD; mean local density 10--90\% transition width; entropy increase across two CFD-centered stagnation-line sample points; and downstream total-enthalpy RMS. Lengths are normalized by $D$. The width and sampling definitions are given in the text.}
 \label{fig:res_geometry}
\end{figure}

The main results show that the adopted FV-CNN fields combine narrower shock transitions and generally smaller CFD discrepancies with smaller closed-band physical-flux defects relative to the archived AD-CNN states. Finite shock displacement, the exceptions in individual variables, and band-width dependence remain material limitations. The subsequent sensitivity and scaling comparisons address which implementation choices are supported by controlled changes to the initial-stage configuration; they do not replace the all-Mach main results reported here.
\end{revisionred}
\FloatBarrier

\clearpage
\providecommand{\FVSensitivityColor}{\color{RevisionColor}}
\providecommand{\FVSensitivityTableReference}{Table~\ref{tab:fv_sensitivity_settings}}
\providecommand{\FVScalingTableReference}{Table~\ref{tab:fv_scaling_settings}}
\providecommand{\FVMainResultsReference}{Section~\ref{sec:baseline_results}}
\providecommand{\FVNetworkReference}{Section~\ref{sec:fv_network}}
\providecommand{\FVSensitivityFigure}[3]{%
\begin{figure}[!htbp]
 \FVSensitivityColor
 \centering
 \includegraphics[width=.97\linewidth]{Figure_#1.eps}
 \caption{#2 All cases use the shared initial-state, seed-42, 15000-update AdamW protocol with no entropy cleanup or L-BFGS. The unchanged sensitivity baseline is black; gray dashed denotes CFD.}
 \label{#3}
\end{figure}}

\section{\textcolor{RevisionColor}{Component Sensitivity and Mach Scaling}}
\label{sec:component_ablations}

Having established the baseline performance, this section asks a different question: which ingredients are supported by controlled changes to that configuration? The tests below address distinct mechanisms and are not offered as a single ranking of all hyperparameters.

\subsection{\textcolor{RevisionColor}{Common initialization, training stage, and profile diagnostics}}
\begin{revisionred}
The $\mathrm{Ma}=15$ suite contains one baseline and 19 alternatives: 18 component variations in \FVSensitivityTableReference{} and the scaling-off case in \FVScalingTableReference. All start from the same saved, untrained parameter tensors with seed 42 and use 15000 AdamW updates. The case records confirm the same initial-state hash, zero entropy weight, no L-BFGS, and the stated intervention in the effective training arguments. The common baseline has $(\alpha_0,\alpha_f)=(1.5,1.5)$, $\lambda_P=50$, $\lambda_{\rm TV}=0.1$, $\lambda_0=10000$, $x_{\rm fix}=-0.777D$, and both directional couplings enabled. These are initial-stage sensitivity states; the continued-training fields in \FVMainResultsReference{} are assessed separately.

Each factor is represented by one stagnation-line figure in Appendix~\ref{sec:factor_figures}, overlaying its variants on this unchanged baseline. CFD is included as an evaluation reference. Temperature is always $T/T_\infty$, with $T_\infty=1/1.4$. To quantify a change without confusing it with an accuracy measure, define
\begin{equation}
 \Delta_{2,q}(a)=100
 \left[\frac{\displaystyle\int_{x_{\min}}^{-D/2}
       [q_a(x,0)-q_b(x,0)]^2\,dx}
 {\displaystyle\int_{x_{\min}}^{-D/2}q_b(x,0)^2\,dx}\right]^{1/2},
 \label{eq:sens_profile_change}
\end{equation}
Here $q$ is any plotted field ($\rho,u,T/T_\infty,p$), $a$ is the tested case, $b$ is the sensitivity baseline, and $x_{\min}=-R_{\rm nose}$ is the outer-axis coordinate for that Mach case. Physical-$x$ trapezoidal integration uses the 160 saved stagnation-line nodes. The stagnation-streamline (SSL) profile discrepancy $e_{2,q}^{\rm SSL}$ replaces $q_b$ by the coordinate-matched CFD profile in both numerator and denominator and omits the factor 100. Superscript SSL denotes the stagnation streamline. These line norms differ from the area-weighted full-field norms in the main validation. A large $\Delta_{2,q}$ measures sensitivity; it does not specify whether the variant is closer to CFD. The same freestream temperature normalization leaves both relative norms unchanged.
\end{revisionred}

\subsection{\textcolor{RevisionColor}{Numerical dissipation and residual regularization}}
\subsubsection{\textcolor{RevisionColor}{Dissipation factor}}
\begin{revisionred}
Figure~\ref{fig:sens_alpha} compares $\alpha=1.0,1.5,2.0$, with $\alpha_0=\alpha_f=\alpha$ throughout each run. This changes the constant FV numerical dissipation, rather than its annealing schedule. Relative to the $1.5$ baseline, the density and pressure changes are $63.3\%$ and $72.2\%$ for $\alpha=1.0$, versus $19.7\%$ and $24.3\%$ for $\alpha=2.0$. Thus reducing the coefficient to one does not simply sharpen an otherwise unchanged shock.

The CFD comparison also has no single-variable ordering that determines a preferred coefficient: $\alpha=2.0$ reduces the density and pressure profile discrepancies from $13.9\%$ and $16.4\%$ to $12.5\%$ and $15.6\%$, while increasing temperature from $11.6\%$ to $23.8\%$ and velocity from $8.7\%$ to $12.4\%$. These results concern the learned state after the fixed optimization budget, including changes of shock position and post-shock profiles.
\end{revisionred}

\subsubsection{\textcolor{RevisionColor}{TV coefficient}}
\label{sec:tv_robust}
\begin{revisionred}
The TV term penalizes angular density variation in the implementation. Figure~\ref{fig:sens_tv} changes its AdamW coefficient from $0.1$ to $0.01$ or zero, retaining the rest of the FV objective. Both smaller coefficients move the compression transition upstream on the stagnation line. Their density changes from the baseline are $35.0\%$ and $39.0\%$, respectively, even though their wall densities remain close to the imposed target. The associated density discrepancies with CFD are $29.9\%$ and $34.3\%$, compared with $13.9\%$ for the baseline.

The small wall deviations therefore conceal a substantial change in the shock-layer profile. This is an observed sensitivity to the TV coefficient around the prescribed baseline; it does not independently select $\lambda_{\rm TV}=0.1$ as necessary or optimal. It does not attribute the displacement uniquely to physical shock broadening: numerical fluxes, regularization, and the optimization trajectory act together. No L-BFGS stage or later change of TV coefficient enters this comparison.
\end{revisionred}

\subsubsection{\textcolor{RevisionColor}{Overall PDE coefficient}}
\begin{revisionred}
Figure~\ref{fig:sens_pde} compares $\lambda_P=25,50,100$. The lower coefficient changes the density and pressure profiles by $31.7\%$ and $38.5\%$ from the baseline. Doubling the coefficient gives smaller changes, $5.5\%$ and $6.8\%$, but does not improve the CFD profile discrepancies: pressure rises from $16.4\%$ to $21.4\%$. The baseline is closer to CFD than either alternative in all four displayed line norms. The effect is therefore asymmetric about the baseline, and a larger multiplier is not a monotonic route to a more accurate learned field. Weighted total losses from these different objectives would not provide a common accuracy scale.
\end{revisionred}
\FloatBarrier

\subsection{Physical Anchors and the Upstream Fixing Region}
\subsubsection{\textcolor{RevisionColor}{Stagnation-point coefficient}}
\begin{revisionred}
Figure~\ref{fig:sens_stagnation} varies $\lambda_0$ over $0,5000,10000,20000$. Removing this prior leaves the wall density and pressure about $55\%$ below their analytical targets. The density profile discrepancy with CFD is then $55.2\%$, compared with $13.9\%$ for the baseline. This is a failure of the remaining constraints to recover the correct stagnation state within the stated training budget.

Among the nonzero coefficients, the response is not monotonic. For the shared seed-42 realization, the $5000$ case is closer to CFD than the baseline in all four profile norms: density, pressure, temperature, and velocity discrepancies are $6.45\%$, $7.37\%$, $4.15\%$, and $2.92\%$. Increasing the coefficient to $20000$ instead gives $31.2\%$ density and $38.1\%$ pressure discrepancies. The wall values of these nonzero-weight cases remain within approximately $1.3\%$ of their targets. Satisfying a localized anchor closely does not determine the complete shock profile, and this sweep does not establish the baseline coefficient as an optimum. The observed improvement at $5000$ is conditional on that initialization; its $6.45\%$ density value lies below the observed baseline-seed range, but no paired multi-seed estimate of this intervention is available (Sec.~\ref{sec:factor_seed_context}).
\end{revisionred}

\subsubsection{\textcolor{RevisionColor}{Position of the fixing region}}
\begin{revisionred}
The soft upstream term is active where $x<x_{\rm fix}$. Moving $x_{\rm fix}$ from $-0.777D$ to $-0.9D$ or $-1.0D$ reduces this constrained part of the domain while leaving its coefficient at $1000$. Figure~\ref{fig:sens_mask_position} shows substantial changes in all four stagnation-line profiles. The density changes from the baseline are $54.8\%$ and $40.2\%$, and their CFD density discrepancies are $51.5\%$ and $44.8\%$. Thus the smaller fixing region does not merely translate the front by the imposed change of $x_{\rm fix}$, nor is the response monotonic in that change.

The fixing boundary is a prescribed physical prior on this finite domain. These tests show dependence on its placement; they do not independently identify the correct shock location. In particular, the wall anchor can remain nearly satisfied while the transition through the surrounding layer changes markedly.
\end{revisionred}

\subsubsection{\textcolor{RevisionColor}{Strength of upstream fixing}}
\begin{revisionred}
Figure~\ref{fig:sens_mask_weight} tests the separate effect of removing the fixing coefficient, retaining $x_{\rm fix}=-0.777D$. With $\lambda_m=0$, the density and pressure profiles change by $37.3\%$ and $52.2\%$ from the baseline; their CFD discrepancies increase to $41.7\%$ and $56.1\%$. Temperature and velocity also deteriorate. The explicit upstream fixing therefore has a substantial effect in this run, in addition to the outer inflow constraint already retained in the objective. Position and coefficient are separate interventions, and their curves are shown in separate figures.
\end{revisionred}

\subsubsection{\textcolor{RevisionColor}{Additional nose residual}}
\begin{revisionred}
The nose multiplier $s_N$ controls the additional residual weight near the stagnation line; it is distinct from the density/pressure anchor. Setting $s_N=0$ in Fig.~\ref{fig:sens_nose} changes the density and pressure profiles by $18.3\%$ and $22.4\%$. In this case their CFD discrepancies decrease from $13.9\%$ and $16.4\%$ to $9.35\%$ and $11.76\%$, while temperature worsens from $11.6\%$ to $17.5\%$ and velocity changes little ($8.73\%$ to $8.99\%$). The extra nose weighting is therefore not uniformly beneficial according to these line metrics. Its removal does not reproduce the strong wall-state failure caused by removing the stagnation anchor.
\end{revisionred}

\subsubsection{\textcolor{RevisionColor}{Total-enthalpy prior}}
\begin{revisionred}
Removing $\lambda_H=0.5$ in Fig.~\ref{fig:sens_enthalpy} increases all four profile discrepancies with CFD. Density rises from $13.9\%$ to $25.2\%$, pressure from $16.4\%$ to $30.4\%$, and temperature from $11.6\%$ to $28.0\%$. This records sensitivity to removal of the enthalpy prior in the present initial-stage configuration. It is a comparison of learned profiles under two objectives; closed-band physical-flux balances of the final main fields remain a separate diagnostic.
\end{revisionred}
\FloatBarrier

\subsection{\textcolor{RevisionColor}{Coordinate encoding and directional coupling}}
\subsubsection{\textcolor{RevisionColor}{Fourier-feature sensitivity within the full hybrid architecture}}
\label{sec:fourier_hybrid_ablation}
\begin{revisionred}
The Fourier-off intervention in Fig.~\ref{fig:sens_fourier} replaces $\sin(W_{\rm emb}f_e\boldsymbol\zeta+b_{\rm emb})$ by the learned linear embedding $W_{\rm emb}\boldsymbol\zeta+b_{\rm emb}$. Both spatial convolutions and the remaining blocks are retained. Removing this coordinate encoding changes the density and temperature profiles by $41.8\%$ and $56.4\%$ from the baseline. Their CFD discrepancies increase from $13.9\%$ and $11.6\%$ to $37.8\%$ and $52.9\%$, with pressure and velocity also worse. This is a shared-initialization, fixed-budget comparison of the complete sinusoidal encoding with a linear embedding. It changes both a nonlinearity and the scheduled feature scaling, so their individual contributions are not isolated.
\end{revisionred}
\begin{revisionred}
Consequently these data support neither a claim that Fourier features are essential nor a claim that removing them is uniformly more accurate.
\end{revisionred}
\begin{revisionred}
The experiment measures the resulting flow profiles, not a Hessian condition number or a general spectral-bias mechanism.
\end{revisionred}

\subsubsection{\textcolor{RevisionColor}{Meaning and effect of the pointwise intervention}}
\begin{revisionred}
The terms radial and azimuthal refer to directions of neighbor coupling in the network of \FVNetworkReference, not to alternative directions of Euler differentiation. Disabling a spatial convolution selects its center tap. With channel indices $c,c'$ and grid indices $j,i$, these local operators are
\begin{align}
 (\mathcal C_r^{(0)}z)_{jic}
   &=w^{r}_{c,0}z_{jic}+b^r_c,\nonumber\\
 (\mathcal C_\theta^{(0)}z)_{jic}
   &=\sum_{c'}w^\theta_{cc',0}z_{jic'}+b^\theta_c.
 \label{eq:sens_center_taps}
\end{align}
The zero denotes the central spatial offset. The radial-only variant retains radial neighbors and uses $\mathcal C_\theta^{(0)}$; azimuthal-only retains angular neighbors and uses $\mathcal C_r^{(0)}$. Pointwise uses both center-tap operators. The latter retains Fourier encoding, channel transformations, nonlinearities, residual connections, and nodewise LayerNorm, but its forward prediction at one node has no dependence on neighboring input nodes. The FV residual still couples adjacent predicted states through face fluxes.

The parameter tensor shapes and saved initial tensors are retained in these interventions; inactive off-center weights no longer contribute to the forward function. Thus active capacity and receptive field change together. The pointwise FV-trained network is the adopted no-neighbor counterpart for this within-family intervention. A separate conventional-MLP architecture ranking is outside the claim supported by this comparison. For this seed-42 realization, Fig.~\ref{fig:sens_spatial} shows larger CFD profile discrepancies for either one-direction variant than for the complete coupling. Density is $18.9\%$ for radial-only and $16.1\%$ for azimuthal-only, versus $13.9\%$ for the baseline. With both couplings removed it rises to $59.2\%$, and pressure to $70.4\%$. Temperature and velocity show the same ordering relative to the complete network. The $5.06$- and $2.26$-point density differences for the one-direction variants are smaller than the measured $10.35$-point baseline-seed standard deviation, and both values lie inside its observed range. This one-seed ordering has no demonstrated robustness across initializations. The pointwise change is substantially larger; all interventions are compared with the complete baseline dispersion in Sec.~\ref{sec:factor_seed_context}.
\end{revisionred}
\FloatBarrier

\subsection{\textcolor{RevisionColor}{Mach-scaling comparisons at $\mathrm{Ma}=2,3,15$}}
\label{sec:mach_scaling_ablation}
\begin{revisionred}
The separate scaling conditions are given in \FVScalingTableReference. Scaling off sets the squared momentum and energy residual coefficients to $(1,1)$. The enabled coefficients are $(4,16)$ at $\mathrm{Ma}=2$, $(1/9,1/81)$ at $\mathrm{Ma}=3$, and $(1/225,1/50625)$ at $\mathrm{Ma}=15$. The implementation applies these component weights consistently to the PDE, additional nose, and symmetry-face residual terms that use them. Thus the intervention is the recorded Mach weighting within the objective, not a rescaling of the plotted flow variables.

Figure~\ref{fig:sens_scaling} compares all three cases with their CFD profiles, and Table~\ref{tab:sens_scaling_profiles} quantifies the line discrepancies. $\mathrm{Ma}=2$ uses the distinct larger domain and inherited nominal settings, including 15000 AdamW updates followed by L-BFGS. With scaling removed, density and pressure discrepancies rise from $4.14\%$ and $4.78\%$ to $17.65\%$ and $22.44\%$. Temperature and velocity also worsen. 

Both $\mathrm{Ma}=2$ runs use seed 42 and the same initialization procedure. Comparison of the recorded source confirms identical seed-setting functions, model definitions, and execution from seed reset through network construction. Reconstructing the original initialization with PyTorch 2.6.0+cu124 gives all 65 state tensors bitwise identical to the saved initial state of the scaling-off run; the complete state hash also matches. The source snapshots, saved scaling-off initialization, hashes, and verification script requiring only a central processing unit (CPU) are supplied in the companion archive.

This check reconstructs the initialization from the recorded original code; the historical scaling-on initial checkpoint itself was not retained for direct file-to-file comparison. In the AdamW stage, both codes retain \texttt{cudnn.deterministic=False} and \texttt{benchmark=True}. The common initialization is therefore supported by code and tensor reconstruction, while exact repeatability of the graphics processing unit (GPU) optimization trajectory has not been measured. The across-seed dispersion at $\mathrm{Ma}=15$ in Sec.~\ref{sec:seed_ensemble} measures a different variation and cannot be substituted for same-seed execution variability in this comparison.

At $\mathrm{Ma}=3$, the enabled and disabled runs have matching saved initial-state hashes, 15000 AdamW updates, and no L-BFGS. The density and pressure discrepancies change from $3.54\%$ and $4.01\%$ to $41.22\%$ and $50.28\%$. However, the overall PDE coefficient is also $60$ versus $1$. These results characterize two final configurations and cannot isolate Mach weighting from the simultaneous change of PDE strength.

$\mathrm{Ma}=15$ provides the direct scaling control in the sensitivity suite: the saved initial tensors, 15000-update budget, and remaining effective arguments are identical. Density and pressure discrepancies increase from $13.87\%$ and $16.41\%$ to $58.05\%$ and $94.97\%$ when the weights are removed; temperature and velocity rise to $89.61\%$ and $49.88\%$. The disabled run retains finite saved values, but its pressure and temperature profiles are severely distorted. This records a large response to removal of the weighting in this controlled run. At fixed $\lambda_P=50$, the intervention changes both relative component weights and the aggregate strength of the PDE contribution against the priors. A separate scalar-strength-matched control has not been performed; component-gradient balancing is therefore not isolated as the mechanism.

\begin{figure}[!htbp]
 \FVSensitivityColor
 \centering
 \includegraphics[width=\linewidth]{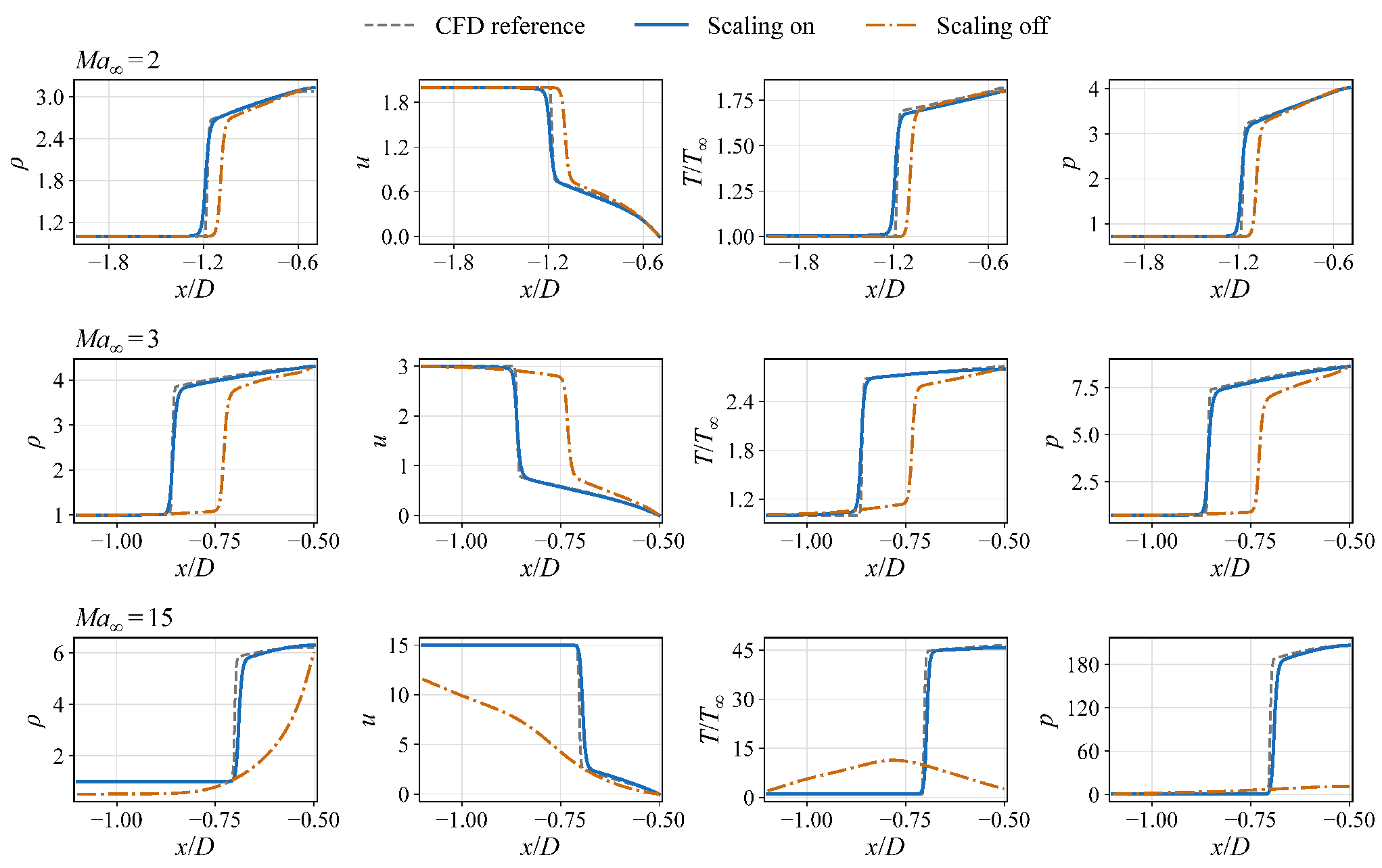}
 \caption{Mach-scaling comparisons on the stagnation streamline. Rows: $\mathrm{Ma}=2,3,15$; columns: $\rho,u,T/T_\infty,p$. Gray dashed: CFD; blue solid: scaling on; orange dash-dot: scaling off. $\mathrm{Ma}=2$ uses its own physical domain; its common seed-42 initialization is verified by reconstruction from the recorded code. GPU trajectory repeatability remains unmeasured. $\mathrm{Ma}=3$ also changes $\lambda_P$ from 60 to 1. $\mathrm{Ma}=15$ changes only Mach weighting within the common-initialization sensitivity protocol. No entropy continuation is used in these scaling tests.}
 \label{fig:sens_scaling}
\end{figure}
\begin{table}[!htb]
\FVSensitivityColor
\centering\small
\setlength{\tabcolsep}{9pt}
\caption{Stagnation-line discrepancies $100e_{2,q}^{\rm SSL}$ with CFD for scaling on/off, in percent. The physical-$x$ trapezoidal norms use the domain of each Mach case. These are profile norms, distinct from the full-field main-result norms. Conditions and comparability limits are specified in the text and the separate scaling-settings table.}
\label{tab:sens_scaling_profiles}
\begin{tabular}{@{}ccrrrr@{}}
\toprule
$Ma_\infty$ & Scaling & $\rho$ & $u$ & $T/T_\infty$ & $p$\\\midrule
2 & on & 4.14 & 5.26 & 3.14 & 4.78 \\
2 & off & 17.65 & 17.89 & 10.67 & 22.44 \\
3 & on & 3.54 & 3.90 & 3.12 & 4.01 \\
3 & off & 41.22 & 51.36 & 32.52 & 50.28 \\
15 & on & 13.87 & 8.73 & 11.61 & 16.41 \\
15 & off & 58.05 & 49.88 & 89.61 & 94.97 \\
\bottomrule
\end{tabular}
\end{table}

\end{revisionred}
\begin{revisionred}
\subsection{\textcolor{RevisionColor}{Initialization sensitivity at $\mathrm{Ma}=15$}}
\label{sec:seed_ensemble}
\subsubsection{Fixed protocol and complete reporting}
Ten additional initializations, seeds 1--10, are evaluated together with the unchanged adopted seed 42. The seed list, training budgets, exclusions policy, and density-dispersion comparison were recorded before the new runs in the plan dated 16 October 2026, 13:31:41 Coordinated Universal Time (UTC). Its exact bytes, 256-bit Secure Hash Algorithm (SHA-256) identity, and matching execution-lock hash are archived. The per-run status record gives the first new run (seed 1) a start time of 16 October 2026, 13:37:26 UTC, 344.4 seconds after the plan timestamp. The additional balance and factor-context checks below are subsequent exploratory postprocessing, rather than additions to the prespecified decision rule. Each new run executes 15000 AdamW updates followed immediately by 10000 entropy-penalized AdamW updates from its own final checkpoint. Only the seed and output path change: $\lambda_P=50$, $\lambda_0=10000$, $(\alpha_0,\alpha_f)=(1.5,1.5)$, $\lambda_{\rm TV}=0.1$, $x_{\rm fix}=-0.777D$, the Mach weights, network, grid, and remaining priors are retained. Continuation uses fresh AdamW moments, learning rate $10^{-5}$, $\lambda_{\rm ent}=20000$ from its first update, and the inherited terminal nose coefficient. No L-BFGS is used.

All new runs use PyTorch 2.6.0+cu124, CUDA 12.4, cuDNN 90100, and the same NVIDIA GeForce GTX 1070. The recorded flags match the baseline: \texttt{cudnn.deterministic=True}, \texttt{benchmark=False}, both TF32 flags false, and \texttt{deterministic\_algorithms=False}. Distinct initial parameter tensors and exact transfer of each initial-stage endpoint to its own continuation are verified from the saved states. All ten runs complete both budgets. All eleven fields, including large-discrepancy outcomes, are retained; no seed is replaced or excluded according to accuracy or shock quality. The original seed 42 remains the main adopted field. Native arrays, model states, per-run information, both loss histories, the fixed experiment plan, and every individual metric are archived.

The five full-domain relative $L_2$ discrepancies use the same area weighting and native $96\times160$ quadrant as Table~\ref{tab:res_field_errors}. The shock locator and angular sampling follow the main geometry diagnostic; the $10$--$90\%$ density width is averaged over 90 rays. The four stagnation-line norms use the physical-coordinate definition in Sec.~\ref{sec:component_ablations}. Temperature is normalized by $T_\infty$. Figure~\ref{fig:seed_full_domain} displays every seed at both stages, together with the mean, sample standard deviation (SD, with $n-1$ denominator), minimum, and maximum. Table~\ref{tab:seed_individual} lists each final density norm, shock-position RMS, transition width, and four closed-band defects. These eleven-run summaries deliberately include the previously adopted seed; they are descriptive summaries of this specified set, not an unbiased sample of historical selection. Summaries of the ten new runs alone are also archived.

\subsubsection{Full-domain discrepancies and the prespecified comparison}
After entropy continuation, density discrepancy ranges from $12.04\%$ to $43.70\%$, with mean $24.34\%$ and sample standard deviation $S=11.78$ percentage points (pp). The range is $R=31.67$ points. The ten new runs alone give mean $25.04\%$ and standard deviation $12.17$ points. Pressure, temperature, streamwise velocity, and transverse velocity also vary appreciably: their final mean $\pm$ standard deviation values are $(29.02\pm15.91)\%$, $(27.29\pm15.11)\%$, $(9.60\pm5.30)\%$, and $(35.48\pm15.89)\%$, respectively. These standard deviations describe dispersion of individual outcomes, rather than uncertainty bars on their means.

The prespecified density comparison uses the historical single-field FV--AD gap $G=20.6894707-17.2928236=3.3966471$ percentage points. Summary statistics are calculated before rounding the displayed extrema. The recorded decision categories are $S<G$ with $R<G$, $S<G$ with $R\geq G$, or $S\geq G$. The observed category is $S\geq G$, with $S/G=3.47$ and $R/G=9.32$. Thus the adopted-field gap is smaller than the measured baseline dispersion. A generic accuracy advantage of FV-CNN over AD-CNN cannot be inferred from those two historical realizations. Conversely, one historical AD field does not determine the AD-CNN outcome distribution or establish its statistical superiority over FV-CNN.

The eleven density outcomes are unevenly grouped: six lie at $12.04$--$19.28\%$, two at $22.25$--$26.28\%$, and three at $38.18$--$43.70\%$. These observed groupings describe the small sample; neither a normal error distribution nor a statistically established two-mode distribution is assumed. Six of eleven outcomes (seeds 42, 1, 3, 6, 7, and 10), including five of the ten new runs, are below the historical AD-CNN value $20.69\%$. This reference value was fixed in the pre-experiment plan. The count describes comparison with that single historical field rather than a method-wide success probability.

The adopted density discrepancy, $17.29\%$, ranks fourth among the eleven fields. New seeds 1, 6, and 7 have smaller discrepancies in all five variables than seed 42. Therefore comparable or better fields are obtained from more than one initialization. This count is a descriptive comparison with the adopted field, not a success rate defined after inspection of the results. All final fields have a pressure compression ratio of at least two on all 90 sampled rays and valid density-width crossings. The observed variability concerns shock geometry and field agreement even when a compression front is detected throughout the sampled arc.

\begin{figure}[!htbp]
 \centering
 \includegraphics[width=\linewidth]{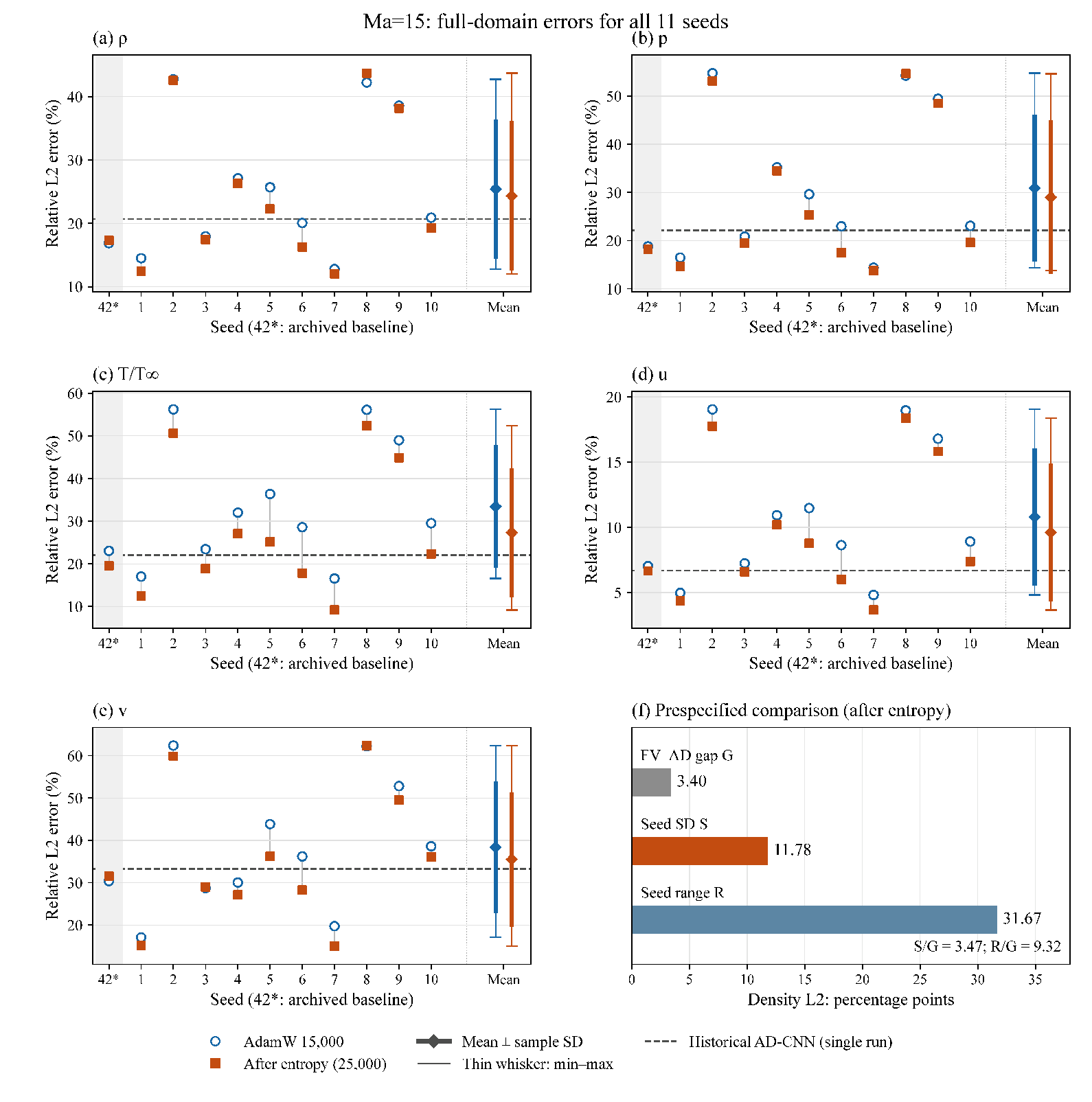}
 \caption{Complete $\mathrm{Ma}=15$ initialization experiment: five area-weighted full-domain discrepancies and the prespecified density-dispersion comparison. Open circles: 15000 AdamW updates; filled squares: the subsequent 10000 entropy-penalized updates. All ten new seeds and the unchanged archived seed 42 (shaded) are shown. Diamonds and thick whiskers give the eleven-run mean and sample standard deviation; thin capped whiskers span the minimum and maximum. Dashed lines are historical single-run AD-CNN values, not AD ensemble statistics. Panel (f) compares the final density dispersion with the historical adopted-field gap, in percentage points.}
 \label{fig:seed_full_domain}
\end{figure}
\FloatBarrier

\subsubsection{Shock geometry, continuation, and component interpretation}
The supplementary geometry and line-norm plots in Appendix~\ref{sec:seed_supplement} retain every seed and both endpoints. Final shock-position RMS ranges from $0.00929D$ to $0.08134D$, whereas the mean density-transition width ranges only from $0.02281D$ to $0.02607D$; the CFD width is $0.01084D$. The historical AD-CNN width is $0.10620D$, giving AD/FV width ratios of $4.07$--$4.66$ over all eleven final fields. The narrower FV transition is therefore maintained across these initializations, despite their different position errors. Across the eleven final fields, density discrepancy and shock-position RMS have Pearson correlation $0.981$. This association supports a connection between field discrepancy and displacement, while it does not partition the error causally or remove post-shock profile differences. Similar local transition widths can coexist with markedly different shock locations and global norms.

For the high-discrepancy seeds 2, 8, and 9, the axis front centers are $-0.75044D$, $-0.75020D$, and $-0.75369D$, respectively, leaving gaps of $0.02656D$, $0.02680D$, and $0.02331D$ from $x_{\rm fix}=-0.777D$. These three centers cluster within $0.00348D$ (about one native axis-grid interval), at distances of $1.02$--$1.13$ times their respective angular-mean transition widths from the fixing boundary; finite offsets alone do not exclude an influence of that region. Table~\ref{tab:seed_individual} lists the axis front centers for all eleven seeds.

Entropy continuation reduces shock-position RMS and transition width in all eleven runs. Full-domain density discrepancy decreases in nine, pressure in ten, temperature and streamwise velocity in all eleven, and transverse velocity in eight. Mean density discrepancy decreases from $25.40\%$ to $24.34\%$, while its standard deviation changes from $11.02$ to $11.78$ points. Continuation therefore sharpens and repositions the fronts under this protocol without eliminating initialization sensitivity or uniformly improving every variable.

At the initial stage used for the factor tests, baseline density $e_{2,\rho}^{\rm SSL}$ has sample standard deviation $10.35$ percentage points. The seed-42 change from $\alpha=1.5$ to $2.0$ is $13.8709\%-12.4805\%=1.3904$ points. Baseline dispersion is substantially larger than that difference, so the one-seed improvement cannot be assumed to persist across initializations. However, only the baseline has been repeated over seeds: this comparison neither estimates the paired treatment-effect distribution nor establishes that the dissipation intervention has zero effect. The component plots remain conditional responses from their common initial state. No auditable same-seed repeated-training records, including execution conditions and an endpoint-matching criterion, are retained. Initial-weight reconstruction and checkpoint-transfer checks do not constitute repeated full training. The retained data therefore do not quantify same-seed execution variability or separate it from the variability associated with changing initialization.

\subsubsection{Closed-band balances across initializations}
\label{sec:seed_band_balances}
The final native fields of all eleven seeds are evaluated with the same self-centered closed-band procedure as Sec.~\ref{sec:res_conservation}: radial half-width $h=0.10D$, 15 angular wedges, 384 Gauss points per face, and the same freestream-flux normalization. Each band follows its own field's located front. Figure~\ref{fig:seed_bands} shows all outcomes against the unchanged historical AD-CNN and CFD fields. The normalized FV-CNN RMS ranges are $0.00238$--$0.00443$ for mass, $0.00475$--$0.00886$ for streamwise momentum, $0.00217$--$0.01492$ for transverse momentum, and $0.00705$--$0.01254$ for energy. All 44 seed--component values are below the corresponding historical AD-CNN values, with FV/AD ratios $0.00914$--$0.194$. The adopted seed 42 gives the smallest mass and both momentum defects among the eleven fields, and the second-smallest energy defect; its values lie at a favorable end of this observed set. Ratios of the eleven-seed mean defects to the fixed historical AD values are $0.01279$, $0.04604$, $0.08750$, and $0.03535$, respectively.

The CFD defects are $0.00465$, $0.00319$, $0.00298$, and $0.00449$ in the same component order. FV/CFD ratios span $0.512$--$0.952$ for mass, $1.49$--$2.78$ for streamwise momentum, $0.727$--$5.00$ for transverse momentum, and $1.57$--$2.79$ for energy. Thus all eleven FV mass defects are below CFD, whereas all streamwise-momentum and energy defects and eight transverse-momentum defects are above it. The CFD values provide reference scales from the first-order CFD field and band quadrature, rather than lower bounds; smaller evaluated defects alone do not establish greater conservation or field accuracy than CFD. Table~\ref{tab:seed_individual} retains every final outcome.

The archived seed-42, AD-CNN, and CFD results reproduce the original diagnostic to roundoff. Selected fields are rechecked with 768 points per face, and uniform-state closure remains below $10^{-10}$. These saved-field checks show that the small self-centered balance defects persist even for the seeds with large full-domain discrepancies. Together with the width results, they distinguish repeatable local sharpness and this balance diagnostic from accurate placement of the complete shock layer. They do not replace the common-volume comparisons in Sec.~\ref{sec:res_conservation} or an AD seed ensemble.
\end{revisionred}

\subsection{\textcolor{RevisionColor}{Scope of the component evidence}}
\label{sec:factor_seed_context}
\begin{revisionred}
Figure~\ref{fig:factor_seed_context} places all 19 seed-42 interventions beside the eleven baseline initial-stage density line norms. Baseline $100e_{2,\rho}^{\rm SSL}$ ranges from $9.85\%$ to $34.77\%$, with sample standard deviation $S_{\rm SSL}=10.35$ percentage points; the seed-42 baseline is $13.87\%$, toward the lower-discrepancy group. The plotted difference $\Delta=100[e_{2,\rho}^{\rm SSL}(a)-e_{2,\rho}^{\rm SSL}(42)]$, in percentage points, is divided by this sample standard deviation solely as a descriptive scale.

Nine interventions exceed the observed baseline maximum: $\alpha=1.0$, pointwise, scaling off, $\lambda_0=0$, both fixing-boundary shifts, $\lambda_m=0$, Fourier off, and $\lambda_P=25$. Their density changes from seed 42 span $2.29$--$5.02$ baseline standard deviations. Pointwise, scaling off, and removal of the stagnation anchor give $4.38$, $4.27$, and $4.00$, respectively. These are large conditional contrasts. Scaling off still changes aggregate PDE strength as well as component proportions.

The TV variants, $\lambda_0=20000$, $\lambda_H=0$, $\lambda_P=100$, radial-only, azimuthal-only, and $\alpha=2.0$ lie inside the observed baseline range. The smaller stagnation coefficient and removal of the additional nose term instead give $6.45\%$ and $9.35\%$, below its minimum; their changes are $-0.72$ and $-0.44$ baseline standard deviations. Thus range membership and change divided by standard deviation are different descriptive summaries.

None of these summaries estimates the distribution of a paired intervention effect. A result beyond eleven baseline observations is not by itself a significance test, and overlap with their range does not establish that an intervention has zero effect or is merely execution noise. In particular, the apparent benefits of $\lambda_0=5000$, $s_N=0$, or $\alpha=2.0$, and the one-direction coupling ordering, remain unverified across initializations. Every factor uses the common saved initial tensors; the retained settings are \texttt{cudnn.deterministic=True}, \texttt{benchmark=False}, disabled TF32, and \texttt{deterministic\_algorithms=False}~\cite{pytorchreproducibility}. Auditable same-seed repeated-training records and paired multi-seed intervention estimates are not available.

The controlled $\mathrm{Ma}=15$ comparisons support conclusions about this network and its initial-stage training budget. The $\mathrm{Ma}=2$ and $3$ scaling comparisons have the distinct limitations stated above. The output lower bound is another common prior. With the explicit criterion $\rho-0.5\leq0.001$, 922 of 15,360 nodes in the $\alpha=1$ case and 4,969 nodes in the scaling-off case lie near that bound. These saturation counts accompany the profile failures; they do not establish their unique cause. The exact counts for all variants accompany the data. The $\mathrm{Ma}=15$ sensitivity-baseline arrays coincide with the main 15,000-update arrays, while the adopted main field includes continuation. Table~\ref{tab:fv_stage_accuracy} compares those stages with identical evaluation metrics. The fixed-budget component tests remain distinct from that final-field assessment.
\end{revisionred}

\FloatBarrier
\providecommand{\FVOptimizationMethodsReference}{Sec.~\ref{sec:fv_protocol}}
\providecommand{\FVOptimizationADReference}{Sec.~\ref{sec:fv_ad_failure}}
\providecommand{\FVOptimizationResultsReference}{Sec.~\ref{sec:results}}
\providecommand{\FVOptimizationSensitivityReference}{Sec.~\ref{sec:component_ablations}}
\providecommand{\FVOptimizationMainTableReference}{Table~\ref{tab:fv_main_settings}}
\providecommand{\FVOptimizationFloatColor}{\color{RevisionColor}}
\section{Optimization Dynamics and Computational Cost}
\label{sec:optimization_dynamics}
\begin{revisionred}
The baseline fields and the component tests address solution quality. The optimization histories identify the stages that produced those fields, and the timing comparison reports initial-training and L-BFGS costs.
\end{revisionred}
\RevisionAddition{The optimization histories and wall-clock accounting are therefore reported separately; the plotted loss is not interpreted as a direct error measure.}

\subsection{Convergence Analysis and Optimization Dynamics Across Mach Numbers}
\begin{figure}[p]
\FVOptimizationFloatColor
\centering
\includegraphics[width=\linewidth]{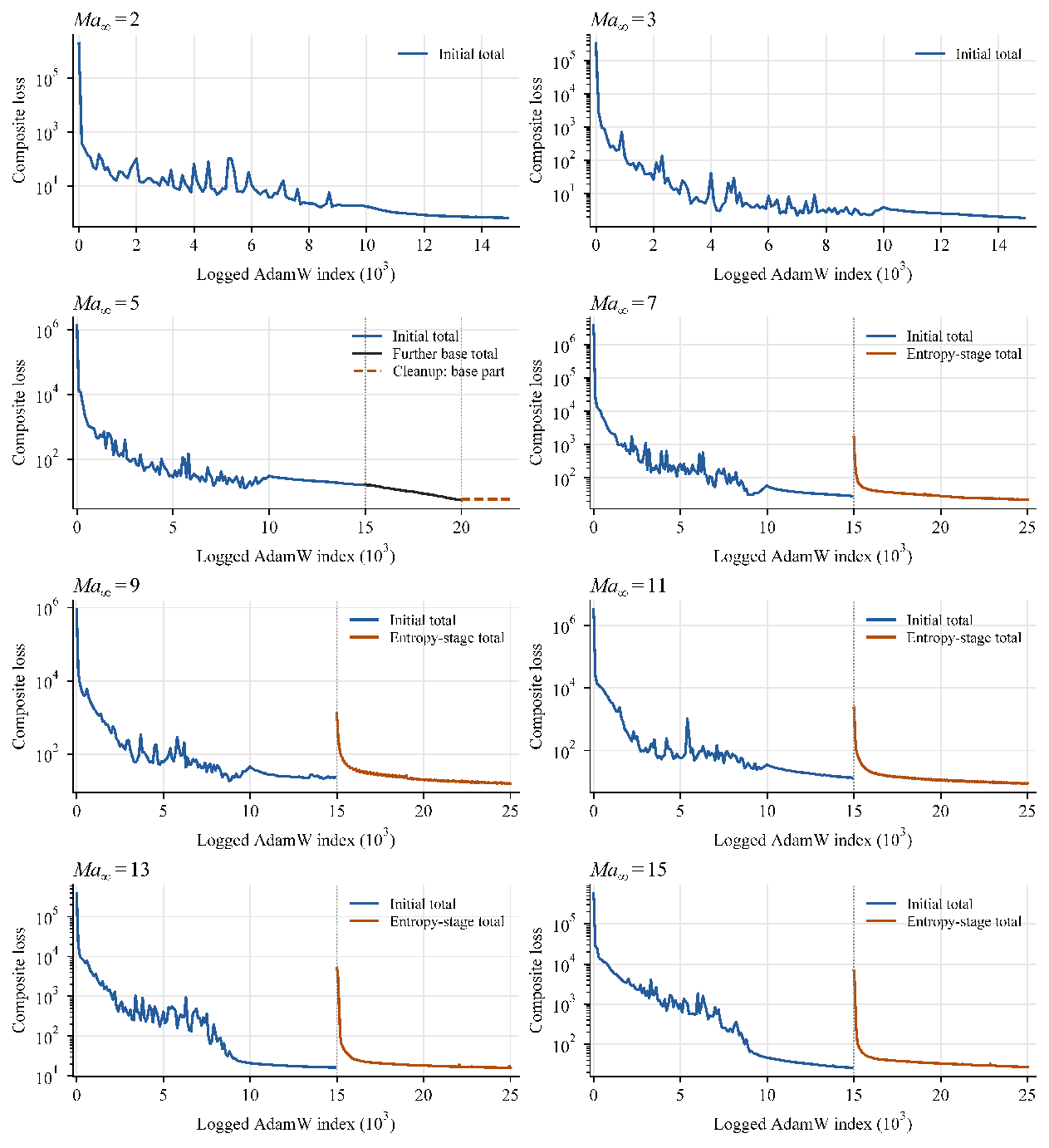}
\caption{Main FV-CNN loss histories. Blue curves show the initial composite total; orange solid curves show the total after adding the upstream entropy penalty in $\mathrm{Ma}=7$--15. The $\mathrm{Ma}=5$ black curve shows the further base-objective stage and its orange dashed curve shows only the base part during entropy continuation, not a combined entropy-stage objective. Vertical dotted lines mark changes of stage. Curves on opposite sides of a stage change are not connected. The abscissa preserves the recorded labels: initial losses have zero-based labels 0--14,900, whereas the inherited entropy logs have labels 15,001--25,000. The last initial point is not an evaluation of the final saved 15,000-update field. The $\mathrm{Ma}=2$ L-BFGS history has a separate counter in Fig.~\ref{fig:fv_optimizer_diagnostics}(a).}
\label{fig:fv_optimizer_histories}
\end{figure}

\begin{revisionred}
Figure~\ref{fig:fv_optimizer_histories} uses the saved FV-CNN histories for all eight main Mach-number cases. The executed stages follow \FVOptimizationMethodsReference\ and are summarized in Table~\ref{tab:fv_optimizer_accounting}. The optimizer used for the initial 15,000 updates is AdamW. There is no common AdamW--L-BFGS sequence for the current main results: only $\mathrm{Ma}=2$ includes L-BFGS, $\mathrm{Ma}=3$ uses the initial AdamW result, and the other cases include further AdamW optimization. In particular, the old AD-CNN L-BFGS histories for $\mathrm{Ma}=3$--15 are not histories of the FV-CNN computations.

The initial logged losses decrease substantially in every case, with non-monotone fluctuations along the way. These are histories of the executed composite objectives. The Fourier scale, the learning rate, the nose coefficient, and, in some cases, the numerical-flux dissipation vary during the initial stage according to the case settings. The objectives also differ between Mach numbers through their residual weights and boundary penalties. A lower plotted total therefore does not rank two cases by physical accuracy, and the initial trajectory is not minimization of one fixed objective throughout training.
\end{revisionred}

\clearpage

\begin{table}[htbp]
\FVOptimizationFloatColor
\centering\small
\caption{Executed optimizer stages for the main FV-CNN fields. Counts in the AdamW columns describe progress through the saved field used in the main comparison. The $\mathrm{Ma}=5$ entropy continuation uses its own optimizer and rollback mechanism; rejected attempts can add work beyond its progress counter.}
\label{tab:fv_optimizer_accounting}
\begin{tabular}{@{}r r r r r@{}}
\toprule
& Initial & Further base & Entropy & L-BFGS\\
$Ma_\infty$ & AdamW & AdamW & continuation & iterations / calls\\
\midrule
2 & 15,000 & 0 & 0 & 8,412 / 10,002\\
3 & 15,000 & 0 & 0 & 0 / 0\\
5 & 15,000 & 5,000 & 2,500 & 0 / 0\\
7 & 15,000 & 0 & 10,000 & 0 / 0\\
9 & 15,000 & 0 & 10,000 & 0 / 0\\
11 & 15,000 & 0 & 10,000 & 0 / 0\\
13 & 15,000 & 0 & 10,000 & 0 / 0\\
15 & 15,000 & 0 & 10,000 & 0 / 0\\
\bottomrule
\end{tabular}
\par\medskip\raggedright\footnotesize
``Calls'' denotes closure evaluations, including line-search and diagnostic evaluations, rather than optimizer iterations. All five $\mathrm{Ma}=7$--15 continuations use AdamW at fixed learning rate $10^{-5}$ and entropy coefficient 20,000. Their inherited base settings remain case-dependent.
\end{table}

\begin{figure}[htbp]
\FVOptimizationFloatColor
\centering
\includegraphics[width=\linewidth]{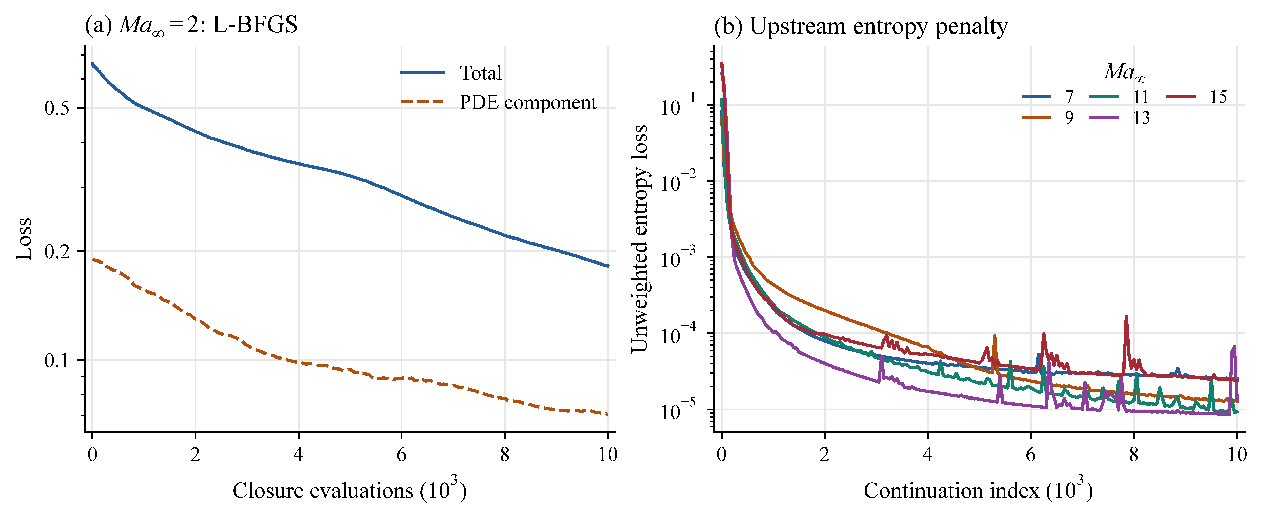}
\caption{Stage-specific optimization diagnostics. (a) $\mathrm{Ma}=2$ accepted-state total and PDE component against closure evaluations; trial line-search losses are not shown. The PDE component already contains the case's Mach-component weights and is not an independent physical Euler error. (b) Unweighted upstream entropy loss during the five fixed-coefficient continuations. The horizontal coordinate is the recorded global label minus 15,000. Neither panel is an error history against CFD.}
\label{fig:fv_optimizer_diagnostics}
\end{figure}

\begin{revisionred}
At $\mathrm{Ma}=2$, the accepted-state L-BFGS record distinguishes 8,412 optimizer iterations from 10,002 closure calls, with float64 arithmetic in this stage. Its recorded total decreases from 0.6612 at entry to 0.1819 at the final state [Fig.~\ref{fig:fv_optimizer_diagnostics}(a)]. The configured evaluation budget is 10,000; the last accepted block reaches call 10,001 and the final diagnostic reaches 10,002. This is consistent with termination at the evaluation budget, not evidence that the gradient tolerance was met: the final recorded maximum parameter-gradient magnitude is about 3.80, above the configured $10^{-7}$ tolerance. The distinction matters both for counting optimization work and for interpreting termination.

For $\mathrm{Ma}=7,9,11,13,15$, 10,000 additional AdamW updates add the upstream entropy penalty with coefficient 20,000 from the start, at fixed learning rate $10^{-5}$. The base settings are inherited from the corresponding completed initial run. Figure~\ref{fig:fv_optimizer_diagnostics}(b) separates the unweighted entropy component from the composite total. Between the first and last recorded continuation evaluations it decreases by factors of approximately $3.18\times10^3$, $5.05\times10^3$, $1.24\times10^4$, $1.92\times10^4$, and $1.42\times10^4$, respectively. The pressure-based upstream mask can change with the predicted field, so these factors describe the executed penalty and mask rather than an error norm on one fixed set of nodes. The initial rise of the total when the new penalty is introduced is a change of objective, not by itself evidence of numerical divergence.
\end{revisionred}

\begin{revisionred}
The $\mathrm{Ma}=5$ main field follows 5,000 further base-objective updates and the saved state at local index 2,500 of the entropy continuation. Its frozen-mask entropy loss decreases from $1.393\times10^{-3}$ to $9.436\times10^{-4}$, while the base loss changes from 5.508 to 5.783 (about $+5.00\%$). The effective entropy learning rate at that saved state is $6.04\times10^{-14}$ after adaptive backoff, despite the much larger nominal schedule. Thus a nearly flat late curve in this stage must not be described as evidence that an unconstrained minimizer has been reached. The curve uses retained states through the main field only; it neither includes subsequent states nor treats rejected and restored update blocks as accepted progress.

Training termination and physical validation remain separate. The CFD discrepancies, smooth-region Euler residuals, and finite-band flux balances in \FVOptimizationResultsReference\ assess the saved fields using the distinct operators and regions specified there. Table~\ref{tab:fv_stage_accuracy} adds direct evaluation of the saved stage endpoints. All stages use the same CFD nodes and the same reference-defined upstream/downstream masks, rather than comparing changing training-mask losses. The upstream RMS of $Z-1$ decreases in every entropy continuation, but other metrics are not uniformly improved. At $\mathrm{Ma}=7$, density $e_2$ changes from $17.30\%$ to $18.36\%$ and pressure from $19.57\%$ to $19.70\%$. At $\mathrm{Ma}=15$, density changes from $16.88\%$ to $17.29\%$, while pressure decreases from $18.78\%$ to $18.19\%$. Self-centered mass imbalance increases at $\mathrm{Ma}=9,11,13$. These tradeoffs prevent interpreting entropy cleanup as general error reduction. At $\mathrm{Ma}=5$, the immediate before/after-entropy density values are nearly unchanged ($5.4284\%$ and $5.4286\%$), whereas the fixed-region entropy diagnostic decreases. The saved $\mathrm{Ma}=15$ sensitivity baseline and main initial-stage arrays match exactly in every stored field; this identity is not an independent repeat-run reproducibility experiment.

\begin{table}[!htb]
\FVOptimizationFloatColor\centering\small
\caption{Evaluation of the recorded continuation stages on identical CFD nodes and fixed physical masks. $e_\rho=e_{2,\rho}$ and $e_p=e_{2,p}$ are the area-weighted relative field discrepancies of Eq.~\eqref{eq:fv_evaluation_norms}; $R_Z$ is area-weighted RMS of $Z-1$ in the fixed reference-defined upstream region. The last column is self-centered mass-balance RMS at $h=0.10D$.}
\label{tab:fv_stage_accuracy}
\setlength{\tabcolsep}{5pt}
\begin{tabular}{@{}r r r r r r r@{}}
\toprule
$Ma_\infty$ & Updates & $100e_\rho$ & $100e_p$ & $100R_Z$ & Shock RMS/$D$ & $\mathcal B_\rho$\\
\midrule
5 & 15,000 & 5.86 & 6.56 & 6.32 & 0.00345 & 0.00411 \\
5 & 20,000 & 5.43 & 6.26 & 3.91 & 0.00359 & 0.00436 \\
5 & 22,500 & 5.43 & 6.25 & 3.16 & 0.00352 & 0.00436 \\
\addlinespace[2pt]
7 & 15,000 & 17.30 & 19.57 & 33.83 & 0.02957 & 0.00406 \\
7 & 25,000 & 18.36 & 19.70 & 2.13 & 0.02985 & 0.00315 \\
\addlinespace[2pt]
9 & 15,000 & 8.93 & 9.87 & 24.90 & 0.00856 & 0.00233 \\
9 & 25,000 & 8.46 & 9.55 & 0.17 & 0.00900 & 0.00252 \\
\addlinespace[2pt]
11 & 15,000 & 9.36 & 10.53 & 39.84 & 0.01110 & 0.00202 \\
11 & 25,000 & 6.27 & 7.73 & 0.19 & 0.00510 & 0.00303 \\
\addlinespace[2pt]
13 & 15,000 & 19.66 & 21.98 & 143.17 & 0.03483 & 0.00419 \\
13 & 25,000 & 18.69 & 19.07 & 5.32 & 0.02813 & 0.00421 \\
\addlinespace[2pt]
15 & 15,000 & 16.88 & 18.78 & 152.56 & 0.02609 & 0.00267 \\
15 & 25,000 & 17.29 & 18.19 & 8.49 & 0.02525 & 0.00238 \\
\addlinespace[2pt]
\bottomrule
\end{tabular}
\par\medskip\raggedright\footnotesize
The entropy stage starts at 20,000 updates for $\mathrm{Ma}=5$ and at 15,000 for $\mathrm{Ma}=7$--15. Percentages apply to the three columns headed by 100; lengths and $\mathcal B_\rho$ are dimensionless. No entropy continuation is used at $\mathrm{Ma}=2,3$. The $\mathrm{Ma}=15$ initial row also describes the saved sensitivity baseline.
\end{table}

Endpoint errors have thus been evaluated directly, while complete intermediate CFD-error histories have not been reconstructed from the loss logs. Neither a decrease of a weighted loss nor completion of its update budget establishes convergence to an entropy-admissible Euler solution, a Hessian condition number, or an error reduction proportional to the loss reduction. The controlled comparisons in \FVOptimizationSensitivityReference\ provide the separate evidence for the tested component changes.
\end{revisionred}
\FloatBarrier

\subsection{Computational Cost and Efficiency Considerations}
\label{sec:comp_cost}
\begin{revisionred}
A single NVIDIA GeForce GTX 1070 GPU is used. Table~\ref{tab:fv_training_time} separates approximate author estimates from archived process timers. Synchronized traces and complete matched-run metadata were not archived for the 210-minute AD-CNN, 16-minute FV-CNN, and contextual 16-minute MLP estimates. The baseline records 1,140.39 seconds (19.01 minutes); the 20 runs (baseline and 19 alternatives) span 18.79--26.10 minutes including setup and output. These timings provide neither repeat-run uncertainty nor a matched training-only or matched-accuracy speedup.
\end{revisionred}

\begin{table}[htbp]
\FVOptimizationFloatColor
\centering\small
\caption{Timing evidence and its recorded scope. Author estimates have no archived synchronized trace. Process timers refer to separate sensitivity runs and include setup and output. None of these entries is a matched-accuracy comparison.}
\label{tab:fv_training_time}
\begin{tabular}{@{}l r l@{}}
\toprule
Record & Time (minutes) & Scope\\
\midrule
Historical AD-CNN, author estimate & $\approx210$ & Initial 15,000 updates\\
FV-CNN, author estimate & $\approx16$ & Initial 15,000 updates\\
FV-CNN $\mathrm{Ma}=2$ L-BFGS, author estimate & $\approx30$ & Nominal budget 10,000; counts in Table~\ref{tab:fv_optimizer_accounting}\\
MLP, contextual author estimate & $\approx16$ & Incomplete run metadata\\
FV sensitivity baseline, archived timer & $19.01$ & Complete launched process\\
FV sensitivity suite, archived timers & $18.79$--$26.10$ & 20 distinct factor runs\\
\bottomrule
\end{tabular}
\end{table}

\begin{revisionred}
The source code identifies a structural reason for the large AD-CNN overhead. Each old residual evaluation calls coordinate AD twice for each of four flux divergences, followed by one further call for the pressure sensor: nine vector--Jacobian-product calls in total. Each call uses \texttt{create\_graph=True} and \texttt{retain\_graph=True}. PyTorch therefore records the derivative computation so that the subsequent parameter-gradient calculation can differentiate through it~\cite{pytorchgrad,pytorchautograd}. For the inviscid part of the old residual in \FVOptimizationADReference, a parameter component $\vartheta_a$ consequently requires mixed derivatives of the form
\begin{equation}
 \frac{\partial\widetilde R_i^{\rm AD}}{\partial\vartheta_a}
 =\sum_j\frac{\partial^2F_{x,j}}{\partial\vartheta_a\partial x_i}
  +\sum_j\frac{\partial^2F_{y,j}}{\partial\vartheta_a\partial y_i}.
 \label{eq:fv_ad_mixed_cost}
\end{equation}
The pressure sensor also uses coordinate AD, whereas the old artificial-viscosity Laplacian is a discrete convolution. Thus AD-evaluated viscous second derivatives are not the source of this cost. The coordinate graphs traverse convolution, padding, mixing, activations, and normalization. A pointwise AD-trained MLP also requires mixed derivatives but has different operations and no neighbor mixing; equal derivative order does not imply equal runtime.

In FV-CNN the spatial operator instead evaluates a discrete map $R^{\rm FV}(U_\vartheta)$ on the fixed grid. Its parameter gradient has the form
\begin{equation}
 \nabla_\vartheta\mathcal L_{\rm FV}
 =\left(\frac{\partial U_\vartheta}{\partial\vartheta}\right)^T
   \nabla_U\mathcal L_{\rm FV},
 \label{eq:fv_cost_chain}
\end{equation}
Here $\mathcal L_{\rm FV}$ denotes the discrete FV-CNN training objective, and $\nabla_U\mathcal L_{\rm FV}$ differentiates the numerical fluxes and other discrete terms. Parameter AD is retained without coordinate AD through the CNN. The old nine calls are vector--Jacobian products, not a full $N\times N$ coordinate Jacobian, and imply neither quadratic cost nor an exact ninefold factor. Their cost contributions remain unprofiled. The old sum derivative also remains a physically mismatched divergence for coupled outputs.

The author reports approximately 30 minutes for the $\mathrm{Ma}=2$ L-BFGS stage with its nominal budget of 10,000. The archived counts remain 8,412 optimizer iterations and 10,002 closure evaluations. The 30-minute value is an author-reported wall-clock estimate without an archived synchronized trace. Elapsed times for the additional continuation stages remain unavailable. The 16-minute initial-stage estimate is not extrapolated: L-BFGS uses float64 and repeated closures, and the $\mathrm{Ma}=5$ continuation can evaluate extra gradients and repeat blocks. Sensitivity runs are separate computations. Controlled speedups, complete-stage totals, and advantages over CFD or other architectures are consequently not quantified.
\end{revisionred}
\FloatBarrier

\clearpage
\providecommand{\FVDiscussionADReference}{Sec.~\ref{sec:fv_ad_failure}}
\providecommand{\FVDiscussionFVReference}{Sec.~\ref{sec:fv_discretization}}
\providecommand{\FVDiscussionCancellationReference}{Eq.~\eqref{eq:fv_telescoping}}
\providecommand{\FVDiscussionMainReference}{Table~\ref{tab:fv_main_settings}}
\providecommand{\FVDiscussionErrorTable}{Table~\ref{tab:res_field_errors}}
\providecommand{\FVDiscussionErrorFigure}{Fig.~\ref{fig:res_common_errors}}
\providecommand{\FVDiscussionResidualFigure}{Fig.~\ref{fig:res_smooth}}
\providecommand{\FVDiscussionBandFigure}{Fig.~\ref{fig:res_band}}
\providecommand{\FVDiscussionBandTable}{Table~\ref{tab:res_band_width}}
\providecommand{\FVDiscussionGeometryFigure}{Fig.~\ref{fig:res_geometry}}
\providecommand{\FVDiscussionSensitivityReference}{Sec.~\ref{sec:component_ablations}}
\providecommand{\FVDiscussionScalingReference}{Fig.~\ref{fig:sens_scaling} and Table~\ref{tab:sens_scaling_profiles}}
\providecommand{\FVDiscussionCostReference}{Sec.~\ref{sec:comp_cost}}
\section{Discussion and Scope}
\label{sec:discussion_scope}
The preceding sections establish the benchmark result and the controlled role of several design choices, but they do not turn the method into a general convergence theory for discontinuous Euler solutions.

\begin{revisionred}
\subsection{Meaning of the finite-volume reformulation}
\label{sec:discussion_fv_meaning}
The central change is the separation of the network representation from the spatial conservation operator. The previous AD-CNN applies a sum-seeded coordinate derivative to coupled outputs; the resulting Jacobian column sums are generally not the spatial divergence of the predicted field (\FVDiscussionADReference). FV-CNN supplies an explicit numerical-flux balance on a fixed mesh, with AD retained for parameter optimization. Shared-face cancellation holds algebraically for any predicted state; obtaining an accurate steady solution additionally depends on residual minimization, boundary treatment, representation, and resolution.

The executed Hpres option replaces $\Delta E$ by $\Delta(E+p)$ in the dissipative jump. Its shared-face cancellation and local enthalpy compatibility remain distinct from entropy stability. The old nonconservative artificial-viscosity term is inactive in FV mode. The mathematical correction of the residual definition is therefore more firmly established than any attribution of all historical field differences to that correction alone: the archived AD fields were trained under different conditions and retain the stated export and interpolation limitations.

\subsection{What the complementary diagnostics resolve}
\label{sec:discussion_validation}
On common physical support, the adopted FV-CNN fields have smaller density, pressure, and streamwise-velocity discrepancies at all eight Mach numbers, and their local density transitions are narrower. Finite discrepancies remain: the upstream-quadrant density and pressure $L_2$ ranges are $5.43$--$18.69\%$ and $6.25$--$19.70\%$. AD-CNN has smaller $\mathrm{Ma}=3$ temperature and transverse-velocity discrepancies, smaller shock-position RMS at $\mathrm{Ma}=11,15$, and smaller downstream total-enthalpy RMS in seven of eight cases. The reference CFD is a residual-monitored numerical solution on the stated grid. Its iterative convergence is distinct from grid convergence, so these comparisons quantify differences from that numerical reference rather than continuum-solution errors.

The centered dual-mesh residual is operator-consistent with the nondissipative part of FV-CNN training; historical AD-CNN is interpolated and assessed with a different operator from its training residual. Its smooth-region ordering is therefore not independent evidence of residual-formulation superiority. The stride-two check tests sensitivity to exact grid alignment, while the closed-band quadrature and CFD norms supply complementary evidence. The revised diagnostics also separate shock displacement from local balance more explicitly. Outside a common mask excluding all three centered bands, FV-CNN has smaller physical-flux RMS in all 32 case--component comparisons at the seven tested widths. The original CFD-only exclusion gives 31 of 32 at $h=0.10D$. For integrated balances, identical CFD-centered bands give 22--31 of 32 smaller FV-CNN values depending on width, whereas each field's self-centered bands give 32 of 32 throughout $0.06\leq h/D\leq0.12$. Reusing the CFD projected-length normalization preserves the latter ordering. These results support smaller integrated defects near the FV-CNN transitions while exposing the geometrical dependence of the original common-band assessment. Self-centering changes the physical volumes and is retained alongside the common-volume comparison.

Continuation and conservation also require separate assessment. The fixed-region upstream entropy diagnostic decreases during all recorded entropy continuations, while field discrepancies and self-centered balance defects can increase (Table~\ref{tab:fv_stage_accuracy}). The upstream penalty targets a selected freestream-like region; it does not impose the entropy jump or a global inequality. Finite-band balances and two-point axis entropy samples consequently complement, rather than establish, entropy admissibility.

\subsection{Interpretation of the component controls}
\label{sec:discussion_components}
The pointwise FV-trained variant supplies the no-neighbor counterpart within the stated network family. Both spatial couplings are removed while the affine channel maps, nonlinearities, residual paths, normalization, FV residual, and saved initial tensors are retained. Its larger profile discrepancies measure the response to that connectivity intervention, including the associated changes of active capacity and receptive field. The large pointwise contrast is conditional on the shared initialization. The much smaller radial-only and azimuthal-only density differences lie within the observed baseline range, so their ordering does not establish an initialization-robust advantage of complete coupling. Separately designed conventional MLPs are not ranked. Fourier-off likewise tests a combined encoding change, removing the sine nonlinearity and frequency schedule together. The explicit output lower bounds supply another common prior and are approached in several failed variants.

The $\mathrm{Ma}=15$ scaling intervention changes the prescribed component weights with the same initial state and remaining arguments. Because the overall PDE multiplier is unchanged, both component proportions and total PDE strength relative to the priors change. Thus the resulting large profile differences do not isolate a gradient-balancing mechanism. $\mathrm{Ma}=3$ additionally changes $\lambda_P$. At $\mathrm{Ma}=2$, common initialization is supported by reconstruction of the recorded original code and exact agreement with the saved scaling-off initial tensors (Sec.~\ref{sec:mach_scaling_ablation}); its distinct protocol retains the opposite Mach-weight convention. This initialization check is separate from the unmeasured same-seed GPU execution variability and the measured across-seed dispersion at $\mathrm{Ma}=15$. The dimensional comparison in Eq.~\eqref{eq:fv_flux_balancing_rationale} motivates high-Mach orders without deriving the executed coefficients. Component-gradient norms or spectra and a systematic near-transition Mach sweep have not been measured.

The fixing boundary is prescribed from a shock-detachment estimate, and the remaining loss coefficients are empirical. The coefficient, seed, and endpoint selection history is incomplete as detailed in Sec.~\ref{sec:fv_protocol}; reference-independent selection is not established. The factor tests quantify sensitivity about the documented baseline; they do not establish that its coefficients are necessary or optimal. The smaller stagnation coefficient improves all four profile norms in the seed-42 realization, and other factors show tradeoffs; the baseline-dispersion comparison in Sec.~\ref{sec:factor_seed_context} provides descriptive context without testing the paired effects. Every factor result is a single run from the shared seed. The recorded cuDNN settings reduce some sources of nondeterminism, but strict deterministic execution was not established. The baseline dispersion is now measured in Sec.~\ref{sec:seed_ensemble}; the interventions still lack paired multi-seed estimates.

\subsection{Initialization sensitivity and the accuracy claim}
\label{sec:discussion_initialization}
The completed initialization experiment changes the interpretation of the selected-field comparisons. The adopted $\mathrm{Ma}=15$ field is reproducible in quality from several distinct initial states: seeds 1, 6, and 7 improve all five CFD norms. At the same time, the fixed-budget protocol also produces substantially displaced fronts and larger discrepancies without any change of coefficients or priors. The measured dispersion exceeds the historical FV--AD density gap. Consequently, the main comparison documents the archived realizations; it cannot support an initialization-robust FV-CNN accuracy advantage, and the original adopted field is retained rather than replaced by a more favorable new seed.

This variability does not alter the algebraic definition or shared-face cancellation of the FV residual, which hold for any network output. It limits how reliably optimization reaches a field close to the specified CFD reference under the chosen budget. No uniqueness or convergence guarantee follows from residual consistency. Across all eleven final fields, AD/FV transition-width ratios remain $4.07$--$4.66$, and all 44 self-centered band-balance comparisons at $h=0.10D$ retain smaller FV defects (Sec.~\ref{sec:seed_band_balances}). Thus local sharpness and this balance diagnostic persist in the sampled initializations even when global placement is inaccurate. The fixed priors constrain the solution but do not remove this optimization sensitivity.

The single-seed component responses are likewise conditional on their initialization. Baseline dispersion is relevant context for the small dissipation-coefficient difference; it is not a substitute for repeating each intervention and evaluating its paired effect. The eleven-run summary includes an historically adopted seed, while the ten new runs alone have density mean $25.04\%$ and standard deviation $12.17$ points. No AD seed ensemble or auditable same-seed repeated-training comparison is available. The archive verifies initial weights and checkpoint transfer; those checks are distinct from replicating complete optimization. These limits prevent a population-level method ranking or a separation of initialization effects from remaining execution nondeterminism.

\subsection{Computational and physical limits}
\label{sec:discussion_limits}
The study reports $2\le\mbox{Ma}_\infty\le15$ only. The fields cover the upstream quadrant, with the larger domain and separate protocol at $\mathrm{Ma}=2$; the wake is excluded. In particular, the adopted $\mathrm{Ma}=2$ field has a partly subsonic outflow edge under a copied-state closure. A characteristic-boundary or domain-extension study would be needed to assess that boundary influence. Its L-BFGS termination is budget-limited with an unmet gradient criterion, and its entropy and enthalpy discrepancies are comparatively large. Its inclusion in the all-Mach table documents the available computation rather than extending the common higher-Mach protocol to $\mathrm{Ma}=2$.

Each adopted field is one run. The $\mathrm{Ma}=11$ seed differs from the other main cases, its selection rationale was not archived, and no matched alternative seed has been evaluated in this revision. The recorded settings and fields support reproduction of the reported comparisons. The completed $\mathrm{Ma}=15$ ensemble supplies the initialization assessment in Sec.~\ref{sec:seed_ensemble}; comparable ensembles at the other Mach numbers remain unavailable. A broad Mach range does not substitute for replication at each condition.

The reported 210- and 16-minute initial-stage times and the approximately 30-minute $\mathrm{Ma}=2$ L-BFGS time are author estimates without archived synchronized traces. The sensitivity baseline has a separate recorded process time of 19.01 minutes, including setup and output. Avoidance of the coordinate-derivative graph is established by the source, but a controlled speedup has not been measured; later continuation elapsed times remain unavailable. Quantitative speed claims have therefore been removed from the abstract and conclusion.

The calorically perfect gas with $\gamma=1.4$ defines the benchmark, including at $\mathrm{Ma}=15$. Applicability to a dimensional gas requires thermodynamic conditions and possible nonequilibrium physics. Three-dimensional geometry, unsteady flow, cross-Mach generalization of one network, mesh-convergent weak-solution accuracy, and global entropy admissibility require separate evidence.
\end{revisionred}

\clearpage
\section{Conclusion}
\label{sec:conclusion}
\begin{revisionred}
FV-CNN combines a convolutional field representation with an explicitly defined shared-face finite-volume residual for steady Euler flow around a cylinder. AD supplies parameter gradients. The earlier sum-based coordinate derivative generally differs from the physical divergence of spatially coupled outputs. Shared-face cancellation is an algebraic property of the revised residual, separate from optimization convergence and accuracy.

For the eight adopted fields, density and pressure relative $L_2$ discrepancies from matched finite-resolution CFD are $5.43$--$18.69\%$ and $6.25$--$19.70\%$. Compared with historical AD-CNN, density, pressure, and streamwise-velocity norms are smaller on common support. AD-CNN retains smaller downstream enthalpy RMS in 7/8 cases, smaller shock-position RMS at $\mathrm{Ma}=11,15$, and smaller $\mathrm{Ma}=3$ temperature and transverse-velocity discrepancies. The adopted-field self-centered and common CFD-centered band comparisons give 32/32 and 22--31/32 smaller FV defects, respectively; the different physical-volume interpretations are retained.

Ten additional $\mathrm{Ma}=15$ seeds and the unchanged adopted seed give density discrepancies of $12.04$--$43.70\%$, mean $24.34\%$, and sample standard deviation $11.78$ percentage points. The ten new runs alone give mean $25.04\%$ and standard deviation $12.17$ points. Six of eleven outcomes, including five new runs, fall below the fixed historical AD density value; three new seeds improve all five field norms over the adopted field. The grouped outcomes and dispersion exceeding the historical $3.40$-point FV--AD gap prevent an initialization-robust accuracy ranking.

Two narrower conclusions persist across all eleven final fields: AD/FV transition-width ratios are $4.07$--$4.66$, and all 44 self-centered seed--component balances at $h=0.10D$ are smaller for FV-CNN. Their largest FV/AD ratio is $0.194$. Local transition sharpness and small balance defects can therefore coexist with substantial position error. Entropy continuation reduces shock-position RMS and width in all eleven runs without eliminating the spread of full-domain discrepancies.

The factor tests remain conditional on their shared initial state. Baseline range comparisons describe their scale, not the significance or absence of paired intervention effects. Shock-location agreement remains conditioned on the Billig-based fixing prior: shifting $x_{\rm fix}$ from $-0.777D$ to $-0.9D$ or $-1.0D$ changes the density profile by $54.8\%$ or $40.2\%$. The distinct, budget-limited $\mathrm{Ma}=2$ protocol, historical selection records, finite-resolution reference, and replication limits are specified in Sec.~\ref{sec:discussion_scope}. The supported contribution is the conservative discrete training residual and these auditable, explicitly scoped saved-field comparisons.
\end{revisionred}

\section*{Data availability}
\begin{revisionred}
The companion archive, \texttt{JSTAT\_FVCNN\_v17\_Reviewer\_Data\_Code.zip}, contains native fields, recorded model weights and run information, CFD sources and complete fields, convergence histories, and evaluation/plotting scripts. File hashes identify the evaluated states. The $\mathrm{Ma}=2$ initialization reconstruction includes both source snapshots, the saved scaling-off initial tensors, and the exact-comparison script and audit. Band sweeps, stride-two checks, stage-endpoint comparisons, and the complete eleven-run initialization evaluation are reproducible without retraining. The seed archive retains both training endpoints, native arrays, weights, all per-run information and loss histories, the exact dated experiment plan with SHA-256 identity and matching execution lock, per-run UTC status records, every individual metric, and the figure scripts. The plan and execution lock are also supplied as exact standalone byte copies with a SHA-256 sidecar; the final individual metrics in Table~\ref{tab:seed_individual} have a companion comma-separated values (CSV) file. Subsequent exploratory checks include all-seed closed-band balances, front-to-fixing-boundary distances, and factor comparisons with baseline dispersion. Auditable same-seed repeated-training endpoint comparisons are not retained. For peer review, this archive is prepared for upload as a journal attachment accessible to the editor and reviewers. No public digital object identifier (DOI) has yet been assigned. Following acceptance, public deposition of the finalized data/code archive is planned; its DOI or persistent uniform resource locator (URL) is to be included in the availability statement before publication. Missing selection records and timing traces, and unperformed ensembles at other Mach numbers or CFD refinement, are identified in the manuscript.
\end{revisionred}

\section*{Artificial-intelligence-assisted preparation}
\begin{revisionred}
OpenAI ChatGPT/Codex was used for language revision, source-code inspection, and preparation of evaluation and document-processing scripts. Anthropic Claude was used for technical review and proposed evaluation-script and code corrections that informed the revision and saved-field diagnostics. The archived script versions and numerical checks identify the calculations used. Responsibility for the manuscript and its scientific conclusions is retained by the sole author.
\end{revisionred}

\clearpage
\appendix
\section{Factor-by-factor stagnation-line comparisons}
\label{sec:factor_figures}
\begin{revisionred}
The figures below accompany the component tests in Sec.~\ref{sec:component_ablations}. Each factor retains its own figure, with the shared baseline overlaid on every variant. The baseline is the initial 15000-update state rather than the continued adopted main field.
\end{revisionred}
\setcounter{figure}{0}
\renewcommand{\thefigure}{A\arabic{figure}}
\renewcommand{\theHfigure}{A.\arabic{figure}}
\FVSensitivityFigure{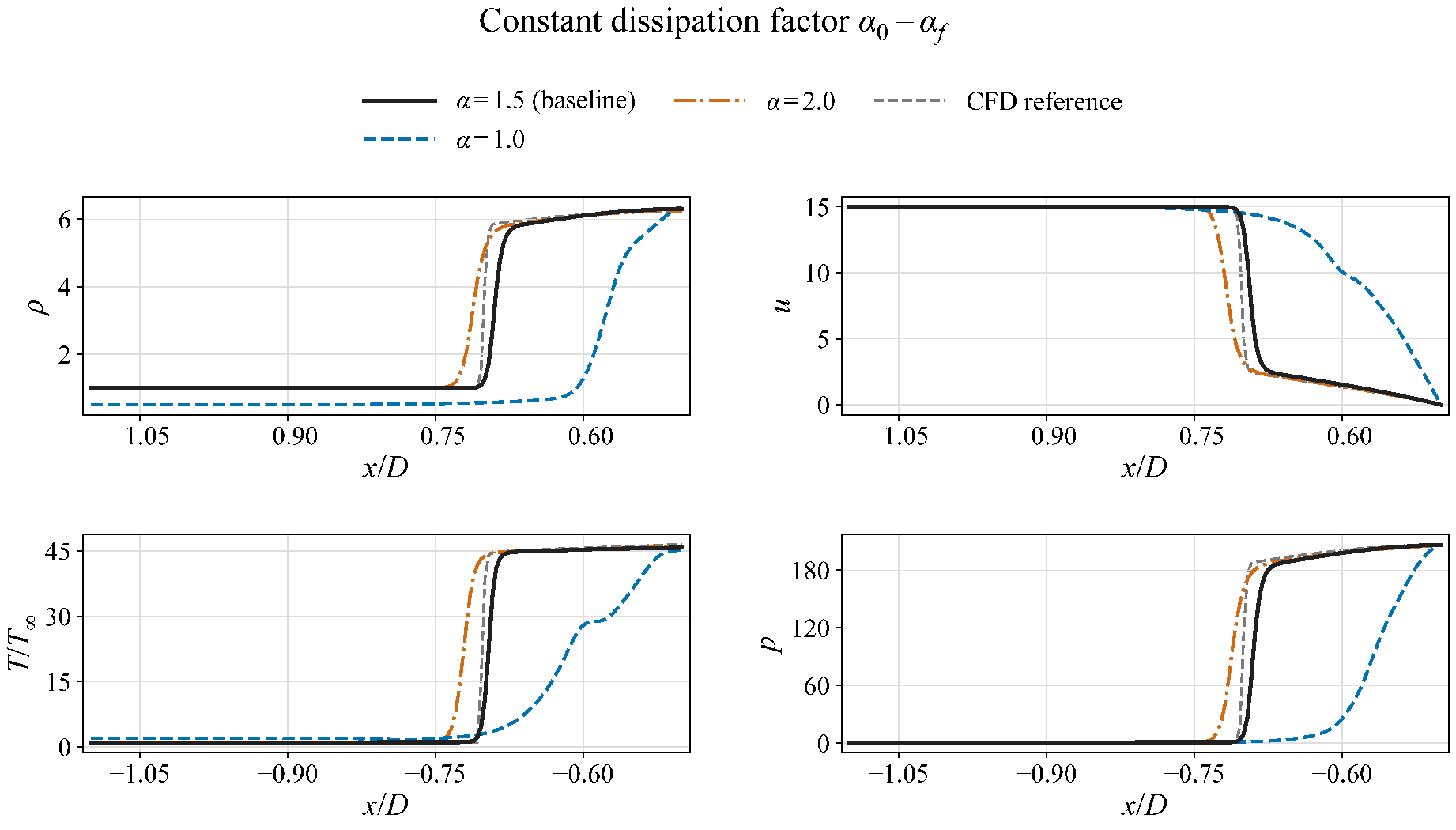}{$\mathrm{Ma}=15$ sensitivity to the constant dissipation factor $\alpha_0=\alpha_f$. The baseline and both variants use the common 15000-update protocol. Gray dashed curves are CFD. Temperature uses $T/T_\infty$; all four axes include vertical margins.}{fig:sens_alpha}

\FVSensitivityFigure{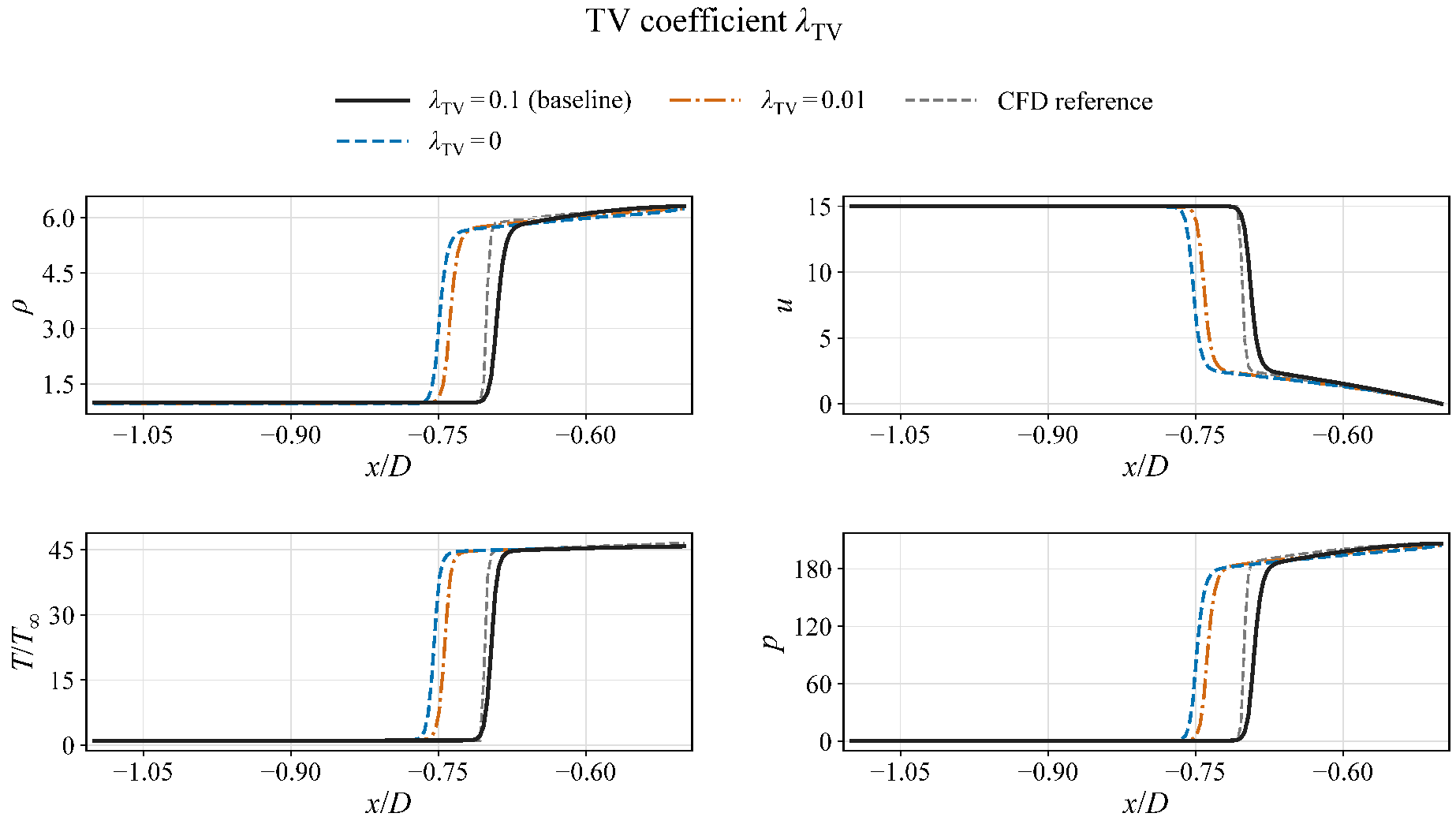}{$\mathrm{Ma}=15$ TV-coefficient sensitivity: $\lambda_{\rm TV}=0.1$ (baseline), $0$, and $0.01$. Each coefficient is retained for the complete 15000-update AdamW run. The stagnation-line displacement is visible despite similar wall values.}{fig:sens_tv}
\FloatBarrier
\FVSensitivityFigure{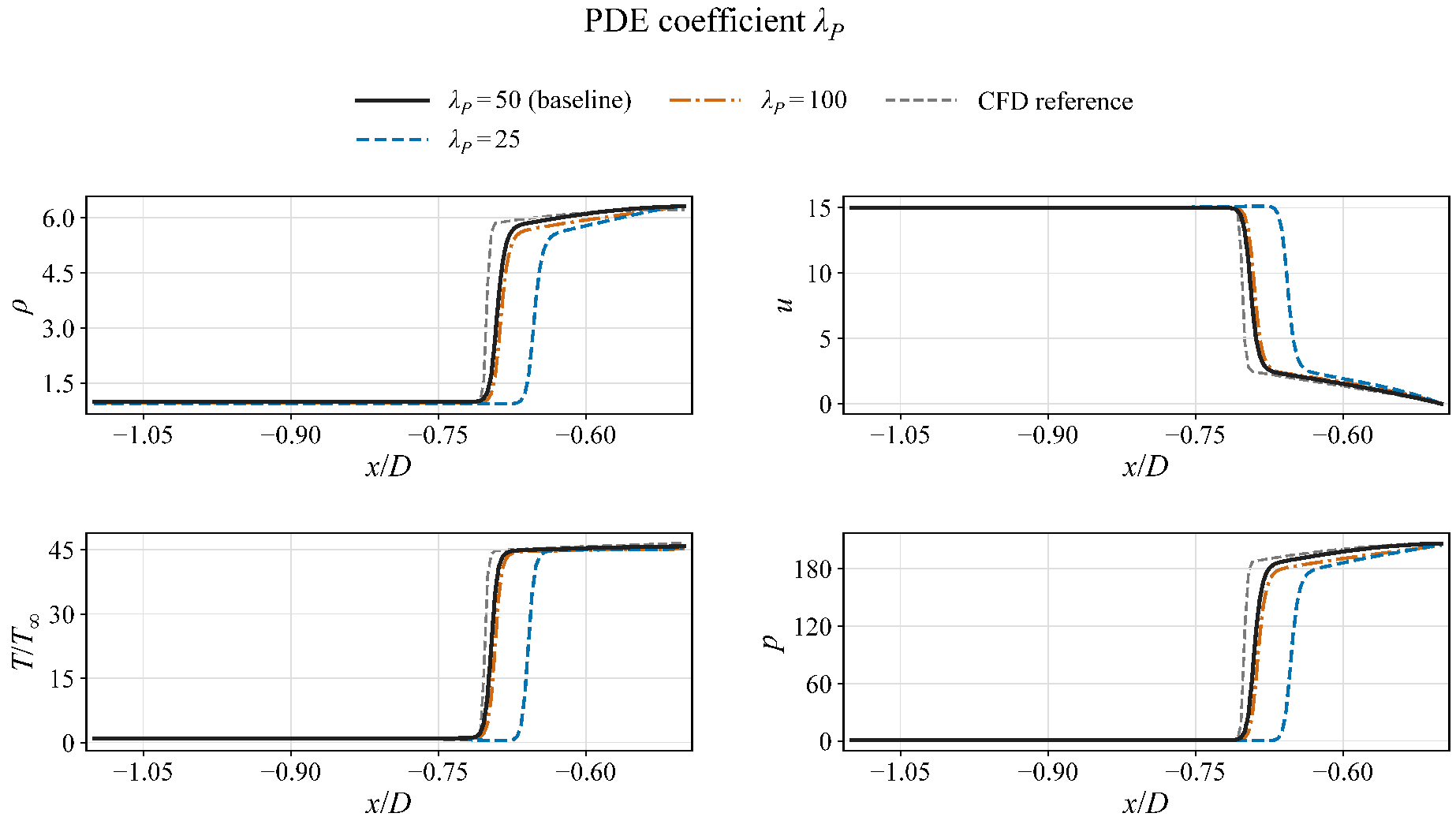}{$\mathrm{Ma}=15$ sensitivity to the overall PDE coefficient $\lambda_P$. The componentwise Mach weights and all other loss coefficients are retained; only the overall PDE multiplier changes.}{fig:sens_pde}

\FVSensitivityFigure{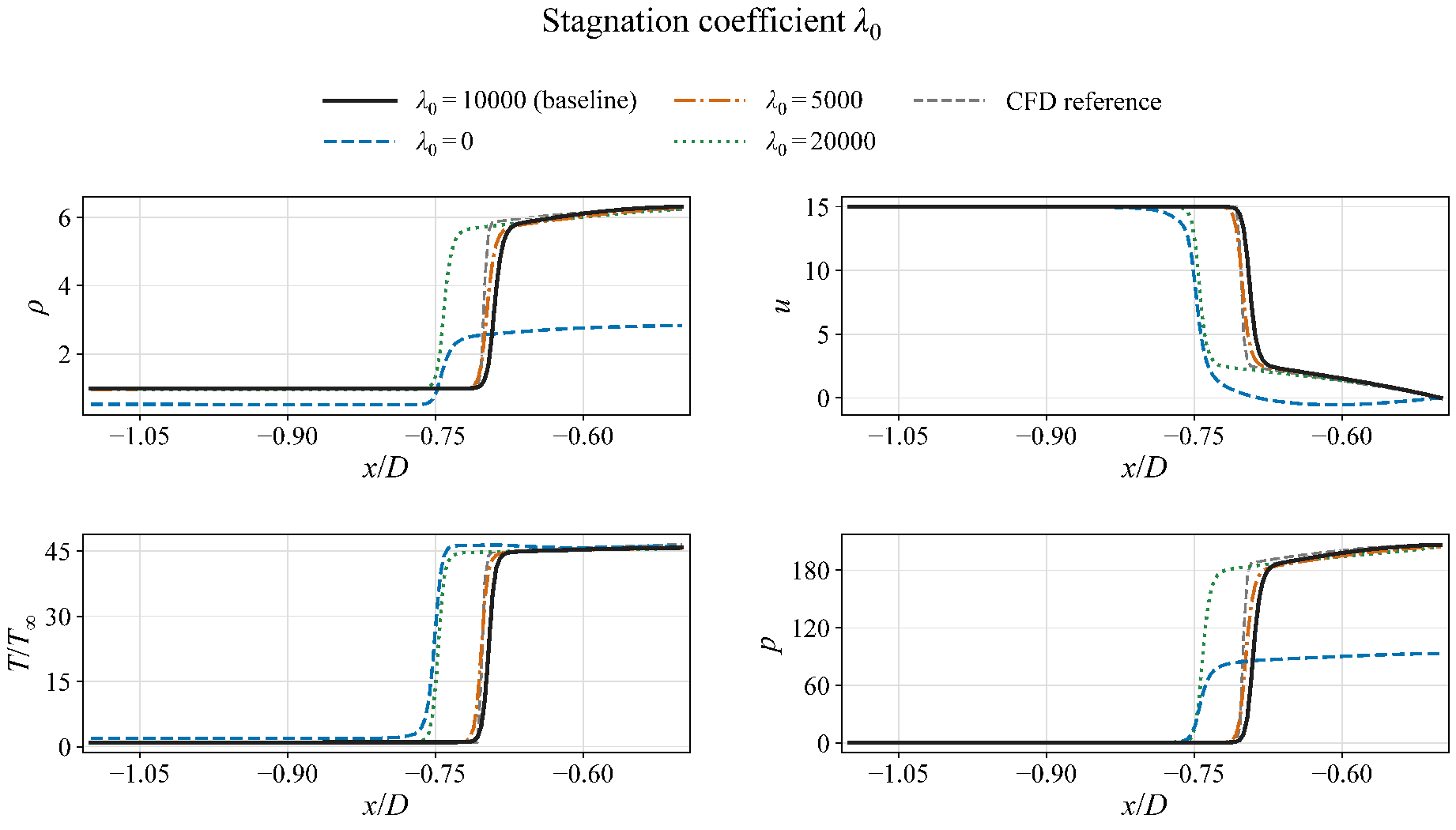}{$\mathrm{Ma}=15$ stagnation-prior sensitivity. The baseline $\lambda_0=10000$ is compared with $0$, $5000$, and $20000$. The soft relative tolerance remains $0.01$ in every case; removing the coefficient disables the prior without changing its analytical target.}{fig:sens_stagnation}
\FloatBarrier
\FVSensitivityFigure{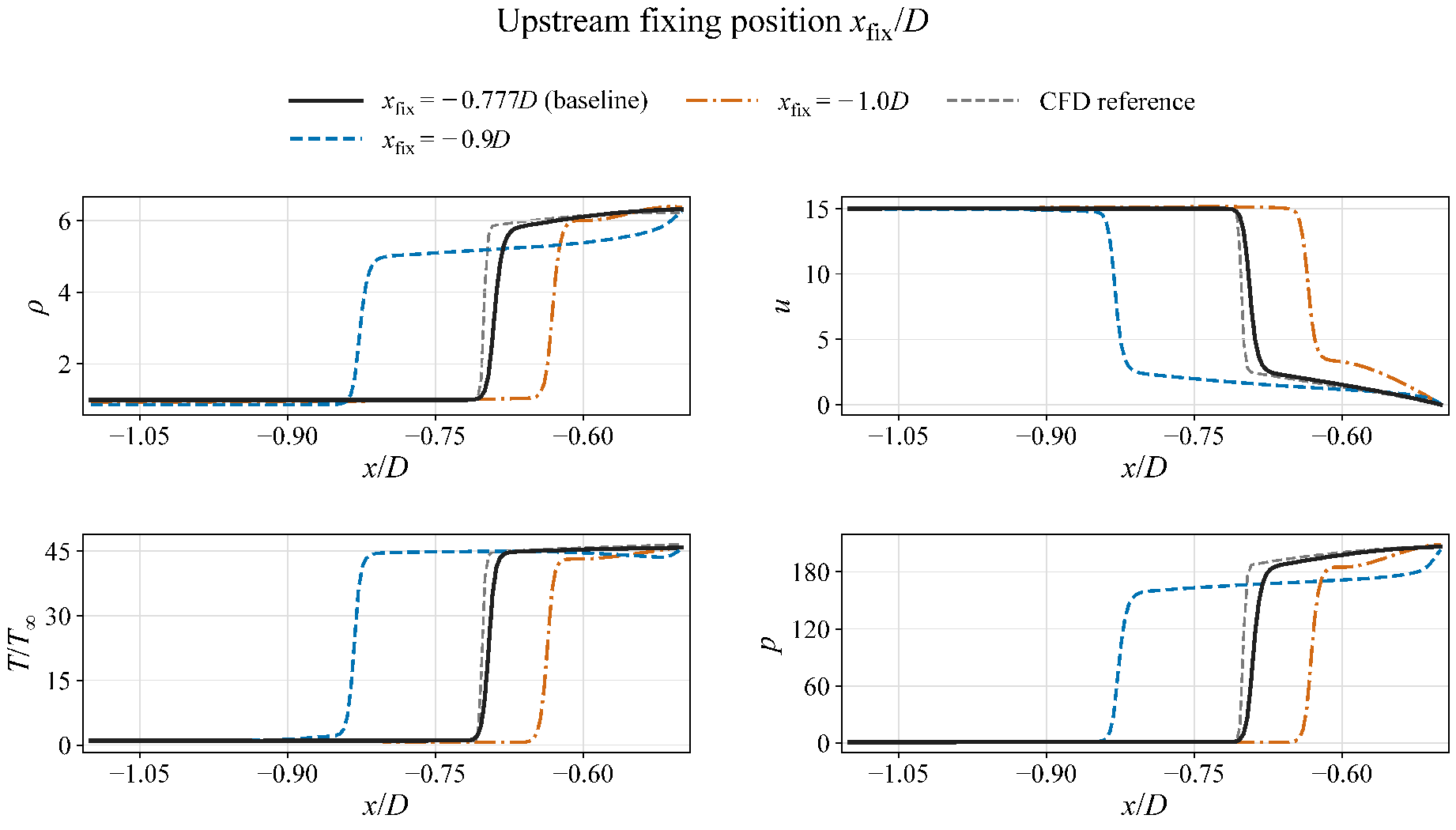}{$\mathrm{Ma}=15$ sensitivity to the fixing-region boundary $x_{\rm fix}=-0.777D$ (baseline), $-0.9D$, and $-1.0D$. The soft penalty remains active only at nodes satisfying $x<x_{\rm fix}$, with the same coefficient and freestream target.}{fig:sens_mask_position}

\FVSensitivityFigure{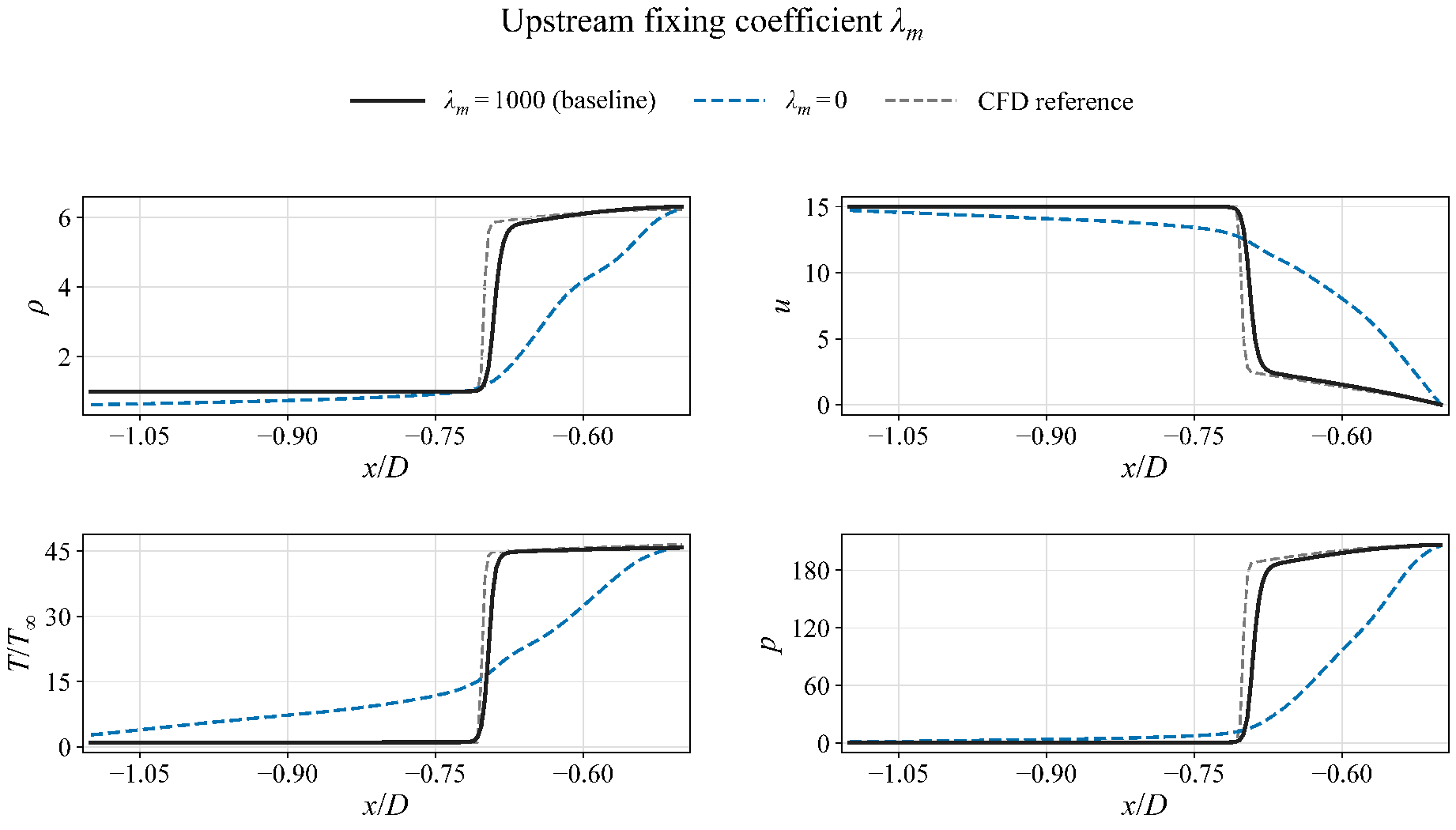}{$\mathrm{Ma}=15$ upstream-fixing coefficient test: $\lambda_m=1000$ (baseline) versus zero. The fixing boundary, outer inflow term, network, and remaining objective are retained.}{fig:sens_mask_weight}
\FloatBarrier
\FVSensitivityFigure{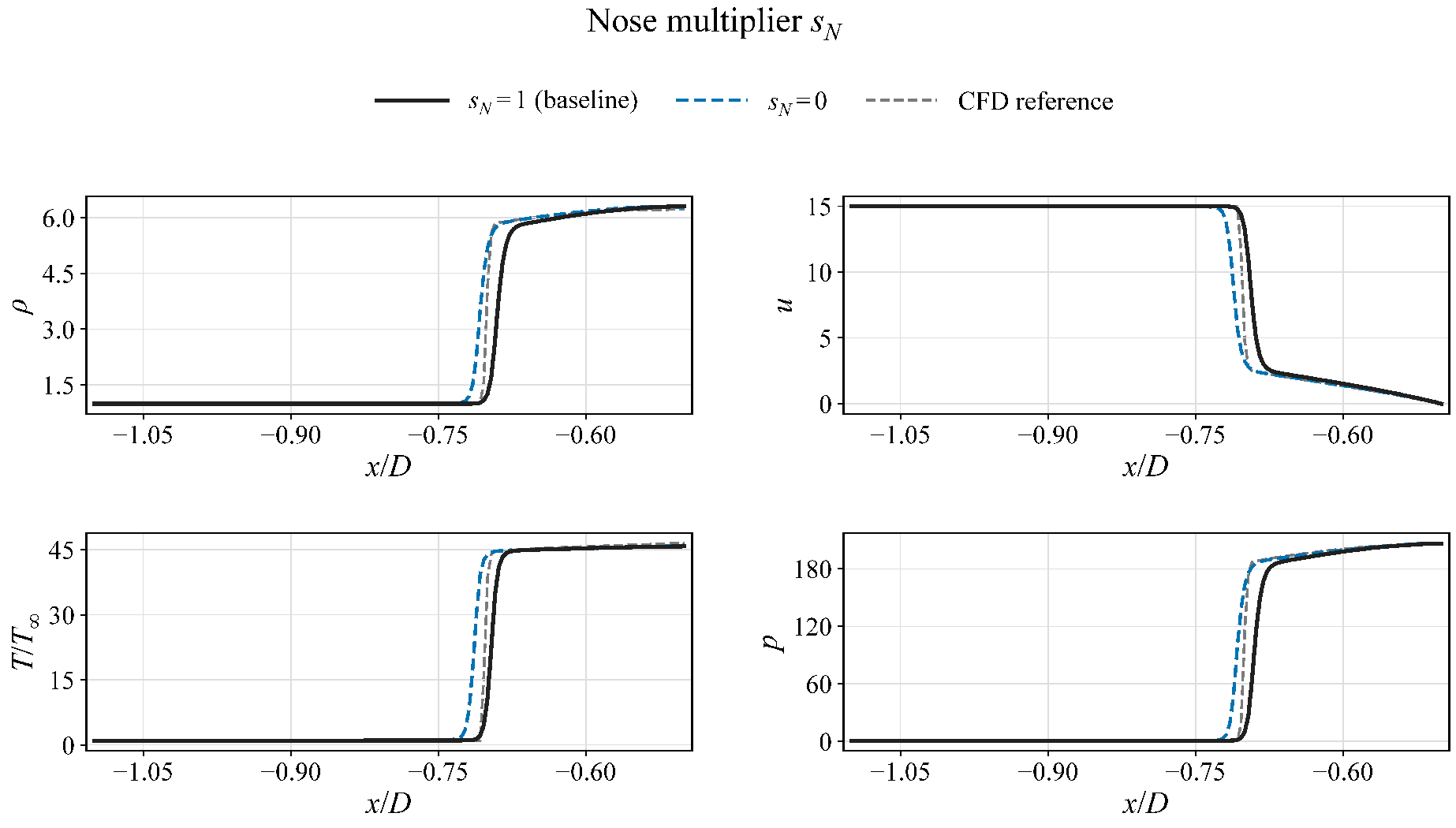}{$\mathrm{Ma}=15$ test of the additional nose residual: $s_N=1$ (baseline) versus zero. The time-dependent coefficient is disabled only in the zero-multiplier run; the stagnation-point density/pressure prior remains active in both.}{fig:sens_nose}

\FVSensitivityFigure{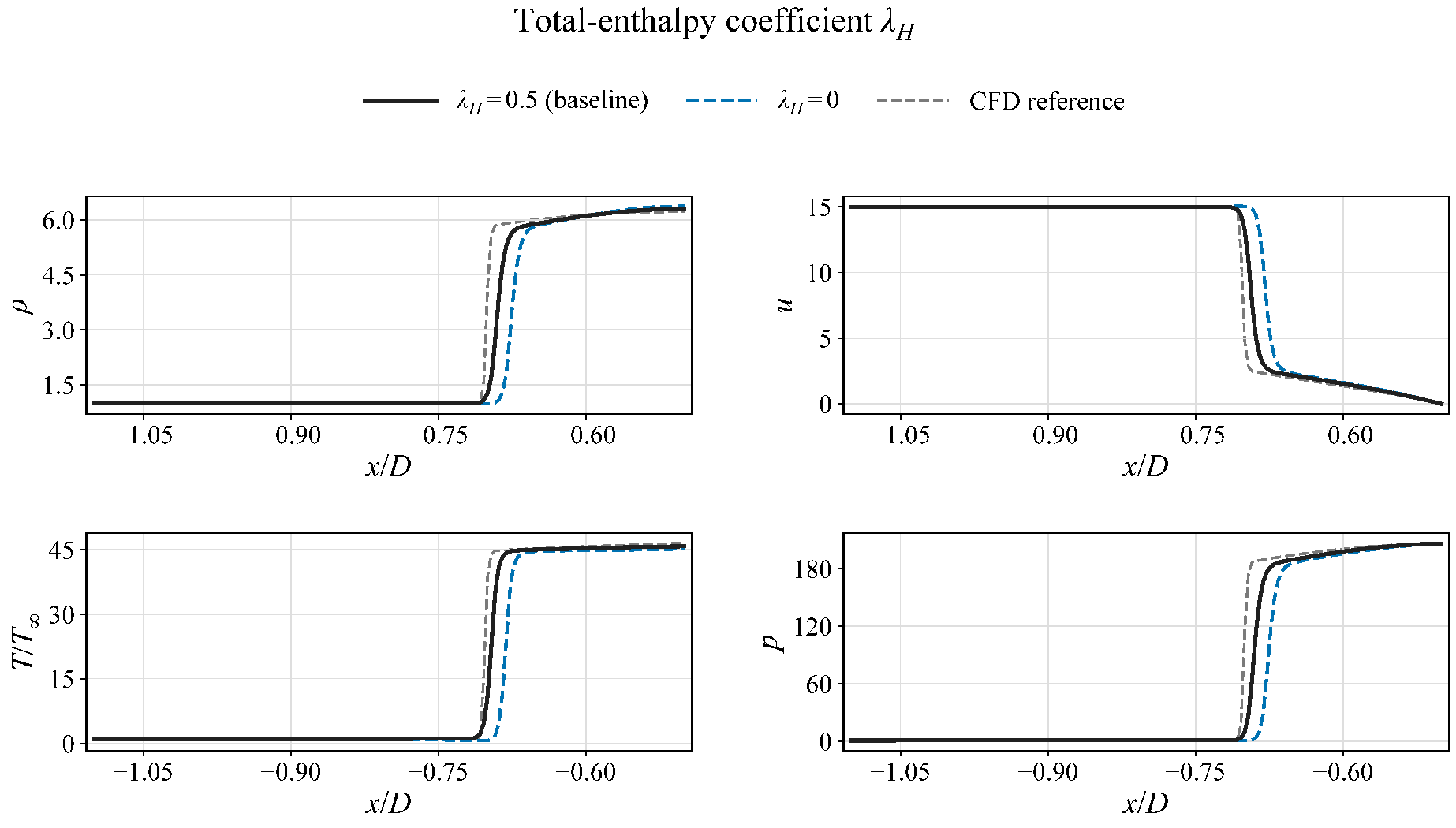}{$\mathrm{Ma}=15$ total-enthalpy coefficient test: $\lambda_H=0.5$ (baseline) versus zero. The energy equation remains in the FV PDE residual in both cases.}{fig:sens_enthalpy}
\FloatBarrier
\FVSensitivityFigure{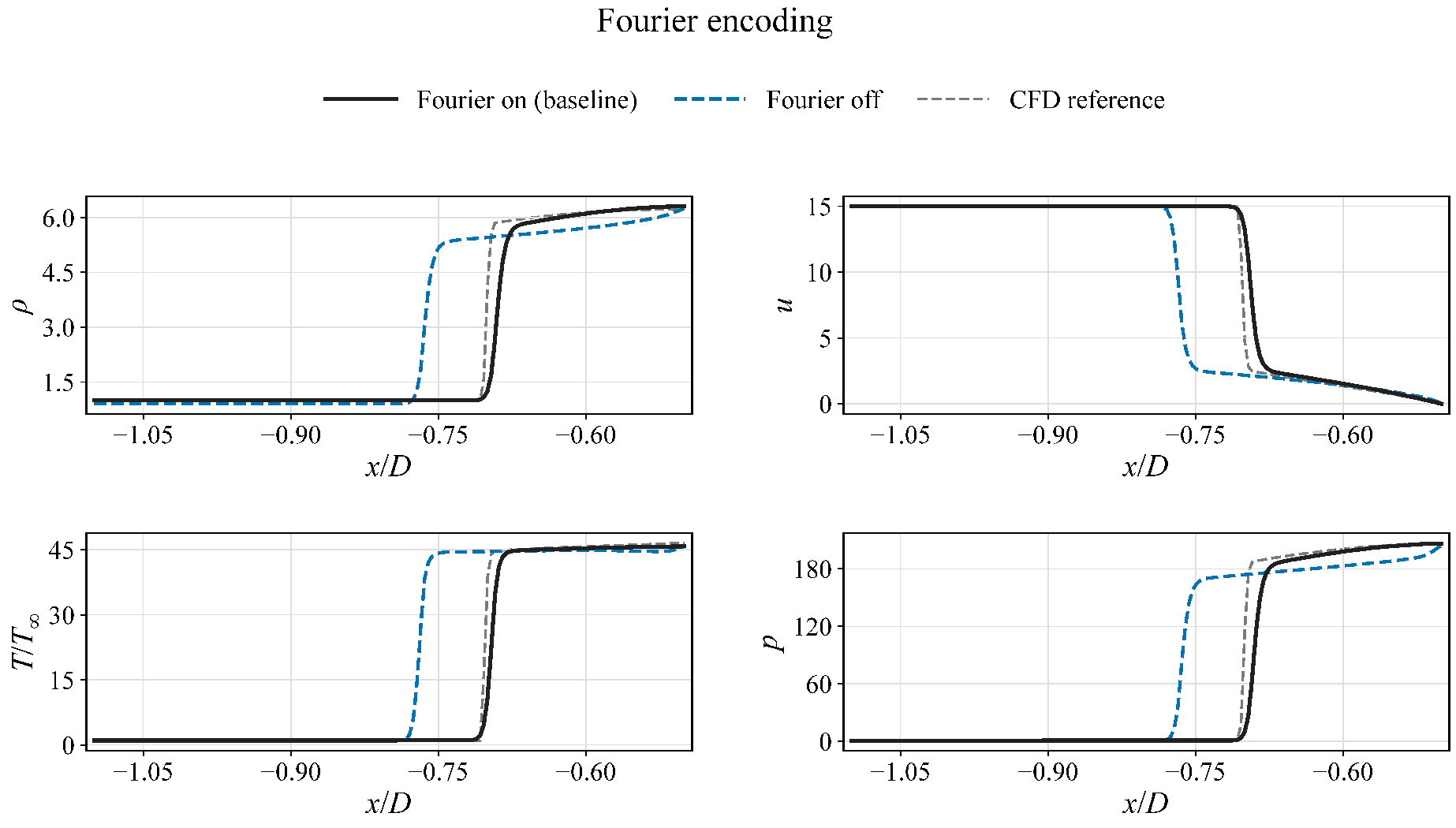}{$\mathrm{Ma}=15$ encoding test. Fourier on is the baseline; Fourier off uses the raw learned linear embedding while retaining both directional convolutional couplings, decoder, objective, and training budget.}{fig:sens_fourier}

\FVSensitivityFigure{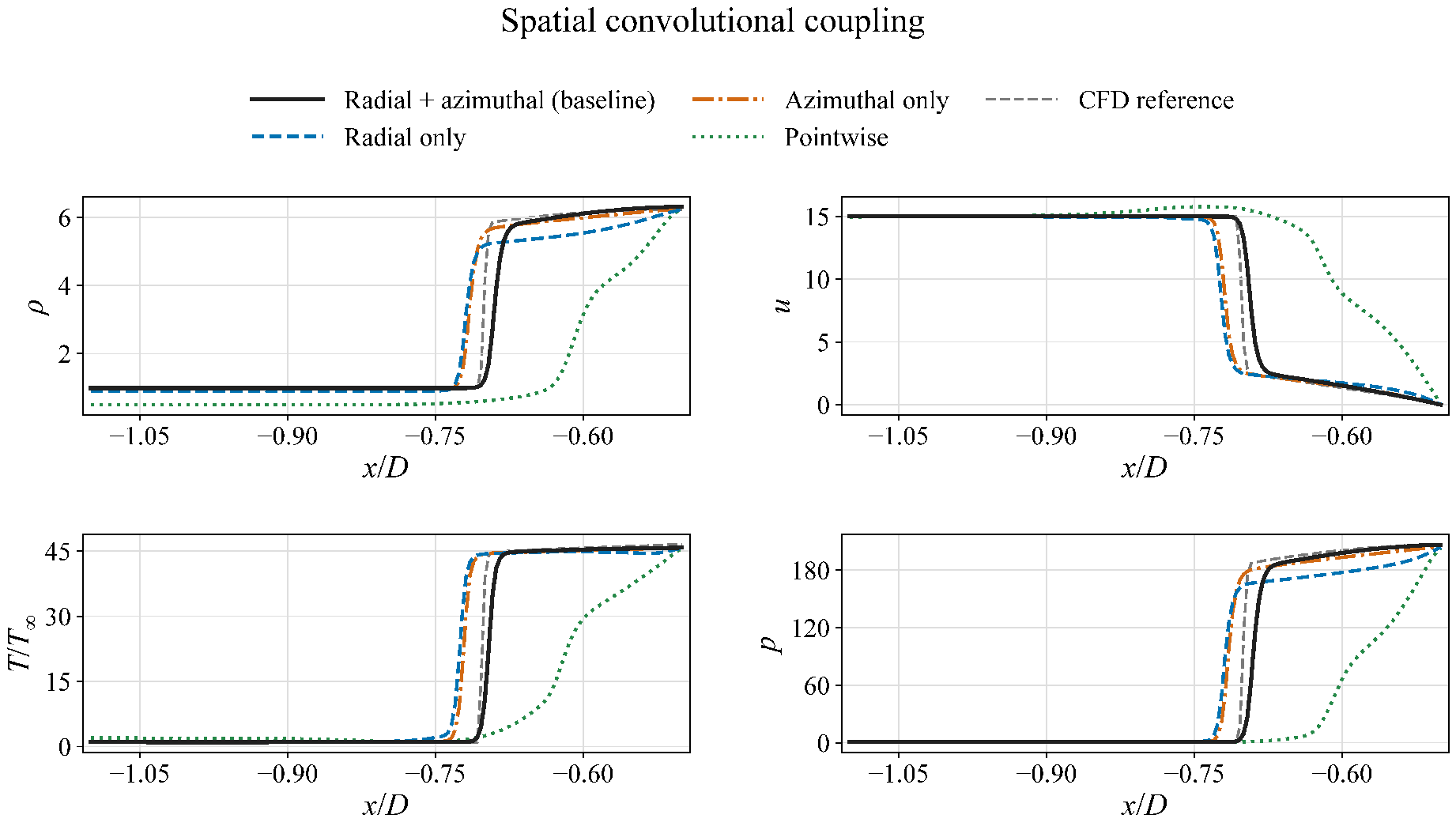}{$\mathrm{Ma}=15$ spatial-coupling tests. Radial-only and azimuthal-only disable the other direction's neighbor coupling, retaining its center-tap operation. Pointwise disables both spatial neighbor couplings while retaining the remaining architecture and Fourier encoding. Every state is trained with the same FV residual formulation.}{fig:sens_spatial}
\FloatBarrier
\FloatBarrier
\clearpage
\section{Centered-flux check on stride-two dual meshes}
\label{sec:coarsened_diagnostic}
\begin{revisionred}
The saved fields are additionally evaluated on four stride-two meshes, retaining even/even, even/odd, odd/even, and odd/odd angular/radial indices. Dual-face geometry and areas are rebuilt on each retained mesh. Native AD-CNN fields are interpolated directly to those nodes; no learned field is retrained or evolved. The common mask excludes boundary nodes, unsupported neighbor stencils, and the union of all three native located-front bands with half-width $0.10D$. Each phase uses identical nodes for all fields, while different phases need not retain identical physical support.

Table~\ref{tab:coarsened_flux} gives the componentwise ordering and the range of FV/AD RMS ratios across components and phases. The ordering is preserved in 128 of 128 coarse case--component--phase comparisons. At $\mathrm{Ma}=2$, the FV/AD ratio reaches $0.960$, corresponding to only about $4\%$ smaller RMS in the closest comparison. This check reduces exact training-grid alignment, but retains the centered physical-flux family. It is a stencil-sensitivity check on fixed fields, not an independent-operator proof or a grid-convergence calculation. All individual values, input hashes, and a uniform-state closure check are supplied with the archive.
\setcounter{table}{0}
\renewcommand{\thetable}{B\arabic{table}}
\renewcommand{\theHtable}{B.\arabic{table}}
\begin{table}[!htb]\centering\small
\caption{Operator-consistent residual check outside the union of all three shock bands. Fine: number of components with smaller FV-CNN than AD-CNN RMS, out of four (mass, $x$-momentum, $y$-momentum, and energy). Coarse: number out of 16 comparisons ($4$ components $\times$ $4$ stride-two phases). The final column spans the 16 coarse component/phase FV/AD RMS ratios at that Mach number. Values below one favor FV-CNN.}
\label{tab:coarsened_flux}
\begin{tabular}{@{}cccc@{}}\toprule
$Ma_\infty$ & Fine & Coarse & Coarse FV/AD ratio\\\midrule
2 & 4/4 & 16/16 & 0.290--0.960\\
3 & 4/4 & 16/16 & 0.199--0.398\\
5 & 4/4 & 16/16 & 0.073--0.118\\
7 & 4/4 & 16/16 & 0.061--0.083\\
9 & 4/4 & 16/16 & 0.060--0.140\\
11 & 4/4 & 16/16 & 0.054--0.106\\
13 & 4/4 & 16/16 & 0.035--0.134\\
15 & 4/4 & 16/16 & 0.025--0.102\\
\bottomrule\end{tabular}\end{table}
\end{revisionred}

\begin{revisionred}
\clearpage
\section{\textcolor{RevisionColor}{Complete initialization diagnostics at $\mathrm{Ma}=15$}}
\label{sec:seed_supplement}
\setcounter{figure}{0}
\renewcommand{\thefigure}{C\arabic{figure}}
\renewcommand{\theHfigure}{C.\arabic{figure}}
Figures~\ref{fig:seed_geometry}--\ref{fig:factor_seed_context} supplement the five full-domain norms in Fig.~\ref{fig:seed_full_domain}. The same eleven fields, symbols, and descriptive whiskers are retained. Table~\ref{tab:seed_individual} supplies individual final density errors, geometry metrics, and four closed-band defects. Every individual value is also supplied in the companion CSV files; no accuracy-based exclusion or post hoc success threshold is applied.
\begin{figure}[!htbp]
 \centering
 \includegraphics[width=\linewidth]{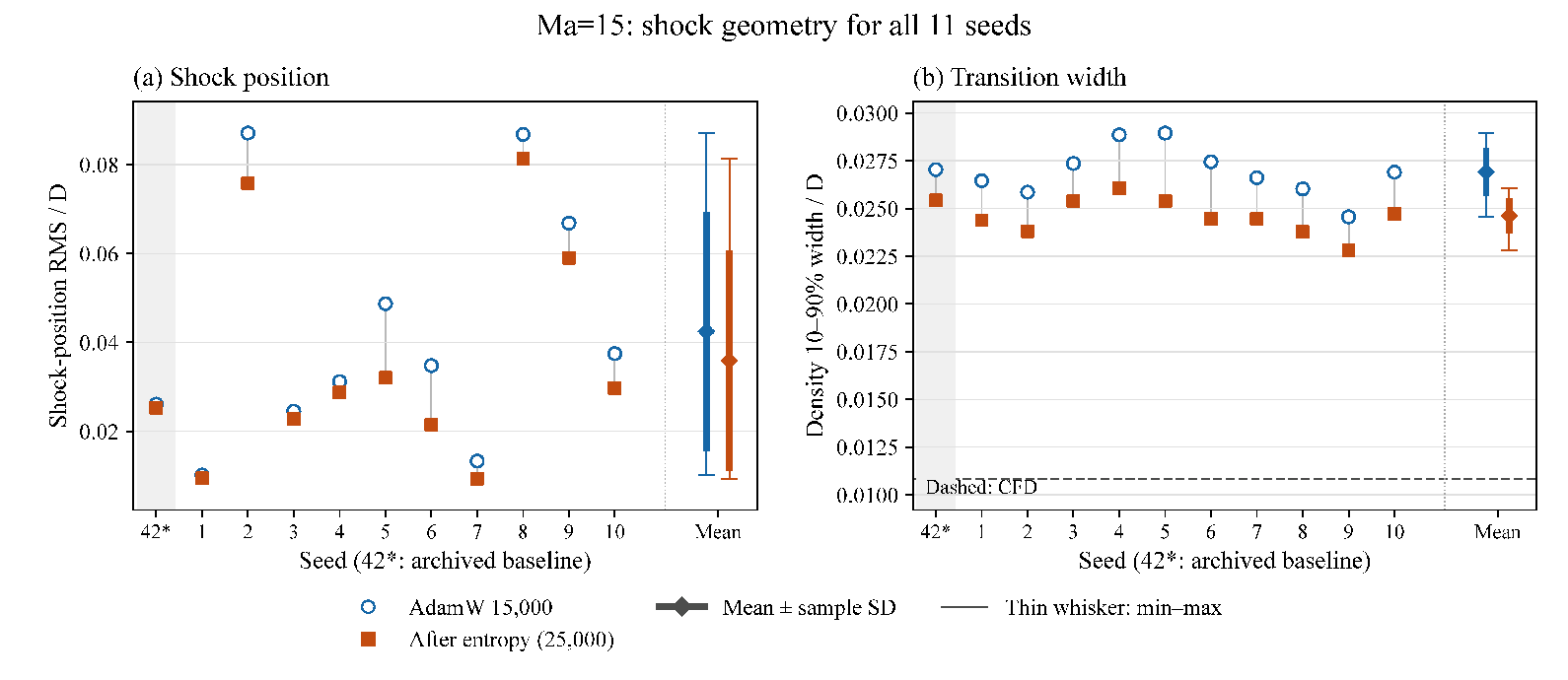}
 \caption{Shock-position RMS relative to CFD and angular-mean $10$--$90\%$ density-transition width for all eleven initializations at both stages. The dashed line in (b) gives the CFD width. Thin whiskers: minimum--maximum; thick whiskers: sample standard deviation about the mean. Entropy continuation reduces both metrics in every run, while appreciable between-seed position differences remain. The historical AD-CNN width is $0.10620D$; all final FV widths are smaller, with AD/FV ratios $4.07$--$4.66$.}
 \label{fig:seed_geometry}
\end{figure}
\FloatBarrier
\clearpage
\begin{figure}[!htbp]
 \centering
 \includegraphics[width=\linewidth]{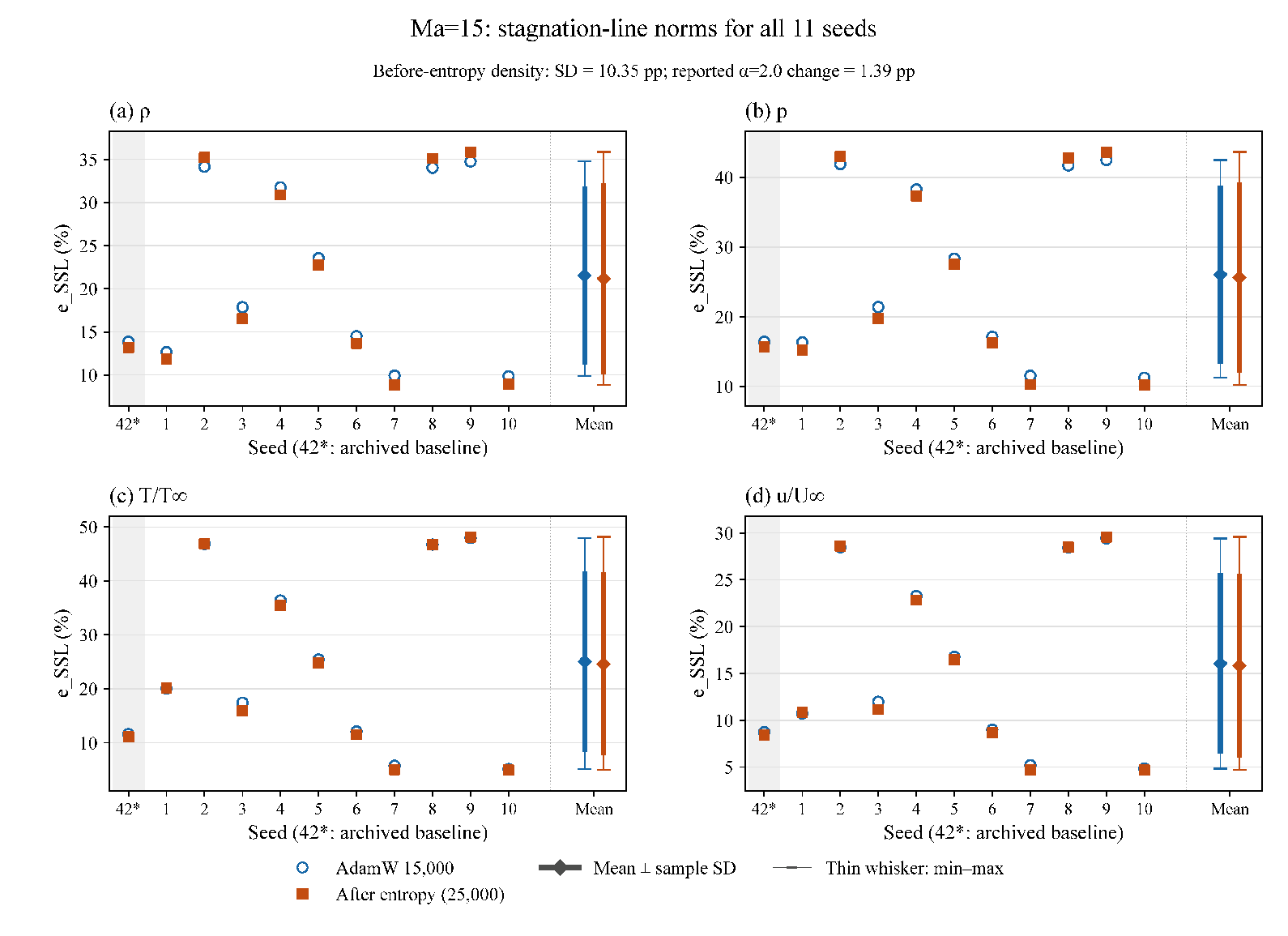}
 \caption{Stagnation-line discrepancies for all eleven initializations at both stages. The axis label $e_{\rm SSL}$ denotes $100e_{2,q}^{\rm SSL}$, and $U_\infty=u_\infty$ is the freestream speed. SD denotes sample standard deviation and pp denotes percentage points. The relative norms are unchanged by the indicated freestream normalization of temperature and velocity. Thick and thin whiskers have the same meanings as in Fig.~\ref{fig:seed_full_domain}. The initial-stage density baseline standard deviation is $10.35$ percentage points, compared with the $1.39$-point seed-42 change for $\alpha=2.0$. A baseline-only ensemble does not estimate the distribution of paired intervention effects.}
 \label{fig:seed_ssl}
\end{figure}
\FloatBarrier

\clearpage
\begin{figure}[!htbp]
 \centering
 \includegraphics[width=\linewidth]{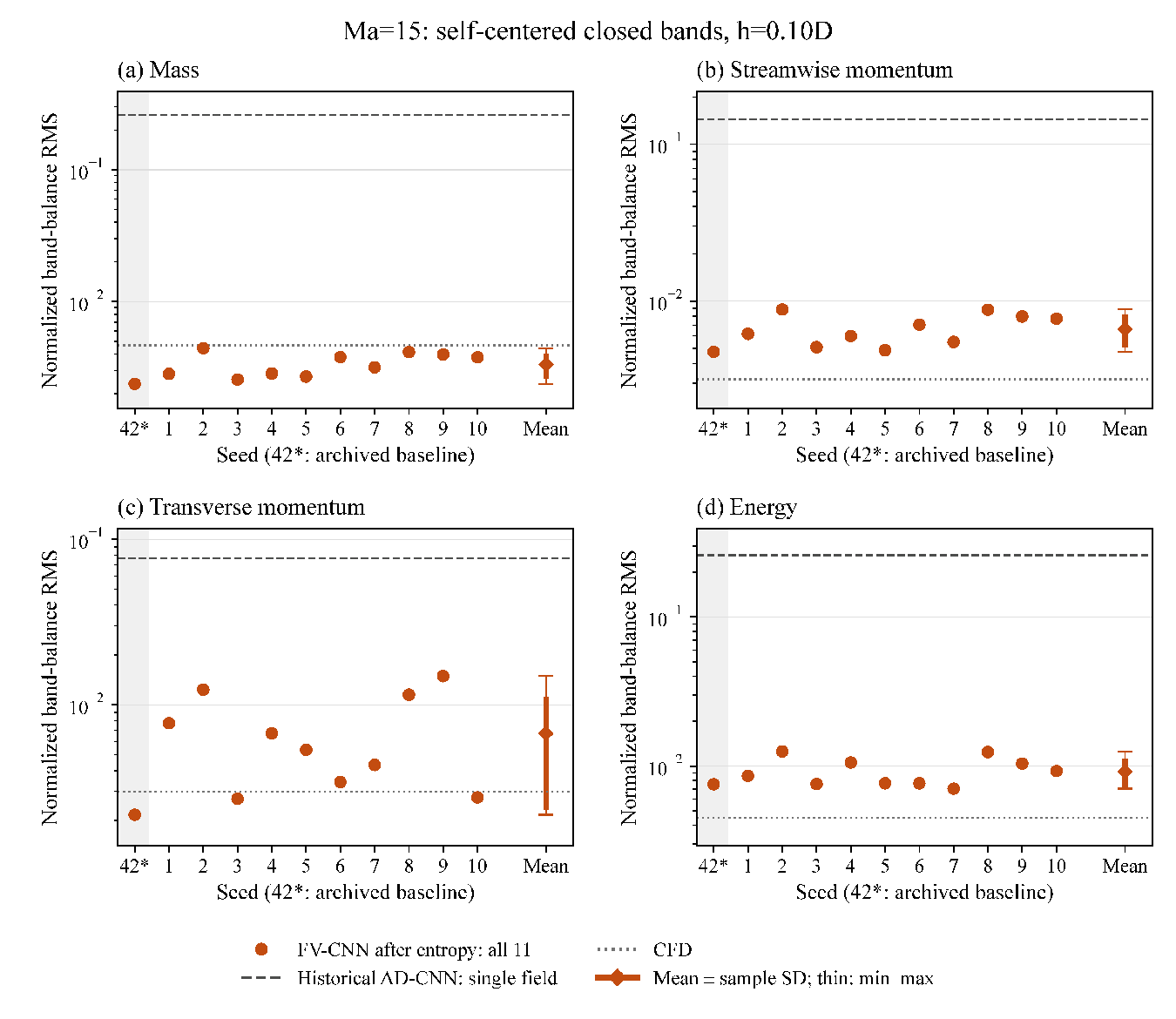}
 \caption{Self-centered closed-band balances at $h=0.10D$ for every final $\mathrm{Ma}=15$ seed. The quadrature, 15 wedges, and normalization follow Sec.~\ref{sec:res_conservation}. The vertical axes are logarithmic. Dashed: the historical AD-CNN field; dotted: CFD. All 44 FV seed--component values lie below the corresponding AD values. Thick whiskers: sample standard deviation; thin capped whiskers: minimum--maximum. Each field's control volume follows its own located front.}
 \label{fig:seed_bands}
\end{figure}
\FloatBarrier
\clearpage
\begin{figure}[!htbp]
 \centering
 \includegraphics[width=\linewidth]{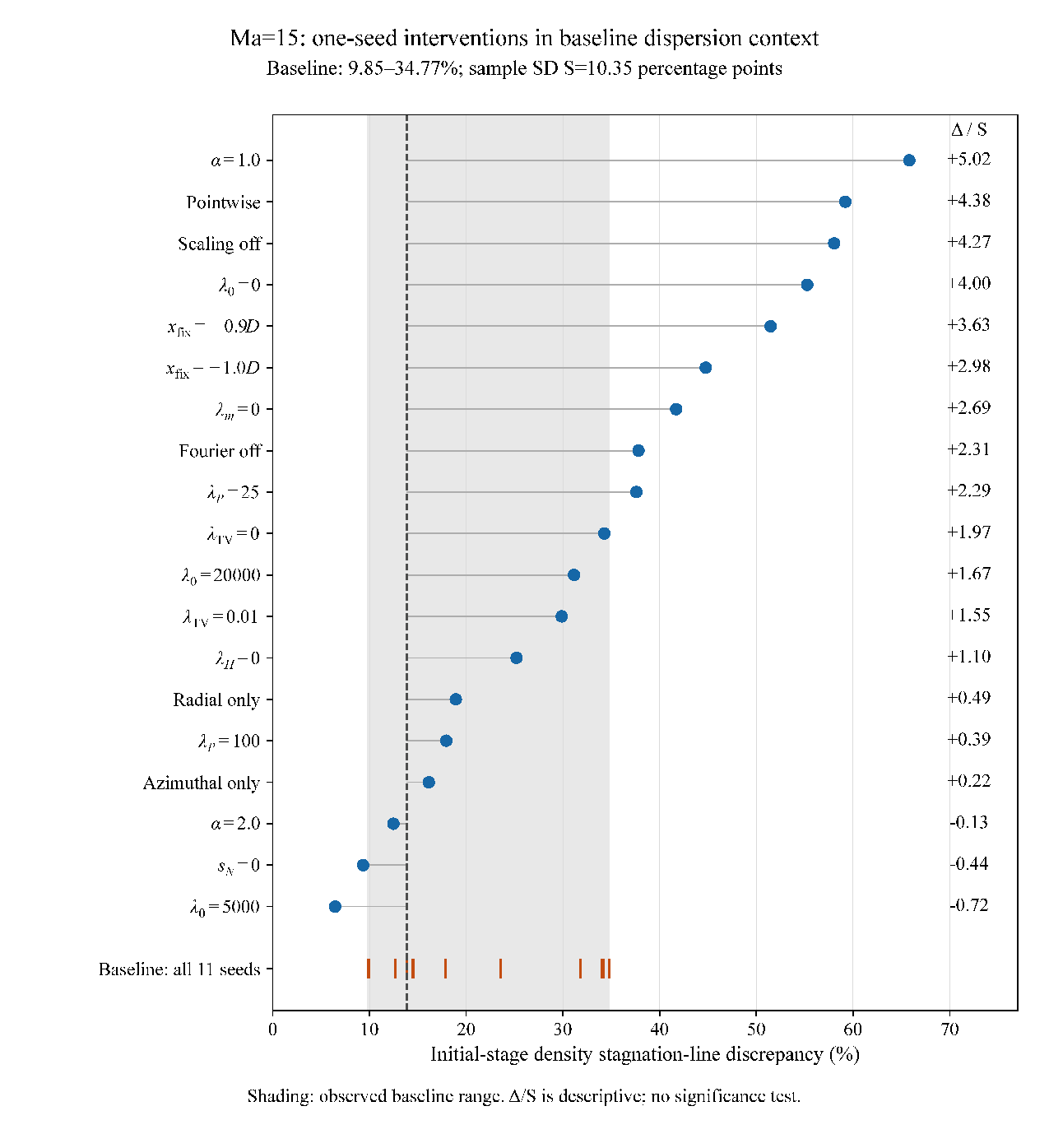}
 \caption{All 19 single-seed interventions in the context of baseline initialization dispersion. Shading spans the eleven initial-stage baseline density line norms, $9.85$--$34.77\%$; the dashed line marks seed 42 at $13.87\%$. Orange ticks retain all eleven baseline outcomes. The right-hand column gives $\Delta/S$, where $\Delta$ is the signed density line-norm change from seed 42 in percentage points and the figure\textquotesingle s $S=S_{\rm SSL}=10.35$ points is the initial-stage baseline sample standard deviation, rather than the final full-domain $S$ in Sec.~\ref{sec:seed_ensemble}. Range membership and this ratio are descriptive; neither tests the paired intervention effect or establishes that a change is execution noise.}
 \label{fig:factor_seed_context}
\end{figure}
\FloatBarrier
\clearpage
\setcounter{table}{0}
\renewcommand{\thetable}{C\arabic{table}}
\renewcommand{\theHtable}{C.\arabic{table}}
\begin{table}[!htbp]
\centering\small
\caption{Final $\mathrm{Ma}=15$ seed outcomes and fixed CFD/historical AD-CNN references. $x_{\rm axis}/D$ is the located axis front center. The density discrepancy is the area-weighted full-domain relative $L_2$ norm, reported in percent; $\mathrm{RMS}_{\rm shock}/D$ is shock-position RMS relative to CFD, and $w_{10-90}/D$ is the angular-mean density-transition width. The last four columns are normalized self-centered closed-band RMS defects at $h=0.10D$ in mass, streamwise momentum, transverse momentum, and energy (Sec.~\ref{sec:seed_band_balances}). CFD discrepancy and position RMS are zero by definition. The two reference rows are excluded from seed statistics. Seed 42 is the unchanged adopted field.}
\label{tab:seed_individual}
\setlength{\tabcolsep}{4.0pt}
\renewcommand{\arraystretch}{1.15}
\begin{tabular}{@{}lrrrrrrrr@{}}\toprule
Case/seed & $x_{\rm axis}/D$ & $100e_{2,\rho}$ & $\mathrm{RMS}_{\rm shock}/D$ & $w_{10-90}/D$ & Mass & $x$-mom. & $y$-mom. & Energy\\\midrule
CFD & $-0.70063$ & $0.00$ & $0.00000$ & $0.01084$ & $0.00465$ & $0.00319$ & $0.00298$ & $0.00449$\\
AD-CNN & $-0.69780$ & $20.69$ & $0.00552$ & $0.10620$ & $0.26043$ & $0.14389$ & $0.07687$ & $0.26035$\\
\midrule
42 & $-0.69154$ & $17.29$ & $0.02525$ & $0.02543$ & $0.00238$ & $0.00475$ & $0.00217$ & $0.00755$\\
1 & $-0.71096$ & $12.46$ & $0.00942$ & $0.02439$ & $0.00283$ & $0.00619$ & $0.00775$ & $0.00858$\\
2 & $-0.75044$ & $42.57$ & $0.07584$ & $0.02381$ & $0.00443$ & $0.00886$ & $0.01234$ & $0.01254$\\
3 & $-0.68802$ & $17.45$ & $0.02275$ & $0.02540$ & $0.00256$ & $0.00509$ & $0.00271$ & $0.00757$\\
4 & $-0.66580$ & $26.28$ & $0.02874$ & $0.02607$ & $0.00286$ & $0.00600$ & $0.00673$ & $0.01056$\\
5 & $-0.67957$ & $22.25$ & $0.03201$ & $0.02539$ & $0.00270$ & $0.00487$ & $0.00534$ & $0.00766$\\
6 & $-0.69125$ & $16.25$ & $0.02151$ & $0.02448$ & $0.00380$ & $0.00708$ & $0.00342$ & $0.00766$\\
7 & $-0.69583$ & $12.04$ & $0.00929$ & $0.02448$ & $0.00318$ & $0.00550$ & $0.00434$ & $0.00705$\\
8 & $-0.75020$ & $43.70$ & $0.08134$ & $0.02379$ & $0.00415$ & $0.00881$ & $0.01151$ & $0.01242$\\
9 & $-0.75369$ & $38.18$ & $0.05902$ & $0.02281$ & $0.00398$ & $0.00798$ & $0.01492$ & $0.01039$\\
10 & $-0.69589$ & $19.28$ & $0.02977$ & $0.02472$ & $0.00379$ & $0.00773$ & $0.00276$ & $0.00926$\\
\bottomrule\end{tabular}
\end{table}

\FloatBarrier
\end{revisionred}

\clearpage
\begingroup\color{OriginalColor}\interlinepenalty=10000

\endgroup
\end{document}